\documentclass[twocolumn,onecolappendix,trackchanges,tighten]{aastex61}
\usepackage{epsf}
\usepackage{graphicx}
\usepackage{color}
\usepackage{txfonts}
\usepackage{tabularx}
\usepackage{here}
\usepackage{float}
\usepackage{threeparttable}
\usepackage{tabu}
\usepackage{dcolumn}
\usepackage[T1]{fontenc}
\usepackage{aecompl}

\newcolumntype{Y}{>{\centering\arraybackslash}X} 
\newcolumntype{Z}{>{\raggedleft\arraybackslash}X}
\newcolumntype{W}{>{\raggedright\arraybackslash}X}

\newcommand{\oi}{[O\,{\sc i}]}
\newcommand{\oii}{[O\,{\sc ii}]}
\newcommand{\oiii}{[O\,{\sc iii}]}
\newcommand{\oiv}{[O\,{\sc iv}]}

\newcommand{\NI}{[N\,{\sc i}]}
\newcommand{\nii}{[N\,{\sc ii}]}
\newcommand{\niii}{[N\,{\sc iii}]}

\newcommand{\sii}{[S\,{\sc ii}]}
\newcommand{\siii}{[S\,{\sc iii}]}

\newcommand{\hei}{He\,{\sc i}}

\newcommand{\Neiii}{[Ne\,{\sc iii}]}

\newcommand{\ariii}{[Ar\,{\sc iii}]}
\newcommand{\ariv}{[Ar\,{\sc iv}]}

\newcommand{\cii}{C\,{\sc ii}}

\newcommand{\ha}{H$\alpha$}
\newcommand{\hb}{H$\beta$}
\newcommand{\hg}{H$\gamma$}

\usepackage{color} 
\newcommand{\kms}{km s$^{-1}$}

\newcommand{\te}{$T_{\rm e}$}
\newcommand{\Ne}{$n_{\rm e}$}

\begin{document}

\defcitealias{Ueta:2014aa}{HerPlaNS1}

\title{The \emph{Herschel} Planetary Nebula Survey
(H\lowercase{er}P\lowercase{la}NS)
\thanks{Herschel is an ESA Space Observatory with science 
instruments provided by European-led Principal Investigator
consortia and with important participation from NASA.}\\
A Comprehensive Dusty Photoionization Model of NGC\,6781}

\shorttitle{Dusty photoionization model of NGC\,6781}
\shortauthors{Otsuka et al.}

\hypersetup{linkcolor=red,citecolor=blue,filecolor=cyan,urlcolor=magenta}

\correspondingauthor{M.~Otsuka}

\email{E-mail: otsuka@asiaa.sinica.edu.tw (MO)}

\author{Masaaki Otsuka} 
 \affil{Institute of Astronomy and Astrophysics, 11F of
 Astronomy-Mathematics Building, AS/NTU. No.1, Sec. 4, Roosevelt Rd,
 Taipei 10617, Taiwan, R.O.C.}

\author{Toshiya Ueta} 
 \affil{Department of Physics and Astronomy, University of Denver, 2112
 E.\ Wesley Ave., Denver, CO 80210, USA}

\author{Peter A. M. van Hoof}
 \affil{Royal Observatory of Belgium, Ringlaan 3, 1180, Brussels, Belgium}

\author{Raghvendra Sahai}
 \affil{Jet Propulsion Laboratory, 4800 Oak Grove Drive, Pasadena, CA
 91109, USA} 

\author{Isabel Aleman}
 \affil{Instituto de Astronomia, Geof\'{\i}sica e Ci\^{e}ncias
 Atmosf\'ericas (IAG-USP), Universidade de S\~{a}o Paulo, 
 Cidade Universit\'aria, Rua do Mat\~ao 1226, S\~{a}o Paulo,
 SP, Brazil, 05508-090}

\author{Albert A. Zijlstra}
 \affil{Jodrell Bank Centre for Astrophysics, Alan Turing Building,
 University of Manchester, Manchester, M13 9PL, UK}
 \affil{Department of Physics \& Laboratory for Space Research,
 University of Hong Kong, Pok Fu Lam Road, Hong Kong}
 
\author{You-Hua Chu}
 \affil{Institute of Astronomy and Astrophysics, 11F of
 Astronomy-Mathematics Building, AS/NTU. No.1, Sec. 4, Roosevelt Rd,
 Taipei 10617, Taiwan, R.O.C.}
 
\author{Eva Villaver}
 \affil{Departamento de F\'{i}sica Te\'{o}rica, Universidad Aut\'{o}noma de
 Madrid, Cantoblanco, 28049, Madrid, Spain}
 
\author{Marcelo L. Leal-Ferreira}
 \affil{Leiden  Observatory, Universiteit Leiden, P.O. Box 9513, NL-2300 RA
 Leiden, Netherlands}

\author{Joel Kastner}
 \affil{Chester F. Carlson Center for Imaging Science and Laboratory for 
 Multiwavelength Astrophysics, Rochester Institute of Technology, 54 Lomb
 Memorial Drive, Rochester, NY, 14623, USA}

\author{Ryszard Szczerba}
 \affil{N. Copernicus Astronomical Centre Rabianska 8, 87-100 Torun, Poland}

\author{Katrina M. Exter}
 \affil{Instituut voor Sterrenkunde, Katholieke Universiteit Leuven, 
 Celestijnenlaan 200D, 3001, Leuven, Belgium}

 \begin{abstract}
  We perform a comprehensive analysis of the planetary nebula (PN) 
  NGC\,6781 to investigate the physical conditions of each of its 
  ionized, atomic, and molecular gas and dust components and the 
  object's evolution, based on panchromatic observational data 
  ranging from UV to radio. 
  Empirical nebular elemental abundances, 
  compared with theoretical predictions via nucleosynthesis models 
  of asymptotic giant branch (AGB) stars, 
  indicate that the progenitor is a solar-metallicity, 
  $2.25-3.0\,M_{\sun}$ initial-mass star. 
  We derive the best-fit distance of 0.46\,kpc by fitting the stellar 
  luminosity (as a function of the distance and effective temperature 
  of the central star) with the adopted post-AGB evolutionary tracks. 
  Our excitation energy diagram analysis indicate high excitation 
  temperatures in the photodissociation region (PDR) beyond the ionized 
  part of the nebula, suggesting extra heating by shock interactions 
  between the slow AGB wind and the fast PN wind.
  Through iterative fitting using the {Cloudy} code with 
  empirically-derived constraints, 
  we find the best-fit dusty photoionization model of the object 
  that would inclusively reproduce all of the adopted panchromatic 
  observational data. 
  The estimated total gas mass ($0.41\,M_{\sun}$) corresponds to 
  the mass ejected during the last AGB thermal pulse event predicted 
  for a $2.5\,M_{\sun}$ initial-mass star.
  A significant fraction of the total mass (about 70\,\%) is found 
  to exist in the PDR, demonstrating the critical importance of the 
  PDR in PNe that are generally recognized as the hallmark of 
  ionized/H$^{+}$ regions.
 \end{abstract}

\section{Introduction}

 
  \begin{table*}
   \footnotesize
   \centering
    \caption{
   \label{T:obslog}
   The log of panchromatic observations of NGC\,6781 adopted
   for the present study.}
   \renewcommand{\arraystretch}{0.85}
   \begin{tabularx}{\textwidth}{@{}l@{\hspace{6pt}}l@{\hspace{6pt}}l@{\hspace{6pt}}c@{\hspace{6pt}}l@{\hspace{6pt}}l@{\hspace{6pt}}l@{}}
    \hline\hline
    \multicolumn{7}{c}{Photometry Observations} \\
    Obs-Date   &Telescope     &Instrument &\multicolumn{1}{c}{Band}
     &Aperture (Nebula+CSPN) &Program-ID/PI&References\\
    \hline
    2011-07-25 &\emph{GALEX}   &GALEX &NUV &180{\arcsec}      &    \\        
    2008-07-31 &ING/INT 2.5-m&WFC &RGO $U$, Sloan $g$ and $r$&320{\arcsec}&I08AN02/P.\,Groot  \\
    2015-05-12 &ESO/NTT 3.6-m     &EFOSC2
	    &Bessel $B$, $V$, $R$ &200{\arcsec}&60.A-9700(D)/Calibration\\ 
    2009-08-09 &ING/INT 2.5-m     &WFC
	    &IPHAS {\ha}            &320{\arcsec}       &C129/J.\,Casare    \\
    1995-07-24 &\emph{HST}    &WFPC2/PC
	   &F555W, F814W (CSPN only)& &GO6119/H.E.\,Bond             \\
    2010-06-26 &UKIRT 3.8-m &WFCAM&$J$, $H$, $K$ &180{\arcsec}\\
    2010-04-13 &\emph{WISE}   &WISE &3.4, 11.6, 22.1\,{\micron}&$220{\arcsec} - 300{\arcsec}$\\
    2004-04-20 &\emph{Spitzer}&IRAC&4.5, 5.8, 8.0\,{\micron}&240{\arcsec}&68/G.\,Fazio\\
    1996-04-28 &\emph{ISO}    &ISOCAM&14.3\,{\micron}&240{\arcsec}&COX\,1/P.\,Cox   \\
    2011-10-17 &\emph{Herschel}& PACS & 70, 100, 160\,{\micron}&240{\arcsec}
		&OT1-tueta-2/T.~Ueta&(1)\\
    2011-10-11 &\emph{Herschel}   &SPIRE & 250, 350, 500\,{\micron}&240{\arcsec}
		&OT1-tueta-2/T.~Ueta&(1)\\
    &Radio telescopes   &Various & 1.4, 5, 22, 30, 43\,GHz &&&(2), (3), (4),
			(5), (6)\\
    \hline
    \multicolumn{7}{c}{Spectroscopy Observations} \\
    Obs-Date   &Telescope &Instrument &\multicolumn{1}{c}{Wavelength}
		&&Program-ID/PI &References\\
    \hline
    1997-08-09  &ING/WHT 4.2-m &ISIS &3600-8010\,{\AA}&
		&W-97B-41/X.-W.\,Liu &(7), (8) \\
    2005-10-19  &\emph{Spitzer}&IRS&5.2-39.9\,{\micron}&
		&1425/IRS-Calibration  \\
    2011-10-14  &\emph{Herschel}&PACS&51-220\,{\micron}&
		&OT1-tueta-2/T.~Ueta &(1) \\
    2012-04-01  &\emph{Herschel} &SPIRE&194-672\,{\micron}&
		&OT1-tueta-2/T.~Ueta &(1)  \\
  \hline
   \end{tabularx}
    \begin{minipage}{\textwidth}
     \tablerefs{\footnotesize
     (1) \citetalias{Ueta:2014aa}; 
     (2) \citet[376.5 $\pm$ 12\,mJy at 1.4\,GHz]{Condon:1998aa}; 
     (3) \citet[323\,mJy at 5\,GHz]{Stanghellini:2010aa};
     (4) \citet[190\,mJy at 22\,GHz]{Petrov:2007aa};
     (5) \citet[264.1 $\pm$ 7.1\,mJy at 30\,GHz]{Pazderska:2009aa};
     (6) \citet[710\,mJy at 43\,GHz]{Umana:2008aa};
     (7) \citet{Liu:2004ab};
     (8) \citet{Liu:2004aa}.
     }
    \end{minipage}
  \end{table*}


 The life cycle of matter in the Universe is intimately connected with
 the stellar evolution because stars are the most fundamental building
 blocks of the Universe. 
 Hence, the chemical evolution of galaxies has always been made possible 
 by stellar nucleosynthesis, convection/dredge-up, and ultimately, 
 stellar mass loss.
 This stellar mass loss becomes significant when stars evolve into the 
 final stage of stellar evolution, i.e.,
 the asymptotic giant branch (AGB) stage for low-mass stars
 ($1-8$\,$M_{\sun}$) and core-collapsed supernovae explosions
 for high-mass stars ($>8$\,$M_{\sun}$).

 Either way, the mass loss process would expel a significant fraction 
 of mass contained in stars as the circumstellar shells, which would
 eventually become part of the interstellar medium (ISM). 
 Besides gas, molecules and solid state particles (i.e., dust grains) 
 participate in the stellar mass loss and make up a significant part of 
 the circumstellar shells as the photodissociation region (PDR). 
 These cold components of the mass loss ejecta will provide the seed 
 material for the formation of the next generation of stars and planets. 
 Hence, understanding of stellar mass loss is important in characterizing 
 the cosmic mass recycling and chemical evolution in galaxies.

 Planetary nebulae (PNe) are low-mass stars that have completed
 mass-loss
 during the preceding AGB phase and consist of a hot central star
 ($\gtrsim 30\,000$\,K;  evolving to become a white dwarf) and an extensive
 circumstellar shell.
 While PNe are famous for their spectacular circumstellar structures
 seen via bright optical emission lines arising from the ionized gas
 component of the nebula, the ionized part of PNe is surrounded by
 the neutral gas and dust components (i.e., the PDR).
 Therefore, being relatively isolated from surrounding objects, PNe
 provide unique laboratories to further our understanding of the
 stellar evolution and the chemical evolution of galaxies, from
 high-temperature fully-ionized plasma to low-temperature dusty molecular gas.

 So far, more than 2000 PNe in the Milky Way have been identified
 \citep{Frew:2008aa,hashpn}. 
 The evolutionary history of the progenitor (the central star of a PN,
 CSPN) is imprinted in the circumstellar shells. 
 Radiation from the CSPN permeates into the circumstellar shells,
 controlling the physical conditions and local structures
 \citep[see e.g.,][]{Villaver:2002aa}.
 Moreover, PNe are in the evolutionary stage in which the circumstellar
 shells would reach their largest extent before the material at the
 periphery begins to dissipate into the ISM. 
 Therefore, by investigating spatially-extended emission from each
 of the ionized, atomic, and molecular gas and dust components, one
 can infer ionic, elemental, and molecular/dust abundances and the
 mass-loss and evolutionary histories of the CSPN.

 Because PNe are H$^{+}$ regions, there is a history of observations
 that has generated a wealth of archival data in the UV and optical. 
 Similarly, the bright ionized gas in PNe is also bright in the
 radio continuum. 
 With the advent of new technologies, PN observations in the X-ray,
 near-IR, and mid-IR follow suit. 
 Recently, a window of opportunity in the far-IR was brought forward
 by a suite of space telescopes, which filled the remaining hole in
 the spectral coverage. 
 We seized this opportunity and initiated the Herschel Planetary
 Nebula Survey \citep[HerPlaNS;][\citetalias{Ueta:2014aa},
 hereafter]{Ueta:2014aa} and its follow-up archival study, HerPlaNS$+$,
 using data collected for a hoard of PNe with the Herschel Space
 Observatory \citep[\emph{Herschel}, hereafter]{hso}.

 In our previous pilot/demonstration study, we focused on the bipolar
 PN NGC\,6781 to empirically characterize its dusty circumstellar
 nebula based mainly on far-IR data.
 We confirmed a nearly pole-on barrel structure of the dust shell
 (of $26-40$\,K, $4 \times 10^{-3}\,M_{\sun}$)
 rich in amorphous carbon via broadband mapping.
 We also determined the physical stratification of the nebular
 gas (of 0.86\,$M_{\sun}$) in terms of the electron density and
 temperature via spatially-resolved far-IR line diagnostics.
 Moreover, we yielded a gas-to-dust mass ratio map by a direct
 comparison between the empirically-derived dust and gas distributions.
 These analyses were made with the adopted distance of 0.95\,kpc.
 Assuming that all mass-loss ejecta were detected and that
 the present-day core mass were $\sim$0.6\,$M_{\sun}$, 
 we concluded that a $1.5\,M_{\sun}$ initial-mass progenitor 
 was about to complete its PN evolution.

 In the present study reported here, we continue our investigation
 of NGC\,6781 by adopting as much panchromatic data as possible
 in addition to our own HerPlaNS far-IR data.
 This time, 
 our focus is to generate a coherent model of 
 NGC\,6781 that would satisfy the adopted panchromatic data 
 as comprehensively as possible.
 To this end, we first
 derive the empirical characteristics of the central
 star and its circumstellar nebula with a greater amount of 
 self-consistently based on the adopted panchromatic data set.
 Then, we use these empirically derived quantities as more constraining 
 input parameters for a dusty photoionization model 
 consisting of ionized, atomic, and molecular gas plus dust grains
 to construct one of the most comprehensive models of the object ever 
 produced.
 In doing so, preference is given to adopting a panchromatic 
 data set rather than exploiting the spatially-resolved nature of 
 the data.
 This is also because, while the existing multi-band images of the 
 nebula certainly help us to empirically establish its 3-D structures, 
 the amount of imaging data (especially emission line maps) still 
 lacks to fit detailed 3-D models of internal stratifications in 
 the nebula.

 The organization of the rest of the paper is as follows.
 We summarize the panchromatic observational data of NGC\,6781
 adopted in the present study (\S\,\ref{S2})
 and review each of the ionized, atomic, and molecular gas 
 and dust components of the nebula and the central star to derive 
 empirical parameters that are pertinent to the subsequent
 dusty photoionization model fitting (\S\,\ref{S3}).
 Then, we present the best-fit
 dusty photoionization model of NGC\,6781 produced with {Cloudy}
 \citep{Ferland:2013aa} while emphasizing improvements made by adopting
 the panchromatic data comprehensively and self-consistently
 (\S\,\ref{S:CLOUDY}),
 before describing conclusions drawn from the empirical analyses 
 and modeling (\S\,\ref{summary}). 
 This study would demonstrate that 
 the derived best-fit model is robust enough to 
 empirically constrain theoretical stellar evolutionary predictions
 and that 
 the cold dusty PDR of PNe is at least equally important
 as the ionized part when characterizing their progenitor's
 evolution and mass-loss histories,
 especially in the context of the cosmic
 mass recycling and chemical evolution of galaxies.


\section{Adopted panchromatic data of NGC\,6781}
\label{S2}

\subsection{Photometry data}

\begin{figure*}
  \centering
  \includegraphics[width=\textwidth]{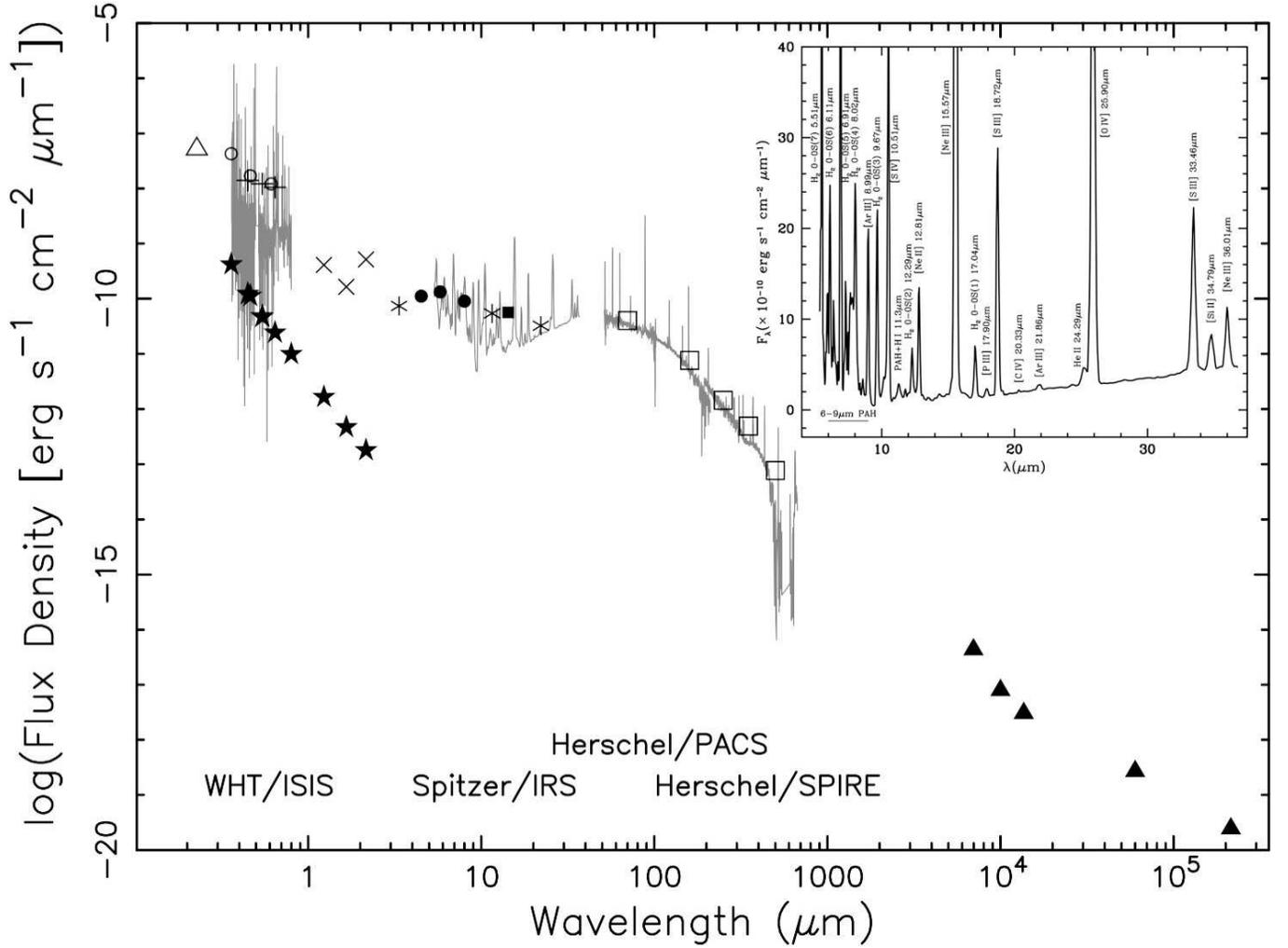}
  \caption{The panchromatic photometric and spectroscopic data
  of NGC\,6781 adopted in the present study.
  Broadband photometry was done over the entire extent of the
  nebula from the following sources:
  \emph{GALEX} (open triangle),
  ING/INT (open circles),
  ESO/NTT (pluses),
  UKIRT (crosses),
  \emph{WISE} (asterisks),
  \emph{Spitzer} (filled circles),
  \emph{ISO} (filled square),
  \emph{Herschel} (open squares), and
  Radio (filled triangles),
  while photometry of the CSPN (filled stars) was also done using
  \emph{HST}/WFPC2 images in addition to the above optical and
  near-infrared $JHK$ sources.
  Spectra (grey lines) were sourced from WHT/ISIS, \emph{Spitzer}/IRS,
  and \emph{Herschel}/PACS and SPIRE. The adopted spectra from four
  instruments are shown in grey lines, with their respective spectral
  ranges indicated at the bottom. 
  The inset displays the
  \emph{Spitzer}/IRS spectra in the mid-IR full of H$_{2}$,
  polycyclic aromatic hydrocarbons (PAHs) and
  ionized gas emission features/lines with the dust continuum
  steadily rising toward longer wavelengths from around 10$\micron$.
  See text for how the data were scaled with respect to each other.
  See also Tables\,\ref{T:phot} and \ref{T:linelist}.}
  \label{S:spt}
 \end{figure*}

 We collect photometry data -- 10 and 27 data points for the CSPN
 alone and the nebula plus the CSPN, respectively -- from previous
 observations made with various ground- and space-based telescopes
 as listed in Table\,\ref{T:obslog} and plotted in Fig.\,\ref{S:spt}. 
 We re-reduce the archived data ourselves to perform photometry 
 measurements unless science grade images are already made available. 
 The diameter of the adopted photometry aperture for the entire 
 nebula (including the CSPN) is indicated in Table\,\ref{T:obslog}. 
 For photometry of the CSPN alone, we use a circular aperture of 
 0.4$\arcsec$
 (\emph{HST}), 1.2$\arcsec$ (EFOSC2), 3.8$\arcsec$ (WFC), and
 2.2$\arcsec$ (WFCAM) centered at the CSPN.
 In Appendix\,\ref{A:PHOT}, we outline the method of data reduction 
 and photometry for the \emph{HST}/WFPC2, INT\,2.5-m/WFC, 
 ESO NTT\,3.6-m/EFOSC2, UKIRT\,3.8-m/WFCAM, and INT\,2.5-m/IPHAS 
 {\ha} broadband images.

 
\subsection{Spectroscopy data \label{S-spec}}

\begin{figure}
  \centering
  \includegraphics[width=\columnwidth]{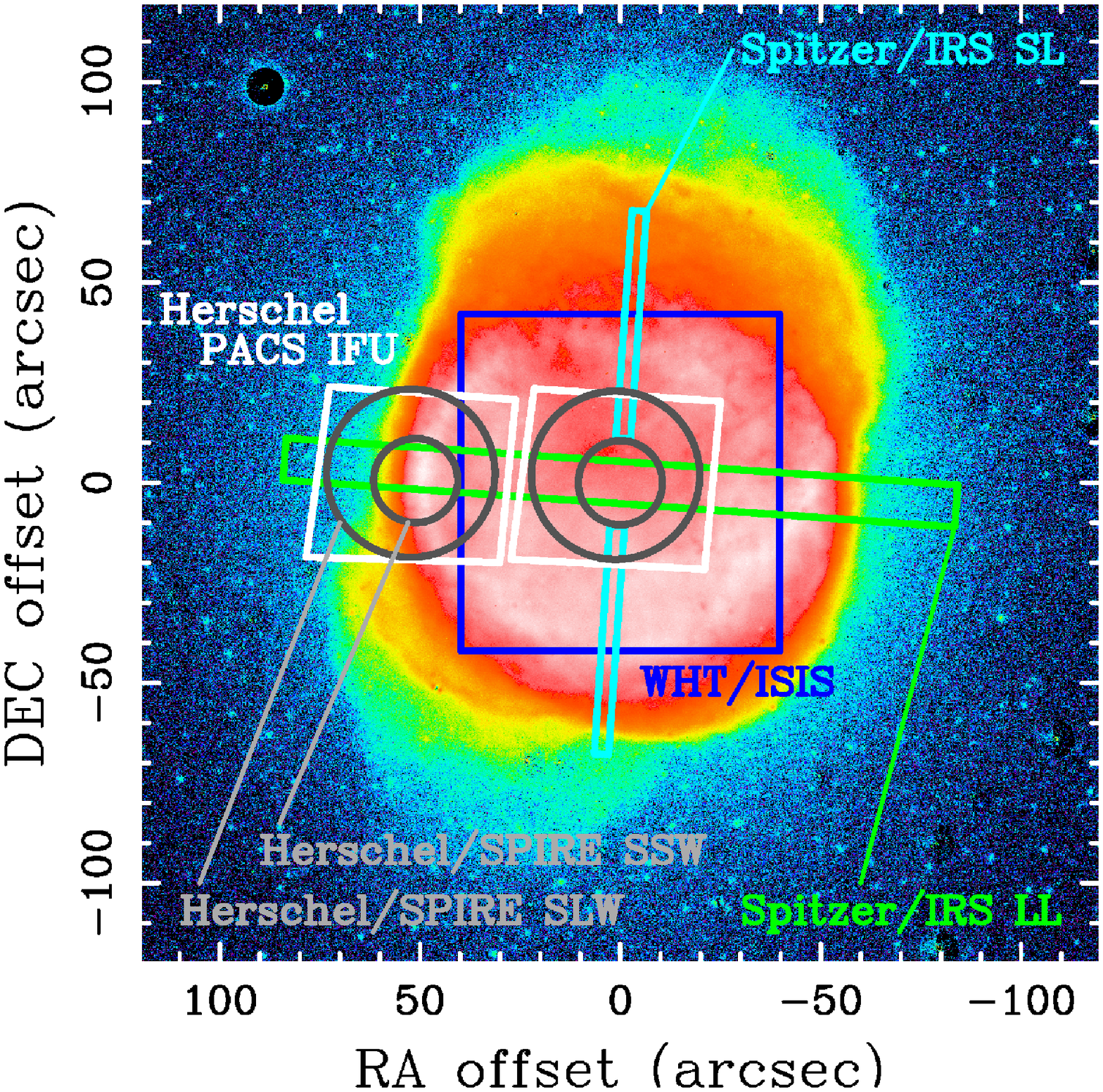}
  \caption{Relative slit positions of previous spectroscopic
  observations with respect to the NGC\,6781 nebula shown on
  the NOT/{\ha} image (previously presented by \citet{Phillips:2011aa}), 
  in which field stars are subtracted by PSF-fitting.
  N is up and E is to the left.}
  \label{S:slit} 
 \end{figure}

 We collected spectroscopy data from previous optical, mid-IR, and
 far-IR observations made as summarized in Table\,\ref{T:obslog} and
 plotted in Fig.\,\ref{S:spt}. 
 Detailed accounts of data reduction and spectroscopic measurements 
 are given in Appendix\,\ref{A:SPEC} for each instrument 
 (WHT/ISIS in the optical, \emph{Spitzer}/IRS in the mid-IR, and 
 \emph{Herschel}/PACS and SPIRE in the far-IR).
 Also given in Appendix\,\ref{A:SPEC} is a detailed description as to 
 how the {\hb} flux of the entire nebula is estimated using the 
 IPHAS {\ha} image. 
 Our choice of the data sets is motivated to ensure that the adopted 
 spectra represent the bulk of the nebula. 
 Fig.\,\ref{S:slit} shows relative slit positions with 
 respect to the entire nebula.

 
\subsubsection{Optical WHT/ISIS spectrum}

 The optical WHT/ISIS spectrum is obtained by scanning the
 nebula along declination during integrations, 
 with the position angle (P.A., defined to be degrees E of N) of 
 the $79.6{\arcsec} \times 1{\arcsec}$ slit is set at 90$^{\circ}$:
 the resulting spectrum, therefore, represents an average of the bulk 
 of the central part of the nebula (X.-W.\ Liu, private communication).
 Fig.\,\ref{S:slit} shows the
 central scanned region of $79.6{\arcsec} \times 84{\arcsec}$
 with a blue box.
 Flux densities of the WHT/ISIS spectrum are scaled to match the
 INT/WFC IPHAS {\ha} band fluxes (see Appendix\,\ref{S:INT2}).

\subsubsection{Mid-IR Spitzer/IRS spectra}
 
 The archival \emph{Spitzer}/IRS \citep{Houck:2004aa} spectra are
 obtained with the SL (5.2-14.5\,{\micron}; a pair of the vertical
 light blue $3.6{\arcsec}\times 57{\arcsec}$ slits at P.A.\ of $-10^{\circ}$
 in Fig.\,\ref{S:slit}) and LL (13.9-39.9\,{\micron}; the horizontal
 $168{\arcsec} \times 10.5{\arcsec}$ slit at P.A.\ of $86^{\circ}$
 in Fig.\,\ref{S:slit}) modules. 
 Only the SL spectrum was previously presented (\citealt{Phillips:2011aa}; \citealt{Mata:2016aa}),
 whereas we include the LL spectrum in our analysis.
 While there is only little flux density offset between the SL and
 LL spectra, we combine the two spectra by scaling the SL spectrum
 to match the LL spectrum so that the combined mid-IR spectrum 
 would represent the central part of the nebula (Fig.\,\ref{S:spt}). 
 Flux densities of
 the combined mid-IR spectrum are then scaled using the results of
 mid-IR photometry (see Appendix\,\ref{S:Spitzer}).

\subsubsection{Far-IR Herschel/PACS and SPIRE spectra \label{S-obs-herschel}}

 Far-IR \emph{Herschel} spectra of the nebula for the present study
 are adopted from those previously presented \citepalias{Ueta:2014aa}. 
 To define a far-IR spectrum representing the bulk of the nebula 
 we combine spectra from all PACS IFU spaxels
 ($5 \times 5$ in the $50{\arcsec} \times 50{\arcsec}$ apertures),
 while a single spectrum from the central bolometer of the SPIRE
 array is included (of $21{\arcsec}$ and $42{\arcsec}$
 diameter in the short and long wavelength band, respectively; 
 at both the ``center'' and ``rim'' positions as depicted as 
 white boxes and gray circles, respectively, in Fig.\,\ref{S:slit}). 
 The combined far-IR spectra are then scaled using the flux density 
 ratios between far-IR lines and {\hb} for the entire nebula with 
 the synthesized {\hb} map constructed from the {\ha} image 
 (see Appendix\,\ref{S:Herschel}).

\subsubsection{Interstellar reddening correction and Flux measurements}
 
 Once we reconstruct spectra in the optical, mid-IR, and 
 far-IR to represent the bulk of the nebula, 
 we measure line fluxes by Gaussian fitting. 
 For the ISIS spectrum, the line fluxes are corrected for the
 interstellar extinction with the following formula:
 \begin{equation}
 I(\lambda) = F(\lambda)\,\cdot\,10^{c({\rm H\beta})(1 + f(\lambda))},
  \label{eq1}
 \end{equation}
 where 
 $I$($\lambda$) is the de-reddened line flux, 
 $F$($\lambda$) is the observed line flux, 
 $c$({\hb}) is the reddening coefficient at {\hb}, and
 $f$($\lambda$) is the interstellar extinction function at $\lambda$
 computed by the reddening law of \citet{Cardelli:1989aa} with $R_{V} = 3.1$.

 We measure the reddening correction factor $c$({\hb}) by comparing 
 the observed Balmer line ratios of {\hg} and {\ha} to {\hb} with the
 theoretical ratios given by \citet{Storey:1995aa} for an electron
 temperature {\te}$ = 10\,000$\,K and an electron density {\Ne}$ =
 200$\,cm$^{-3}$ under the assumption that
 the nebula is optically thick to Ly$\alpha$
 (so called ``Case B''; e.g.\ see \citealt{Baker:1938aa};
 also see \S\,\ref{S-plasma} for the bases
 of these {\Ne} and {\te} values).
 The measured $c$({\hb}) turns out to be $0.951 \pm 0.091$ from
 the $F$({\hg})/$F$({\hb}) ratio and $1.014 \pm 0.033$ from the
 $F$({\ha})/$F$({\hb}) ratio. 
 Thus, we adopt $c$({\hb}) of
 $1.007 \pm 0.031$, which is a weighted-mean of the above values.
 We do not correct for the interstellar extinction at longer 
 wavelengths than $K$-band because extinction would be negligible 
 at those wavelengths. 
 The final de-reddening
 line fluxes measured in the adopted spectra are listed in
 Table\,\ref{T:linelist}. 
 The quoted fluxes are normalized with
 respect to $I$({\hb}) = 100.

 While we adopt these reprocessed 1-D panchromatic spectra
 and duly-measured de-reddened line fluxes 
 as representative of the bulk of the nebula, 
 a word of caution appears appropriate at this point.
 As Fig.\,\ref{S:slit} shows, the spatial coverage of the nebula
 by various spectroscopic apertures is not complete and uniform. 
 As would become apparent later from the model fitting (\S\,\ref{S:CLOUDY}),
 there would be some inconsistencies in line emission strengths,
 especially in neutral and low-excitation lines such as
 [N\,{\sc i}], {\oi}, and [S\,{\sc ii}]
 (see \S\,\ref{S-ionic}).
 This is primarily because the highest surface brightness regions
 (the E and W end of the central ring structure; Fig.\,\ref{S:slit})
 are missed in the optical data and 
 may be less strongly weighted than they should be in the far-IR data.
 We would return to these issues 
 when we discuss model fitting in \S\,\ref{S:CLOUDY}.

\section{Anatomy of NGC\,6781}
\label{S3}

\subsection{The ionized/neutral gas component \label{S-3.1}}

\subsubsection{Plasma diagnostics \label{S-plasma}}

 We determine the {\Ne} and {\te} pairs for the ionized/neutral gas
 component\footnote{Strictly speaking, we expect two kinds of ionized 
 (ionized atomic and ionized molecular) gas, and two kinds of neutral 
 (atomic and molecular) gas. In the present study, however, 
 we almost exclusively mean ionized atomic gas when we refer to 
 ionized gas and neutral atomic gas when we refer to atomic gas.} 
 of NGC\,6781 for a few temperature/ionization regions based
 on various collisionally-excited lines (CELs) and recombination lines
 (RLs) detected in the adopted panchromatic spectra. 
 In the present plasma diagnostics and the subsequent ionic abundance 
 derivations, we adopt the effective recombination coefficients, transition
 probabilities, and effective collisional strengths listed
 in their Tables\,7 and 11 of \citet{Otsuka:2010aa}, in which all
 the original references to all the atomic data are found.
 The diagnostic line ratios used in the present analysis and the
 resultant {\Ne} and {\te} values are summarized in
 Table\,\ref{T:diagno_table}.

 
 \begin{figure}
  \includegraphics[width=\columnwidth]{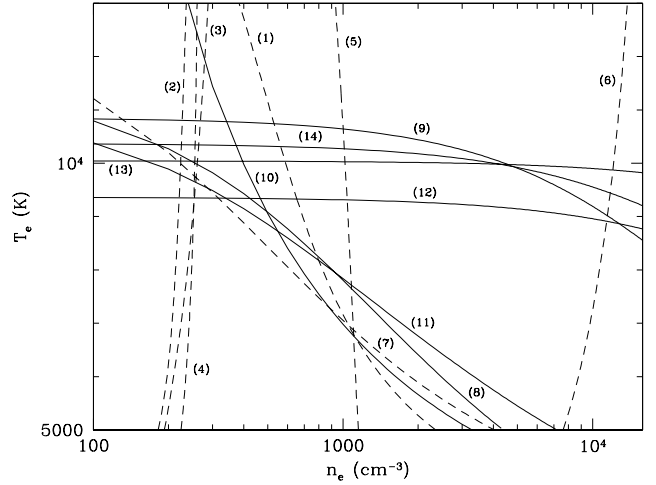}
  \caption{\label{F:diagno}
  The {\Ne}-{\te} diagram based on CEL diagnostic lines. 
  The dashed and solid
  lines are the {\Ne}
  and {\te} indicators, respectively.
  The ID numbers indicate the corresponding line ratios
  listed in Table\,\ref{T:diagno_table}.}
 \end{figure}


 
 \begin{table}
  \centering
  \footnotesize
  \renewcommand{\arraystretch}{0.85}
  \caption{\label{T:diagno_table}
  Summary of plasma diagnostics using nebular lines.}
  \begin{tabularx}{\columnwidth}{@{}l@{\hspace{3pt}}l@{\hspace{2pt}}l
  @{\hspace{2pt}}D{p}{\pm}{-1}@{\hspace{0pt}}D{p}{\pm}{-1}@{}}
   \hline\hline
   ID &Ion& {\Ne}-diagnostics   & \multicolumn{1}{c}{Ratio}
   & \multicolumn{1}{c}{Result (cm$^{-3}$)}    \\ 
   \hline
   1 & {\oi} &$I$(63\,{\micron})/$I$(146\,{\micron})         &
	       11.423~p~2.039 &\multicolumn{1}{c}{590$^{+1190}$}      \\ 
   2 & {\sii}& $I$(6716\,{\AA})/$I$(6731\,{\AA})
	   & 1.201~p~0.048 & 230~p~60  \\ 
   3 & {\oii}& $I$(3726\,{\AA})/$I$(3729\,{\AA})
	   & 0.848~p~0.035 & 270~p~50  \\ 
   4 & {\nii}& $I$(122\,{\micron})/$I$(205\,{\micron})
	   & 4.902~p~0.991 & 280~p~120   \\ 
   5 & {\siii}& $I$(18.7\,{\micron})/$I$(33.5\,{\micron})
	   & 0.939~p~0.092 & 1020~p~300   \\ 
   6 & {\Neiii}& $I$(15.6\,{\micron})/$I$(36.0\,{\micron})
	   & 13.789~p~1.471 & 12\,600~p~7590   \\ 
   7 & {\oiii}& $I$(4959\,{\AA})/$I$(88.3\,{\micron})
	   & 1.438~p~0.178 & 220~p~50 \\ 
   \hline
   ID &Ion &{\te}-diagnostics   & \multicolumn{1}{c}{Ratio}
	       & \multicolumn{1}{c}{Result (K)}   \\ 
   \hline
   8 & {\sii}&  $I$(6716/31\,{\AA})/$I$(4069\,{\AA}) & 14.891~p~3.270
	       & 10\,520~p~1820  \\ 
   9 & {\nii}&  $I$(6548/83\,{\AA})/$I$(5755\,{\AA}) & 81.931~p~2.956
	       & 10\,800~p~170  \\ 
   10 & {\nii}&  $I$(6548/83\,{\AA})/ & 57.325~p~6.201 & 12\,360~p~980   \\ 
     &       & $I$(122\,{\micron}+205\,{\micron})\\
   11 & {\oii}&  $I$(3726/29\,{\AA})/$I$(7320/30\,{\AA}) &
	       50.262~p~1.949
	       & 9650~p~200  \\ 
   12 & {\ariii}&  $I$(7135\,{\AA}+7751\,{\AA})/$I$(9.0\,{\micron})
	   & 1.211~p~0.098 & 9350~p~400   \\ 
   13 & {\oiii} & $I$(4959\,{\AA})/$I$(4363\,{\AA}) & 52.943~p~3.584
	       & 10\,050~p~210 \\ 
   14 & {\Neiii} &  $I$(3868\,{\AA}+3967\,{\AA})/$I$(36.0\,{\micron})
	   & 9.578~p~0.806 & 10\,340~p~250   \\ 
   & {\hei} &$I$(7281\,{\AA})/$I$(6678\,{\AA}) & 0.156~p~0.021 & 7070~p~1880\\
   \hline
  \end{tabularx}
 \end{table}
 

 The {\Ne}-{\te} plot shown in Fig.\,\ref{F:diagno} summarizes how {\Ne}
 and {\te} relate to each other in the regions of the nebula, from which
 the particular CELs involved in the diagnostic line ratios would arise:
 the solid lines are the {\Ne}-{\te} curves derived from the
 {\te}-sensitive ratios, while the dashed lines are those from the
 {\Ne} sensitive-line ratios.
 Strictly speaking, the diagnostic lines labeled as (1), (7), (8), (10) and (11)
 in Fig.\,\ref{F:diagno} are sensitive to both {\Ne} and {\te}. 
 In the present work, however,
 we used the lines (1) and (7)
 as {\Ne} indicators and (8), (10)\footnote{
 One might think
 that the {\nii} $I$(6548/83\,{\AA})/$I$(122\,{\micron}+205\,{\micron})
 ratio is sensitive to {\Ne}, compared with the diagnostic labeled with
 IDs (8) and (11). For that case, we calculate {\Ne} = $400 \pm
 50$\,cm$^{-3}$ at {\te} = 10$^{4}$\,K using this {\nii} diagnostic
 ratio. Adopting this {\Ne} for the following analyses does not
 change our conclusions.},
 and (11) as {\te} indicators, respectively. By doing
 so, we estimated {\te}({\oiii}), {\te}({\oii}),
 {\te}({\nii}), and {\te}({\sii}) by adopting {\Ne}({\oiii}),
 {\Ne}{\oii}, {Ne}({\nii}), and {\Ne}({\sii}). 
 Since we could not de-blend
 {\NI}\,5198/5200\,{\AA} (its ratio is a density indicator for
 the neutral region), we used the far-IR {\oi} ratio.

 \citet{Liu:2004aa} reported five {\Ne} and four {\te} values based
 on the CELs seen in the ISIS spectra augmented by lines detected in
 the \emph{ISO} spectra (see their Table\,7). 
 Taking advantage of the
 fine-structure lines detected at higher sensitivity and better spatial
 resolution in the \emph{Spitzer} and \emph{Herschel} spectra, we
 calculate seven {\Ne} and eight {\te} values. 
 Our values of the CEL
 {\Ne} and {\te} are generally consistent with those determined by
 \citet{Liu:2004aa}.

 The {\Ne}-{\te} diagnostic diagram (Fig.\,\ref{F:diagno}) suggest that
 the bulk of the ionized gas appears to have {\te} between
 $\sim$6\,000\,K and $\sim$12\,000\,K. 
 Thus, we adopt a constant
 {\te}\,$= 10\,000$\,K to derive {\Ne} values.
 The derived {\Ne}({\Neiii}) value is more than one order of magnitude
 larger than the other {\Ne} values. 
 To double-check the above, we analyze the \emph{Spitzer}/IRS spectra of
 the nebula nearby the central star obtained with the
 higher-dispersion SH and LH modules
 (of the
 $4.7{\arcsec}\times11.3{\arcsec}$
 and
 $11.1{\arcsec}\times22.3{\arcsec}$ slit dimensions, respectively;
 not shown in Fig.\,\ref{S:slit}).
 From the SH and LH spectra alone, we derive
 \mbox{{\Ne}({\Neiii}) = $4930 \pm 2780$\,cm$^{-3}$} and 
 \mbox{{\Ne}({\siii}) = $1240 \pm 60$\,cm$^{-3}$}. 
 Because the spatial coverage of the SH and LH modules is very 
 restrictive around the central star, the higher 
 {\Ne} and {\te} values may be influenced heavily by the conditions
 in the vicinity of the central star. 
 Previously, the {\oiii} 52/88\,{\micron} ratio in the central part
 of the cavity yielded 350\,cm$^{-3}$. 

 Next, we calculate {\te} based on the derived {\Ne} values.
 The average of {\Ne} = 260\,cm$^{-3}$ among {\Ne}({\sii}, {\oii},
 {\nii}) is adopted to calculate {\te}({\sii}) and {\te}({\nii}) (ID: 10). 
 To compute {\te}({\ariii} and {\Neiii}), {\Ne}({\oiii}) of
 220\,cm$^{-3}$ is adopted.
 To calculate {\te}({\oiii}), 
 {\te}({\oii}), and {\te}({\nii}) (ID: 9) accurately, we subtract
 contributions from O$^{3+}$, O$^{2+}$, and N$^{2+}$ RLs to 
 the {\oiii}\,4363\,{\AA}, {\oii}\,7320/30\,{\AA}, and
 {\nii}\,5755\,{\AA} lines, respectively, i.e.,
 $I_{\rm R}$({\oiii}\,4363\,{\AA}),
 $I_{\rm R}$({\oii}\,7320/30\,{\AA}), and 
 $I_{\rm R}$({\nii}\,5755\,{\AA}).

 We calculate $I_{\rm R}$(\oiii\,4363\,{\AA}) with 
 \begin{equation}
 \frac{ I_{\mathrm{R}}([\mathrm{O}\,\mbox{{\scriptsize
  III}}]\,4363\,{\AA}) }{I(\mathrm{H}\beta)}
  = 12.4\left(\frac{T_{\mathrm{e}}}{10^{4}}\right)^{0.59}
  \frac{\mathrm{O}^{3+}}{\mathrm{H}^{+}},
 \end{equation}
 \citep[eqn\,3,][]{Liu:2000aa} for which the
 \mbox{O$^{3+}$/H$^{+}$} ratio (3.02(--5), see \S\,\ref{S-ionic}) is
 computed using the \mbox{$I$({\oiv}\,25.9\,{\micron})/$I$({\hb})} ratio
 assuming  {\te}({\Neiii}) and {\Ne}({\oiii}). 
 In the end,
 $I_{\rm R}$({\oiii}\,4363\,{\AA}) turns out to be
 0.73\,$\%$ of the observed $I$({\oiii}\,4363\,{\AA}).
 After we subtract $I_{\rm R}$({\oiii}\,4363\,{\AA})
 from the observed $I$({\oiii}\,4363\,{\AA}),
 we obtain {\te}({\oiii}) by adopting {\Ne}({\oiii}).

 \mbox{$I_{\rm R}$({\oii}\,7320/30\,{\AA})} is calculated with
 \begin{equation}
  \frac{I_{\rm R}({\rm [O\,\mbox{\scriptsize II}]\,7320/30\,{\AA}})}
   {I({\rm H\beta})} 
   = 9.36\left(\frac{T_{\rm e}}{10^{4}}\right)^{0.44}\frac{\rm O^{2+}}
   {\rm H^{+}},
 \end{equation}
 \citep[eqn\,2,][]{Liu:2000aa} 
 where we adopt the
 \mbox{O$^{2+}$/H$^{+}$} ratio ($2.78(-4)$, see \S\,\ref{S-ionic})
 derived from the \mbox{$I$({\oiii}\,88.3\,{\micron})/$I$({\hb})}
 ratio assuming {\te}({\oiii}) and {\Ne}({\oiii}). 
 \mbox{$I_{\rm R}$({\oii}\,7320/30\,{\AA})} turns out to be
 2.19\,$\%$ of the observed $I$({\oii} 7320/30\,{\AA}).
 After we subtract the recombination contribution from
 the observed $I$({\oii} 7320/30\,{\AA}),
 we obtain {\te}({\oii}) by adopting \mbox{{\Ne} = 260\,cm$^{-3}$}.

 Finally, we estimate $I_{\rm R}$({\nii}\,5755\,{\AA}) using 
 \begin{equation}
  \frac{I_{\rm R}({\rm [N\,{\scriptsize II}]\,5755\,{\AA}})}
   {I({\rm H\beta})} = 
   3.19\left(\frac{T_{\rm e}}{10^{4}}\right)^{0.33}
   \frac{\rm N^{2+}}{\rm H^{+}},
 \end{equation}
\citep[eqn\,1,][]{Liu:2000aa}
 where the
 \mbox{N$^{2+}$/H$^{+}$} ratio (7.01(--5), see \S\,\ref{S-ionic}) was
 calculated using the \mbox{$I$({\niii}\,57\,{\micron})/$I$({\hb})} ratio 
 assuming {\te}({\oiii}) and {\Ne}({\oiii}).
 $I_{\rm R}$({\nii}\,5755\,{\AA}) is 0.54\,$\%$ of the observed
 $I$({\nii}\,5755\,{\AA}). 
 After we subtract
 $I_{\rm R}$({\nii}\,5755\,{\AA}),
 we obtain {\te}({\nii}) (ID: 9) by adopting {\Ne} = 260\,cm$^{-3}$.
 In \S\,\ref{S:CLOUDY} below, 
 we verify the above estimates of the RL contributions 
 based on the best-fit modeling results.

 We also determine {\te}({\hei}),
 which is necessary to estimate He$^{+}$ and He$^{2+}$ abundances,
 using the {\hei} \mbox{$I$(7281\,{\AA})/$I$(6678\,{\AA})}
 ratio with the He\,{\sc i} recombination coefficients 
 in the case of {\Ne} = 100\,cm$^{-3}$
 given by \citet{Benjamin:1999aa}.
 The {\Ne} and {\te} pairs derived and adopted from 
 the present plasma diagnostics are summarized in 
 Appendix Table\,\ref{T:teane}.

\subsubsection{Ionic abundance derivations \label{S-ionic}}

 We calculate CEL ionic abundances by solving the equation of
 population at multiple energy levels with the adopted {\Ne} and {\te}
 (Appendix Table\,\ref{T:teane},
 which also lists the adopted 
 {\Ne} and {\te} to calculate RL He$^{+,2+}$ and C$^{2+}$ abundances):
 the resulting ionic abundances are listed in Appendix Table\,\ref{T:ionic}. 
 We give the 1-$\sigma$ uncertainty for each ionic abundance estimate, 
 which is propagated from 1-$\sigma$ uncertainties of line fluxes, 
 $c$({\hb}), {\Ne}, and {\te}. 
 Ionic abundances are derived for each of the detected line intensities 
 when more than one lines for a particular target ion is detected. 
 In such cases, we adopt the weighted-average of all of the derived 
 abundances listed at the last line for that particular ion
 in \textit{Italics}.

 The resulting ionic abundances based on different lines in the optical
 nebular, auroral, and trans-auroral transitions to IR fine-structure lines
 turn out to be generally consistent with each other within the
 associated uncertainties for most of the cases. 
 This indicates 
 that our choice of the {\Ne}-{\te} pair for each ionic species is
 robust and that the adopted scaling of the mid- and far-IR line fluxes
 to the optical {\hb} line flux via the adopted photometry data is reasonable.
 However, there are a few exceptions, which we briefly discuss below.

 As pointed out above, 
 the spatial coverage of the nebula in spectroscopic observations is 
 not complete and uniform: 
 especially, the ISIS spatial coverage in the optical missed the
 brightest E and W ``rim'' regions, 
 in which low-excitation and neutral
 lines are particularly strong (Fig.\,\ref{S:slit}). 
 This explains why 
 the O$^{0}$ abundances derived from optical lines are much
 smaller than the abundance based on the {\oi}\,145\,{\micron} line 
 (by a factor of $7.5 \pm 4.8$).
 Hence, if we were to assume 
 O$^{0}$/H$^{+}$ = $(5.38 \pm 1.05)(-4)$
 based solely on the {\oi}\,145\,{\micron} line,
 we would have 
 N$^{0}$/H$^{+}$ = $(3.69 \pm 2.34)(-4)$
 and 
 S$^{+}$/H$^{+}$ = $(8.97 \pm 6.09)(-6)$ by adopting a factor of
 $7.5 \pm 4.8$.
 Nevertheless, for the O$^{0}$/H$^{+}$ abundance we adopt the average of 
 the observed two optical (6300 and 6364\,{\AA}) and single far-IR
 (145\,$\micron$) lines,
 because there is no way to ascertain how much line flux is missed 
 by incomplete spatial coverage of the nebula.

 To determine the He$^{+}$ abundance, we do not include the
 {\hei}\,4712\,{\AA} line because the blue wing of this line 
 seems to be contaminated by the {\ariv}\,4711\,{\AA} line. 
 Assuming that He$^{+}$ is indeed $1.08$($-1$), 
 $I$({\hei}\,4712\,{\AA}) and $I$({\ariv}\,4711\,{\AA})
 have to be \mbox{$0.47 \pm 0.18$} and
 \mbox{$0.87 \pm 0.27$}, respectively\footnote{%
 Our best fit model using {Cloudy} predicts $I$({\hei}\,4712\,{\AA}) = 0.600
 and $I$({\ariv}\,4711\,{\AA}) = 0.982. See \S\,\ref{S:CLOUDY}}.
 The Ar$^{3+}$ abundance derived from this expected
 $I$({\ariv}\,4711\,{\AA}) is \mbox{$1.99(-7) \pm 6.41(-8)$},
 which is consistent with the Ar$^{3+}$ abundance derived
 from $I$({\ariv}\,4740\,{\AA}).

 To derive the RL C$^{2+}$ abundance, 
 we use the C\,{\sc ii}\,4267\,{\AA} line 
 with its effective recombination coefficient 
 in Case\,B for \mbox{{\Ne} = 10$^{4}$\,cm$^{-3}$}
 defined as a polynomial function of {\te} 
 by \citet{2000A&AS..142...85D}.
 This is justified because 
 while the effective recombination coefficient is
 not available for the case of {\Ne} = 100\,cm$^{-2}$
 that is more appropriate here, 
 the RL abundances are in general insensitive to {\Ne}
 for \mbox{$\lesssim$\,10$^{8}$\,cm$^{-3}$}.
 As for {\te}, we adopt {\te}({\ariii}) 
 because the ionization potential (I.P.)
 of C$^{2+}$ is similar to that of Ar$^{2+}$.

 Overall, 
 we conclude that our derived ionic abundances are improved 
 with new CEL detections in the mid- and far-IR 
 (such as Ne$^{+,2+}$, S$^{2+}$, Si$^{+}$, Cl$^{3+}$, and Ar$^{2+}$)
 made with \emph{Spitzer} and \emph{Herschel} observations, 
 more robust adaptation of {\Ne} and {\te} for targeting ions, 
 and the use of a larger number of lines in various ionization stages, 
 compared with those calculated previously 
 by \citet{Liu:2004ab}.

\subsubsection{\label{S:Abund}%
 Elemental abundance}

 By introducing the ionization correction factor (ICF; 
 see, e.g., \citealt{Delgado-Inglada:2014ab} for more detail),
 we infer the nebular abundances of the observed nine elements
 in the ionized part of the nebula
 based on their observed ionic abundances. 
 In Appendix Table\,\ref{T:ionic},
 the ICF(X) value of the element ``X'' and the resulting elemental abundance, 
 \mbox{X/H = ICF(X)\,$\times (\Sigma_{\rm m=1}$X$^{\rm m+}$/H$^{+})$}, 
 are listed \textbf{in bold} at the last line for each element.
 Here, we exclude C$^{+}$, N$^{0}$, and O$^{0}$ from abundance
 calculations for the respective elements, 
 as these ions are considered to be present mostly in the PDR 
 surrounding the ionized part of the nebula.
 In Table\,\ref{T:abund2}, we compare the derived elemental abundances 
 $\epsilon$(X) corresponding to $\log_{10}(\mathrm{X/H}) + 12$, 
 where $\log_{10}({\rm H}) = 12$ (in column 2) and the relative Solar 
 abundances (X/H; in column 3).

 We perform an ionization correction using the ICF based on the I.P.\
 of the element in question, except for He, O, Ne, and S 
 (i.e., ICF for these four elements is taken to be unity 
 because unobserved high excitation lines are considered negligible).
 We will compare these ICFs based on the I.P.\
 and the predicted ICFs by the best-fit modeling in \S\,\ref{S:CLOUDY}.

 In performing ionization correction,
 the ICF for N, Si, Cl, and Ar is set as follows.
 We assume that the N abundance is the sum of N$^{+,2+,3+}$,
 and adopt ICF(N) $\approx$ ICF(O), which is equal to the
 ${\rm O}/({\rm O}^{+}+{\rm O}^{2+}$) ratio.
 Similarly, we assume that the Si abundance is the sum of
 Si$^{+,2+,3+}$, and adopt ICF(Si) $\approx$ ICF(S),
 which corresponds to the S/S$^{+}$ ratio.
 For Cl and Ar, we assume that the Cl and Ar abundances are
 the sum of Cl$^{+,2+,3+}$ and Ar$^{+,2+,3+}$, respectively. 
 Then, we adopt ICF(Cl) $\approx$ ICF(Ar) $\approx$ ICF(S), which
 corresponds to the ${\rm S}/({\rm S}^{2+} + {\rm S}^{3+}$) ratio.

 As for the ICF(C), we originally adopt ICF(C) $\approx$ ICF(N)
 corresponding to the \mbox{N/N$^{2+}$} ratio.
 With this ICF(C), the derived RL C abundance using the
 RL {\cii} 4267\,{\AA} line
 would come out to be \mbox{$4.06(-3) \pm 1.19(-3)$}.
 Note that we do not include the CEL C$^{+}$ abundance for the
 elemental C abundance because 
 (1) the [C\,{\sc ii}]\,157\,{\micron} line arises mostly from the PDR
 as stated above and 
 (2) the nature of these lines is different (C$^{2+}$ is of RL while
 C$^{+}$ is of CEL). 
 
 However, this RL C abundance would be extremely unlikely for NGC\,6781.
 The average abundance between [Cl/H] and [Ar/H] derived for
 NGC\,6781 suggests that the metallicity ($Z$) of the object is close to 
 the solar metallicity (see also \S\,\ref{abundmass}).
 Then, such a high RL C abundance is very difficult to explain 
 by current AGB nucleosynthesis models \citep[e.g.,][]{Karakas:2010aa} 
 for stars with the solar metallicity ($Z \sim 0.02$, $Z_{\sun}$). 
 Hence, the derived RL C abundance
 of \mbox{$4.06(-3) \pm 1.19(-3)$} appears to be overestimated.

 It is known that C, N, O, and Ne ionic abundances derived from RLs
 are sometimes found to be larger than the corresponding abundances 
 obtained from CELs in PNe and H\,{\sc ii} regions.
 This issue is known as the abundance discrepancy problem.
 (see, e.g., \citet{Liu:2006aa}, for more detail).
 In spite of a number of attempts to explain such ionic/elemental
 abundance discrepancies, no consensus has been reached yet.
 Thus, we need other options to estimate the C abundance in
 light of the abundance discrepancy problem. 
 One option is to compute the {\it expected} CEL C abundance 
 by scaling the {\it measured} RL C abundance with the average
 C$^{2+}$(RL)/C$^{2+}$(CEL) ratio
 because no UV spectrum is available for NGC\,6781.
  
 Previously, \citet{Delgado-Inglada:2014aa} showed general 
 agreement between {\it measured} and {\it scaled} CEL abundances,
 the latter of which was scaled from {\it measured} RL abundances
 with the average C$^{2+}$(RL)/C$^{2+}$(CEL) ratio
 of \mbox{$4.41 \pm 0.81$} among 37 Galactic PNe
 (their Table 5).
 While it is yet unknown whether there is a correlation between
 the RL and CEL C abundances, the relatively small standard 
 deviation of the measured ratios would indicate that this option
 has some merit.
 Because there are no other alternatives, 
 we adopt this option for the present study 
 and use the average C$^{2+}$(RL)/C$^{2+}$(CEL) 
 ratio of \mbox{$4.10 \pm 0.49$} found among 58 PNe 
 in Milky Way and Magellanic Clouds \citep{Otsuka:2011aa}
 to obtain 
 the scaled {\it expected} CEL C of \mbox{$9.89(-4) \pm 3.14(-4)$}.
 
 This {\it expected} CEL C of \mbox{$9.89(-4) \pm 3.14(-4)$}
 ($\epsilon({\rm C}) = 9.00$)
 would be more reasonable than
 the measured RL C abundance of $4.06(-3) \pm 1.19(-3)$
 with respect to current AGB nucleosynthesis models for
 the solar abundance stars \citep[e.g.,][]{Karakas:2010aa}. 
 In addition,
 \citet{Delgado-Inglada:2014aa} reported a
 C$^{2+}$(RL)/C$^{2+}$(CEL) ratio of 3.63 for NGC\,6720,
 which possesses the central star and nebula properties
 very similar to those of NGC\,6781 (see \S\,\ref{S:D}).

\subsubsection{%
   Further on the C and Cl abundances
   \label{S-compabun}}

 Because our present analysis and the previous analysis
 done by \citet[][listed in Table\,\ref{T:abund2}, column 4]{Liu:2004ab} 
 are based on the same ISIS optical spectrum, 
 both results should be consistent with each other.
 However, this is not the case for C and Cl.

 The discrepancy in $\epsilon$(Cl) arises
 because we adopt the Cl$^{2+,3+}$ abundances of $1.07(-7)$ and $1.57(-8)$ 
 and the corresponding ICF(Cl) value of 1.17,
 while \citet{Liu:2004ab} used the Cl$^{2+}$ abundance of $7.92(-8)$ only 
 with the corresponding ICF(Cl) of 3.394.
 In addition, the adopted {\te} could contribute to the discrepancy 
 because the Cl ionic abundances are determined using their CEL lines, 
 whose emissivities are sensitive to {\te}.
 Overall, we would argue again that our $\epsilon$(Cl) value is more
 improved than the previous estimate because we have more robust 
 {\te} for the ionic Cl abundances and we derive a Cl$^{3+}$
 abundance that would reduce uncertainties in ICF(Cl).

   
 \begin{table}
  \centering
  \footnotesize
  \renewcommand{\arraystretch}{0.85}
  \caption{\label{T:abund2}%
  Elemental abundances $\epsilon(\mathrm{X})$ of NGC\,6781
  derived in the present analysis, compared with the solar
  abundances (column 3; $\mathrm{[X/H]}= \epsilon(\mathrm{X})
  - \epsilon_{\sun}(\mathrm{X})$, where $\epsilon_{\sun}(\mathrm{X})$
  is taken from \citealt{Lodders:2010aa}), previous empirical analysis
  (column 4; by \citealt{Liu:2004ab}), and model predictions
  (columns 5 and 6; by \citealt{Karakas:2010aa}; see \S\,\ref{abundmass}).
  }
  \begin{tabularx}{\columnwidth}{@{}lrcccc@{}}
   \hline\hline
   X  &\multicolumn{1}{c}{$\epsilon$(X)} &[X/H]
   &\multicolumn{1}{c}{$\epsilon$(X)}&\multicolumn{1}{c}{$\epsilon$(X)}
   &\multicolumn{1}{c}{$\epsilon$(X)}\\
   (1)   &\multicolumn{1}{c}{(2)} &(3)
	   &\multicolumn{1}{c}{(4)}&\multicolumn{1}{c}{(5)}
		   &\multicolumn{1}{c}{(6)} \\  
   \hline
   He & 11.06 $\pm$ 0.17 & +0.13 $\pm$ 0.17 & 11.08     &11.05& 11.06 \\ 
   C(RL) & 9.61 $\pm$ 0.29 & +1.22 $\pm$ 0.30 & \,\,9.17  &\,\,8.52&\,\,9.06 \\ 
   C(CEL) & 8.56 -- 9.00  & +0.17 -- 0.61
	 &\,\,\,$\cdots$&\,\,\,$\cdots$&\,\,\,$\cdots$   \\ 
   N & 8.15 $\pm$ 0.09 & +0.29 $\pm$ 0.15 & \,\,8.38      &\,\,8.39&\,\,8.42\\ 
   O & 8.76 $\pm$ 0.04 & +0.03 $\pm$ 0.08 & \,\,8.65      &\,\,8.94&\,\,8.94\\ 
   Ne & 8.15 $\pm$ 0.05 & +0.10 $\pm$ 0.11 & \,\,8.22     &\,\,8.12&\,\,8.27 \\ 
   Si & 7.03 $\pm$ 0.27 & --0.50 $\pm$ 0.28
	   &\,\,\,$\cdots$&\,\,7.57&\,\,7.59  \\ 
   S & 6.91 $\pm$ 0.06 & --0.25 $\pm$ 0.06 & \,\,6.97     &\,\,7.42&\,\,7.44 \\ 
   Cl & 5.16 $\pm$ 0.42 & --0.09 $\pm$ 0.42 & \,\,5.43 &$\cdots$&$\cdots$\\ 
   Ar & 6.49 $\pm$ 0.10 & --0.01 $\pm$ 0.14 & \,\,6.35 &$\cdots$&$\cdots$\\ 
   \hline
  \end{tabularx}
  \begin{minipage}{\columnwidth}
   \tablecomments{\footnotesize 
   The number density ratio relative to hydrogen is
   $\epsilon(\mathrm{X})=\log_{10}(\mathrm{X/H}) + 12$, 
   where $\log_{10}({\rm H}) = 12$. 
   The CEL C abundance, C(CEL), is an {\it expected} value.}
  \end{minipage}
   \end{table}


 The discrepancy in RL $\epsilon$(C) is due to 
 different values of $I$(C\,{\sc ii}\,4267\,{\AA}) 
 (might be caused by different adopted $c$({\hb})) and adopted ICF(C):
 our $\epsilon$(C) and ICF(C) values are $2.0(-3)$ and 2.03 
 whereas theirs are $9.05(-4)$ and 1.624, respectively.
 In general, C is a very important element as a coolant of
 the ionized gas component and also a source of C-based molecules
 in PNe. 
 Thus, we would discuss the C abundance further in this section.

 Our expected C(CEL) of $9.89(-4) \pm 3.14(-4)$ ($\epsilon$(C)$=9.00$) 
 adopted in the previous section,
 in comparison with the observed O(CEL) of $5.81(-4) \pm 2.19(-5)$
 ($\epsilon$(O)$=8.76$), would suggest a slightly C-rich nature 
 for NGC\,6781 (C/O number density ratio of $1.70 \pm 0.54$).
 Indeed, the \emph{Spitzer}/IRS mid-IR spectrum (Fig.\,\ref{S:spt}, inset) 
 shows polycyclic aromatic hydrocarbon (PAH) emission at $6-9$\,{\micron}
 (mostly from ionized PAH) and at 11.3\,{\micron} (from neutral PAH) 
 and dust continuum due to amorphous carbon, while the spectrum does
 not clearly show any O-rich dust features such as amorphous
 silicates at $\sim$9\,{\micron} and $\sim$18\,{\micron}
 and crystalline silicates around 30\,{\micron}.

 \citet{Guzman-Ramirez:2014aa} reported detection of  
 PAH emission in O-rich PNe in the Galactic bulge
 and suggested that PAHs could be formed in the compact/dense
 torus (i.e., the ``waist'' region of bipolar PNe)
 using C atoms liberated from CO molecules by photodissociation.
 At this point, there is no clear evidence 
 to suggest this possibility for NGC\,6781
 based on the spatially-resolved spectroscopic data.

 If we adopt RL C$^{2+}$ of $9.05(-4)$ and ICF(C) of 1.634 as
 previously used by \citet{Liu:2004ab} 
 and convert the RL C abundance to the CEL C abundance 
 by the average C$^{2+}$(RL)/C$^{2+}$(CEL) ratio of 4.10 \citep{Otsuka:2011aa}, 
 we would obtain the expected
 CEL C abundance of 3.61(--4), which would correspond to $\epsilon$(C) of 8.56.
 This would result in a C/O ratio of $\sim$0.76, 
 indicating that NGC6781 is slightly O-rich. 
 Hence, the possibility of NGC\,6781 being O-rich is not completely ruled out.

 As seen above, the C abundance depends on many factors, 
 from the $I$(C\,{\sc ii}\,4267\,{\AA}) measurements 
 to the ICF(C) and C$^{2+}$(RL)/C$^{2+}$(CEL) values adopted.
 Therefore, in the present work,
 we opt to allow a range of the expected CEL abundance
 for NGC\,6781 
 as $\epsilon(\mathrm{C}) = 8.56  - 9.00$  
 (correspondingly, $[\mathrm{C/H}] = 0.17 - 0.61$)
 based on the arguments presented above.

\subsubsection{Comparison with the previous model predictions
   \label{abundmass}}

 We compare the derived $\epsilon(\mathrm{X})$ with the values 
 predicted by AGB nucleosynthesis models.
 As for the metallicity $Z$ of the progenitor of NGC\,6781,
 it is best to reference elements that can never be synthesized 
 within AGB stars. 
 Thus, we adopt Cl and Ar as good $Z$ indicators. 
 The average between the observed [Cl/H] and [Ar/H] values of $-0.05$ 
 corresponds to $Z \sim 0.018$.

 However, the S abundance (${\rm [S/H] = -0.25}$)
 suggests a much lower $Z$.
 So far, this S abundance anomaly has been found in many 
 Milky Way and M31 PNe \citep[][see their Fig.\,1]{Henry:2012aa}. 
 \citet{Henry:2012aa}
 concluded that the sulfur deficit in PNe is generally reduced 
 by increasing the S$^{3+}$ abundance and selecting a proper ICF(S).
 Such an S depletion may indicate that
 a significant part of the atomic S mass
 is locked up as sulfide grains in the nebula 
 (e.g., MgS and FeS in C- and O-rich environments, respectively).
 However, the \emph{Spitzer}/IRS spectrum displays neither
 the broad 30\,{\micron} feature often attributed to MgS 
 nor narrower emission features around 30\,{\micron} attributed to FeS.
 The discrepancy between the observed and the AGB model S abundances
 may thus be related to the adopted reaction rates;
 \citet{Shingles:2013aa} demonstrated a possibility
 that the S depletion could be explained by introducing a large
 $^{22}$Ne($\alpha$,$n$)$^{25}$Mg reaction rate.
 Here, we propose that the apparently low [S/H] abundance is attributed to
 missing fluxes of low-excitation {\sii} lines as discussed above 
 (by adopting the revised S$^{+}$/H$^{+} = 8.97(-6)$ in \S\,\ref{S-ionic}, 
 we would obtain $\epsilon(\mathrm{S}) = 7.20$, which is consistent with
 $\epsilon$(S)$_{\sun}$).

 Now, we compare our empirically-derived elemental abundances 
 with those predicted with AGB nucleosynthesis models o
 f $Z = 0.02$ stars \citep{Karakas:2010aa} in Table\,\ref{T:abund2}:
 the values in columns (5) and (6) are the
 predicted values for initially 2.25\,$M_{\sun}$ and
 3.0\,$M_{\sun}$ stars, respectively.
 To assess the goodness of fit of the model prediction, we evaluate
 chi-square values ($\chi^2$) between our derived abundances and the 
 model-predicted abundances for stars in the initial mass range 
 from 1.5 to 4.0\,$M_{\sun}$. 
 Adopting the lower CEL abundance 
 limit of $\epsilon(\mathrm{C}) = 8.56$, a good fit 
 to the observed $\epsilon(\mathrm{X})$ is achieved with the 2.25\,$M_{\sun}$
 model (reduced $\chi^{2}$ = 15.5).

 Meanwhile, adopting the upper CEL abundance limit of 
 $\epsilon(\mathrm{C}) = 9.00$, the $\chi^2$ values suggest that the 
 observed $\epsilon(\mathrm{X})$ is most consistent with the 
 2.5\,$M_{\sun}$ model (reduced $\chi^{2}$ = 16.15).
 The reduced $\chi^{2}$ value = 17.5 of the 3.0\,$M_{\sun}$ model
 is equally good.
 Therefore, based on these results 
 we conclude that the initial mass of the CSPN is between 
 2.25 and 3.0\,$M_{\sun}$.

\subsection{The molecular gas component}

 Given the number of molecular lines seen in the spectra,
 especially with the rare OH$^{+}$ detection
 \citep{Aleman:2014aa},
 NGC\,6781 has to be treated as a PN rich in neutral gas.
 Then, it is critical to include the PDR
 of the nebula for a complete understanding
 of all of its components (ions, atoms, molecules, and dust). 
 In this section, therefore, we investigate the physical conditions
 of the most abundant species in the PDR, H$_{2}$,
 to articulate our understanding of the PDR in NGC\,6781.

\subsubsection{Physical conditions: spatial distribution}


\begin{figure*}
\centering
\includegraphics[width=\textwidth,clip]{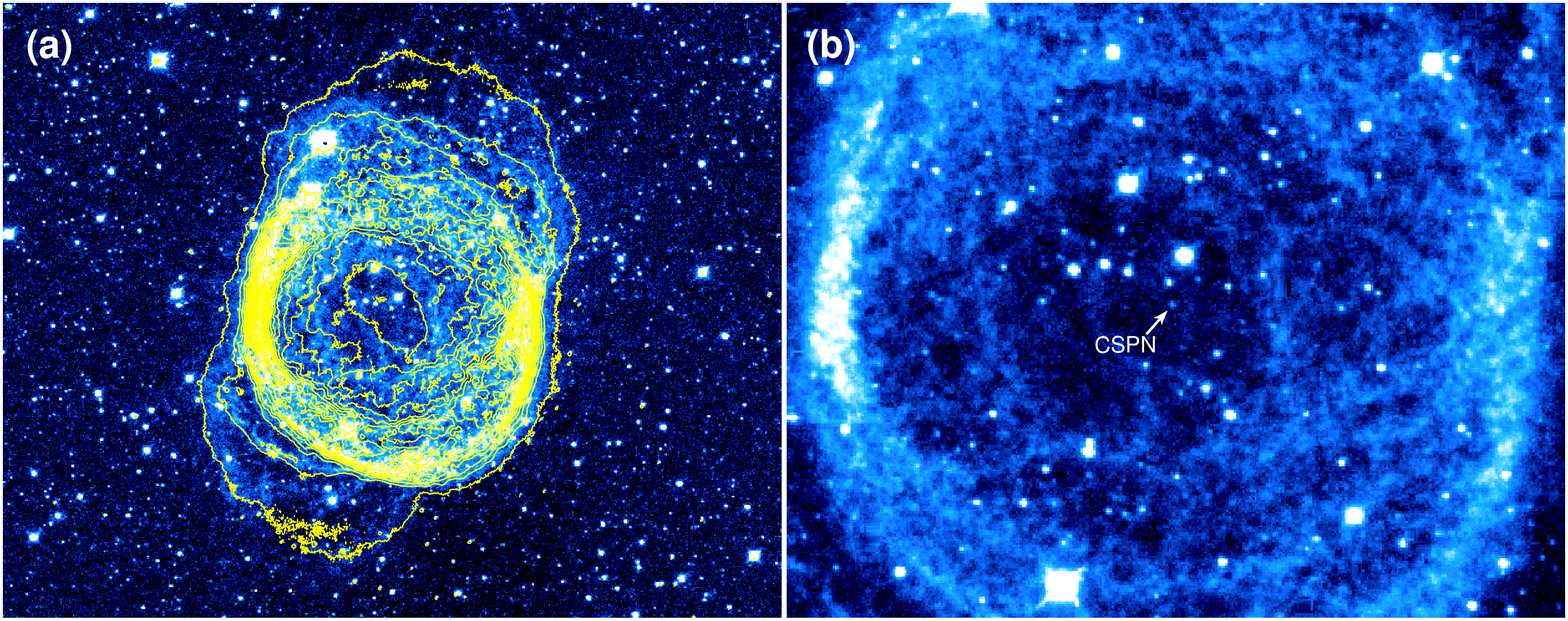}
 \caption{\label{F:H2CFHT}
 (a) A narrow band image of \mbox{H$_{2}$ $v = 1-0$ S(1)} at
 2.122\,{\micron} taken with the 3.6-m CFHT/WIRCAM, overlaid
 with yellow contours of {\nii}\,6583\,{\AA} emission taken with
 the 2.5-m NOT/ALFOSC (\citealt{Phillips:2011aa}; provided to
 us by M.~A.~Guerrero). (b) A close-up of the central part of
 the nebula, showing the filamentary structure of the H$_{2}$
 emission in the central region, from which the adopted spectra arose.
 The location of the central star is also indicated.}
\end{figure*}


 We obtain the H$_{2}$ image taken with the Wide-field Infrared Camera
 \citep[WIRCAM,][]{Puget:2004aa} on the 3.6-m Canada France Hawaii
 Telescope (CFHT) from the Canadian Astronomy Data Centre (CADC).
 The observations were done on 2006 April 14 (PI: S.~Kwok, Prop.\ ID:
 06AT03) through Taiwan CFHT time. 
 The basic calibrated data set
 retrieved from the CADC archive is reduced into a single image after
 bad pixel masking and geometric distortion correction using IRAF.
 Fig.\,\ref{F:H2CFHT} shows the \mbox{H$_{2}$ $v=1-0$ S(1)} image
 at 2.122\,{\micron} overlaid with contours of {\nii}\,6583\,{\AA}
 emission and the close-up of the central region from which emission of
 the spectra adopted in the present study arose (cf.\ Fig.\,\ref{S:slit}).
 Fig.\,\ref{F:H2CFHT}a shows that the spatial distribution of the
 molecular gas component in NGC\,6781 seen via H$_{2}$ emission is
 very similar to that of the 
 cool low I.P.\ gas component 
 seen via {\nii} emission
 (and also via {\ha} emission; Fig.\,\ref{S:slit}). 
 The same
 similarities in the spatial distributions are seen between the
 dust and ionized gas components
 delineating the nearly pole-on cylindrical barrel structure
  (Fig.\,3 of \citetalias{Ueta:2014aa}).
 Highly localized distributions of the molecular gas component 
 are apparent from the filamentary appearance of
 the H$_{2}$ emission (Fig.\,\ref{F:H2CFHT}b). 
 These H$_{2}$ filaments
 (and maybe clumps, too) are patches of H$_{2}$ survived in
 the ionized region.

   \subsubsection{Physical conditions: shocks vs.\ UV radiation
   \label{S:H2}
   }

 Table\,\ref{T:H2} summarizes near- and mid-IR H$_{2}$ lines detected
 in NGC\,6781. As reported by \citet{Phillips:2011aa} and
 \citet{Mata:2016aa}, pure rotational
 H$_{2}$ lines are detected in the \emph{Spitzer}/IRS spectra
 (Fig.\,\ref{S:spt}, inset).
 Observations made by \citet{Arias:2002aa} show that the
 intensity of H$_{2}$ $v = 2-1$ S(1) at 2.248\,{\micron}
 is much fainter than that of H$_{2}$ $v = 1-0$ S(1)
 at 2.122\,{\micron}, which indicates collisional excitation.
 The kinematic studies of \citet{Hiriart:2005aa} pointed to 
 a post-shock origin for the H$_{2}$ emission.
 If the observed H$_{2}$ lines are radiatively excited through
 the absorption of far-UV
 photons ($\sim$11 -- 13\,eV) in PDRs, the upper vibrational level
 would have to have a larger population, resulting in a relatively
 high H$_{2}$ \mbox{$I$(2.248\,{\micron})/$I$(2.122\,{\micron})}
 via UV fluorescence \citep[e.g.,][]{Kwok:2007aa}. 
 Collisional excitation, on the other hand, can occur in both shocks and PDRs.
 Excitation mechanisms of H$_{2}$ in PNe are examined by
 evaluating H$_{2}$ \mbox{$I$(2.248\,{\micron})/$I$(2.122\,{\micron})}
 ratio \citep[e.g.,][]{Otsuka:2013aa}, even though it is not easy to do 
 with $K$-band data alone.

 Interestingly, the expansion velocity of H$_{2}$
 \citep[\mbox{$\sim22$\,{\kms}},][]{Arias:2002aa,Hiriart:2005aa} is
 found to be greater than the expansion velocity measured from the
 {\oiii} line \citep[\mbox{10\,{\kms}},][]{Weinberger:1989aa} and
 {\nii} line \citep[\mbox{12\,{\kms}},][]{Arias:2002aa}.
 \citet{Hiriart:2005aa} concluded that the average
 \mbox{H$_{2}$ $v=1-0$ S(1)} surface brightness could be explained
 by shocks at \mbox{$10-24$\,{\kms}} heading into the pre-shock region
 of the H$_{2}$ density at $3400-14\,900$\,cm$^{-3}$.

 We investigate the conditions in the H$_{2}$ emitting regions
 by comparing the flux ratios of mid-IR H$_{2}$ lines to the
 \mbox{$v=0-0$ S(3)} line at 9.67\,{\micron} with the theoretical
 continuous shock (C-shock) models by \citet{Flower:2010aa}.
 The observed $I$(17.04\,{\micron})/$I$(9.67\,{\micron}) ratio
 suggests a match for a model with the shock velocity of
 \mbox{$V_{s} = 10$\,{\kms}} and pre-shock hydrogen density of
 $n_{s}(\mathrm{H}) = 200\,000$\,cm$^{-3}$, while the observed
 $I$(12.29, 8.02, 6.91, 6.11, 5.51\,{\micron}) to $I$(9.67\,{\micron})
 ratios point to a model with \mbox{$V_{s} = 20$\,{\kms}}
 and $n_{s}(\mathrm{H}) = 20\,000$\,cm$^{-3}$.  
 Here, the possible line flux contamination from the 
 H\,{\sc i} 12.3\,{\micron} line
 to the H$_{2}$\,12.29\,{\micron} line, estimated to be
 $I$(H\,{\sc i}\,12.3\,{\micron}) = 0.971 when \mbox{$I$({\hb})\, = 100}
 in the case of \mbox{{\te}  = 10$^{4}$\,K}
 and \mbox{{\Ne} = 200\,cm$^{-3}$}, is removed.

 \citet{Bachiller:1993aa} reported a CO expansion
 velocity of \mbox{22\,{\kms}}. 
 Recently, \citet{Bergstedt:2015aa}
 reported a velocity of \mbox{16\,{\kms}} via 3-D structure modeling
 using CO velocity maps. 
 A model by \citet{Flower:2010aa} with
 a shock velocity of \mbox{$V_{s} = 30$\,{\kms}} and pre-shock
 hydrogen density of \mbox{$n_{s}(\mathrm{H}) = 20\,000$\,cm$^{-3}$}
 would explain the observed far-IR CO line flux ratios with respect
 to the \mbox{CO $J=7-6$} line at 371.6\,{\micron} obtained from
 our \emph{Herschel} PACS and SPIRE spectra \citepalias{Ueta:2014aa}.

 Based on the arguments above, excitation of H$_{2}$ and
 CO lines in NGC\,6781 appears to be caused by thermal shocks
 at a velocity in the range of $10-30$\,{\kms} impinging onto 
 the pre-shock region at \mbox{$n_{s}(\mathrm{H})
 \sim 20\,000-200\,000$\,cm$^{-3}$}.
 These shocks may be be the consequence of interactions 
 between the slow AGB wind and fast PN wind emanating from 
 the CSPN in the context of the PN evolution.
 The slow-fast wind interactions could cause diffuse X-ray
 emission in the interaction regions. 
 No X-ray detection in NGC\,6781 may thus be because of 
 extinction (see \S\,\ref{tempfloor}).
 Together with the filamentary/clumpy
 appearance of the H$_{2}$ emission regions (Fig.\,\ref{F:H2CFHT}),
 we would conclude that these structures represent high-density
 regions delineating the locations of thermal collisional excitation
 embedded in an lower density ionized gas.
 Such high H$_{2}$ clumps \citep[so called ``cometary knots'',]
 []{Odell:1996aa} within the ionized gas are detected in nearby PNe
 \citep[see e.g.,][]{ODell:2002aa}. 
 Recently, \citet{Manchado:2015aa} detected cometary H$_{2}$ knots 
 within the ionized gas region in the bipolar PN NGC\,2346.

 One might think that the H$_{2}$ distribution in NGC\,6781 is similar
 to that in NGC\,7293 (Helix nebula), in which
 the H$_{2}$ emission is considered to arise from H$_{2}$ clumps. 
 For NGC\,7293, there is no evidence to suggest that 
 the H$_{2}$ emission from its cometary knots is due to shocks
 \citep[][reference therein]{Aleman:2011aa}.
 Another possible
 H$_{2}$ excitation mechanism is due to the structure and steady
 state dynamics of advective
 ionization front/dissociation front \citep{Henney:2007aa}. However,
 our Cloudy models with turbulence velocity of $\le 10$\,{\kms} in the nebula
 by following \citet{Henney:2007aa} failed to reproduce the observed
 H$_{2}$ line fluxes.
 While these are definitely issues that needs to be resolved in
 future, we tentatively conclude that the observed
 H$_{2}$ emission in NGC\,6781 has a shock origin 
 based on the arguments presented above.

\subsubsection{\label{exdiag} Physical conditions: H$_{2}$ excitation diagram}


 \begin{figure}
  \centering
  \includegraphics[width=\columnwidth,clip]{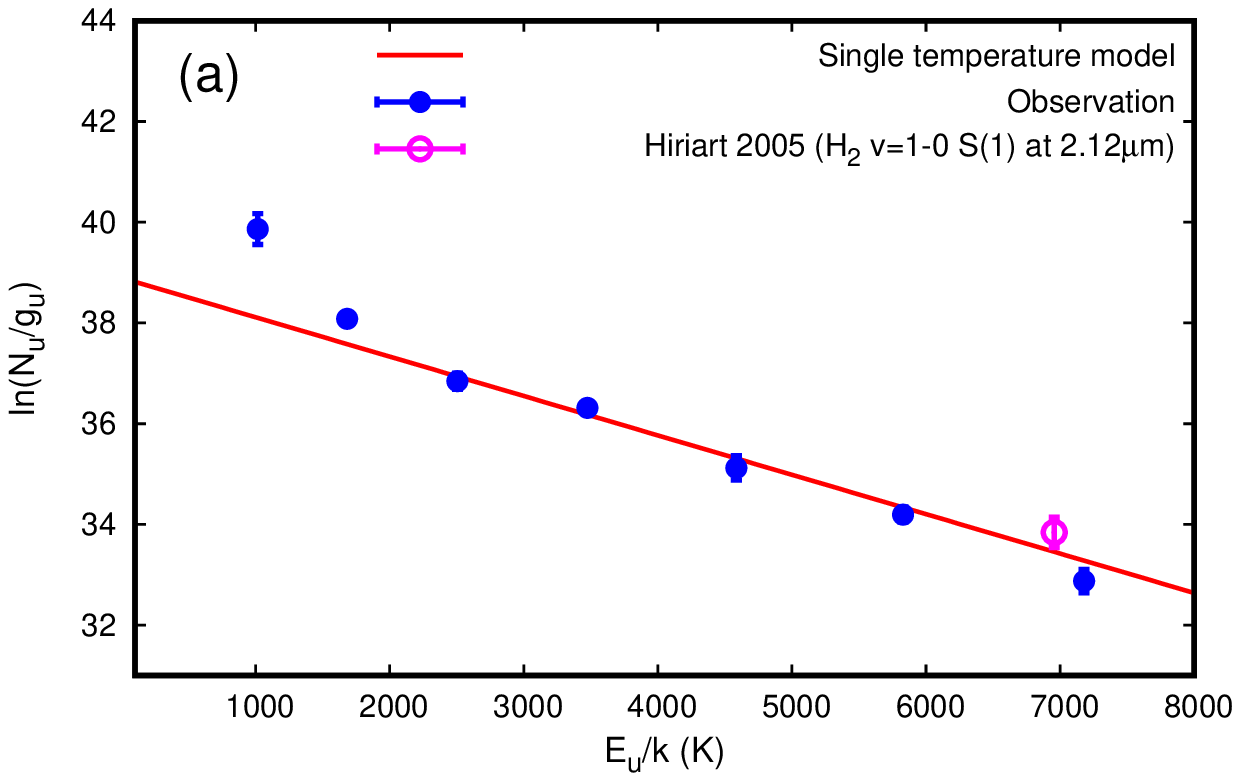}
  \includegraphics[width=\columnwidth,clip]{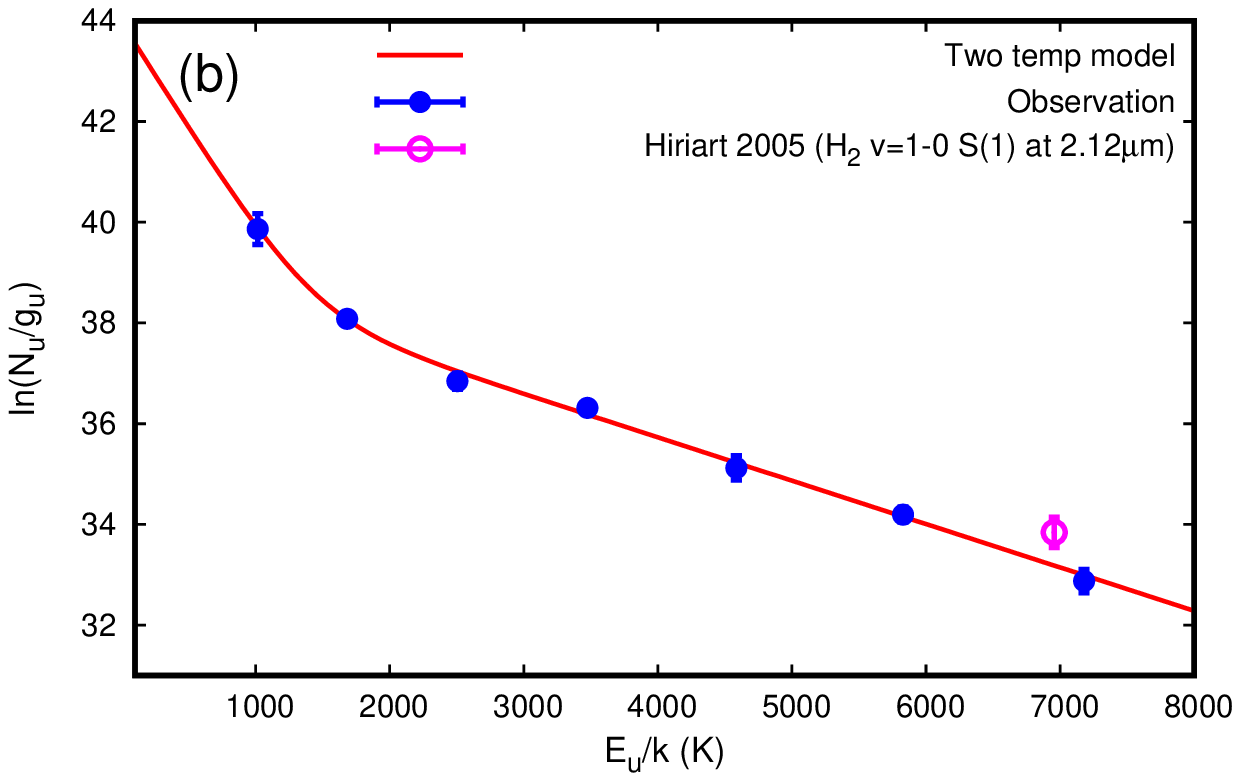}
  \caption{The excitation diagram of pure rotational
  transitions of H$_{2}$ lines. We fit the observed data (Table\,\ref{T:H2})
  with a single excitation temperature
  (top frame; a; with all but the 17.04\,{\micron} line; 
  $T(\mathrm{H}_{2}) = 1279 \pm 109$\,K) and 
  with two excitation temperatures (bottom frame; b; with all lines;
  $T(\mathrm{H}_{2}) = 1161 \pm 72$\,K and $236 \pm 50$\,K).}
  \label{F:h2}
 \end{figure}



 \begin{table}
  \centering
  \footnotesize
  \renewcommand{\arraystretch}{0.85}
  \caption{\label{T:H2}
  The average H$_{2}$ intensities of NGC\,6781
  measured with \emph{Spitzer}/IRS. See, also, Fig.\,\ref{S:spt}.}
  \begin{tabularx}{\columnwidth}{@{}D{.}{.}{-1}YD{p}{\pm}{-1}@{}}
   \hline\hline
   \multicolumn{1}{c}{$\lambda$}  &Transition
   &\multicolumn{1}{c}{Average intensity}\\
   \multicolumn{1}{c}{({\micron})}&
     &\multicolumn{1}{c}{(erg s$^{-1}$ cm$^{-2}$ sr$^{-1}$)}\\
   \hline
   17.04 &0-0 S(1)&1.90(-5) ~p~ 5.83(-6)\\
   12.29 &0-0 S(2)&1.10(-5) ~p~ 1.08(-6)\\
   9.67  &0-0 S(3)&5.31(-5) ~p~ 8.50(-6)\\
   8.02  &0-0 S(4)&4.00(-5) ~p~ 4.90(-6)\\
   6.91  &0-0 S(5)&1.08(-4) ~p~ 2.63(-5)\\
   6.11  &0-0 S(6)&3.56(-5) ~p~ 5.34(-6)\\
   5.51  &0-0 S(7)&6.19(-5) ~p~ 1.43(-5)\\
   2.12^{\dagger}  &1-0 S(1)&\multicolumn{1}{c}{2.70(--4)} \\
   \hline
  \end{tabularx}
      \begin{minipage}{\columnwidth}
     \tablewidth{\columnwidth}
  \tablenotetext{\dagger}{\footnotesize 
  The \mbox{H$_{2}$ v=1-0 S(1)} data is from \citet{Hiriart:2005aa}.}
   \end{minipage}
 \end{table}


 Assuming that H$_{2}$ lines are thermally excited and are in
 local thermodynamic equilibrium (LTE), the H$_{2}$ excitation
 temperature and column density can be estimated via an excitation diagram.
 The H$_{2}$ column density $N_{u}$ in the upper state is written as 
 \begin{equation}
  N_{u} = \frac{4 \pi I({\rm H_{2}})}{A} \cdot \frac{\lambda}{hc},
   \label{h2_eq1}
 \end{equation}
 where $I$(H$_{2}$) is the H$_{2}$ line intensity
 in \mbox{erg\,s$^{-1}$\,cm$^{-2}$\,sr$^{-1}$}, $A$ is the transition
 probability taken from \citet{Turner:1977aa}, $h$ is the Planck constant, and 
 $c$ is the speed of light. In LTE, the Boltzmann equation relates
 $N_{u}$ to the excitation temperature $T$(H$_{2}$) via
 \begin{equation}
  {\rm ln}\,\bigl(\frac{N_{u}}{g_{u}}\bigr) = -\frac{E_{u}}{kT({\rm H}_{2})} 
   + {\rm ln}\,\bigl[N({\rm H}_{2}) \cdot \frac{hcB}{2kT({\rm H}_{2})}\bigr],
   \label{h2_eq2}
 \end{equation}
where $g_{u}$ is the vibrational degeneracy,
 $E_{u}$ is the energy of the excited level taken from
 \citet{Dabrowski:1984aa}, $k$ is the Boltzmann constant, and
 $B$ is the rotational constant (60.81\,cm$^{-1}$).

 In Fig.\,\ref{F:h2}, we plot the $\ln(N_{u}/g_{u})$ vs.\
 $E_{u}/k$ for each of the H$_{2}$ lines detected in NGC\,6781
 (Table\,\ref{T:H2}). 
 The $E_{u}/k$ of the \mbox{H$_{2}$ $v=2-1$ S(1)}
 (magenta circle) is calculated using the average line intensity
 of \mbox{$2.7(-4)$\,erg\,cm$^{-2}$\,s$^{-1}$\,sr$^{-1}$}
 \citep{Hiriart:2005aa}.
 The rotational diagram suggests that
 the bulk of the H$_{2}$ 17.04\,{\micron} line emission is
 produced in a region with different physical conditions 
 from the other H$_{2}$ line emitting regions.

 First, we determine the conditions of the H$_{2}$ emitting region 
 by fitting the line fluxes at 12.29, 9.67, 8.02, 6.91, 6.11, 5.51, 
 and 2.12\,{\micron} (i.e., all but 17.04\,{\micron}) with 
 Equation\,\ref{h2_eq2} using a single excitation
 temperature (Fig.\,\ref{F:h2}a):
 \mbox{$T(\mathrm{H}_{2}) = 1279 \pm 109$\,K} and 
 \mbox{$N(\mathrm{H}_{2}) = (2.28 \pm 0.49)(18)$\,cm$^{-2}$}.
 The derived $T(\mathrm{H}_{2}$) is 
 comparable to $978$\,K and \mbox{$880 \pm 70$\,K},
 previously derived by \citet{Phillips:2011aa} and \citet{Mata:2016aa},
 respectively
 (with a single temperature model using all but 
 the 2.12\,{\micron} and 17.04\,{\micron} lines).

 Next, we fit all H$_{2}$ lines (including 17.04\,{\micron}) 
 using two excitation temperatures (Fig.\,\ref{F:h2}b). 
 The warm component is found to have
 $T(\mathrm{H}_{2}) = 1161 \pm 72$\,K and 
 $N(\mathrm{H}_{2}) = (2.72 \pm 0.53)(18)$\,cm$^{-2}$,
 whereas the cold component is found to have
 $T(\mathrm{H}_{2}) = 236 \pm 50 $\,K and
 $N(\mathrm{H}_{2}) = (6.67 \pm 4.89)(19)$\,cm$^{-2}$
 (while lack of the H$_{2}$ 0-0 S(0) line at 28.2\,{\micron}
 makes the fitting results relatively less certain).
 Nonetheless, the 17.04\,{\micron} line is expected to 
 arise from such colder and denser regions.

 \subsubsection{Empirically determined molecular gas mass%
 \label{S-molmass}}

 To conclude this subsection,
 we estimate the mass of the molecular gas component in the nebula
 by adopting the distance of 0.46\,kpc (\S\,\ref{S:D}).
 Based on the H$_{2}$ and CO emission maps 
 (\citealt{Hiriart:2005aa} and \citealt{Bachiller:1993aa}, respectively),
 we see that molecular emission increases at $\sim$54-55{\arcsec} 
 away from the CSPN with the thickness of 12{\arcsec}. 
 Using H$_{2}$ densities of the warm and cold components
 as derived above
 ($N(\mathrm{H}_{2}) = 2.72(18)$\,cm$^{-2}$ and 
 $6.67(19)$\,cm$^{-2}$, respectively),
 we estimate the H$_{2}$ gas mass of 
 $2.5(-3)\,M_{\sun}$ 
 and $6.2(-2)\,M_{\sun}$ for the warm and cold components, respectively.

 Previously, we derived
 $N(\mathrm{CO}) = 10^{14.70-15.08}$\,cm$^{-2}$ 
 (excitation temperature at $\sim$60\,K)
 based solely on our \emph{Herschel} spectra (\citetalias{Ueta:2014aa}).
 \citet{Bachiller:1997aa} measured
 $N(\mathrm{CO}) = 10^{16.16}$\,cm$^{-2}$ (excitation temperature at
 $\sim$25\,K) based on sub-millimeter data.  
 Assuming that each of the above $N(\mathrm{CO})$ estimates 
 based on data in the different wavelength/temperature realms 
 would represent the warm and cold component, respectively,
 the warm and cold CO gas masses
 are estimated to be $4.6(-6)\,M_{\sun}$ and $6.6(-5)\,M_{\sun}$,
 respectively.
 These estimates are combined to yield the total molecular gas mass
 (of H$_{2}$ and CO) of $6.46(-2)\,M_{\sun}$.

 The empirical $N$(CO)/$N$(H$_{2}$) ratio turns out to be 
 $2.19(-4)$ and $4.37(-4)$
 for the warm and cold temperature regions, respectively. 
 Assuming that the $N$(CO)/$N$(H$_{2}$) ratio translates roughly 
 to $2\times n({\rm C})/n ({\rm H})$, we can estimate
 $\epsilon$(C) of $8.04 - 8.34$ for the molecular gas component. 
 Compared with the adopted CEL {\it expected} $\epsilon$(C) of 
 $8.56 - 9.00$ for the ionized gas component,
 $\sim$11-60\,\% of the C-atoms were estimated to be locked in 
 as molecules.

  \subsection{\label{dust}
  The dust component: summary of HerPlaNS I}

 The surface brightness distribution of thermal dust continuum
 emission from NGC\,6781 is spatially resolved in far-IR \emph{Herschel}
 broadband images \citepalias[see Fig.\,3 of][]{Ueta:2014aa}.
 The bright ring structure with 
 $\sim$60{\arcsec} outer radii represents the bulk of the nearly pole-on
 cylindrical barrel structure \citep[originally proposed by][]{Schwarz:2006aa}, 
 and the elongated nebula of $\sim$200{\arcsec} in the total
 north-south extent indicates the distribution of dust along the polar
 axis of the nebula.
 The spatial extent of thermal dust continuum emission in far-IR
 wavelengths is nearly identical with that of atomic gas and molecular
 emission lines in optical and near-IR wavelengths.

 Previously, we performed spectral energy distribution (SED)
 fitting of the \emph{Herschel} 70/160/250/350/500\,{\micron} images
 using a modified blackbody function, and found that dust
 grains are composed mostly of amorphous-carbon based material
 (i.e., the power-law dust emissivity index $\beta$ is $\sim$1
 across the nebula) having the dust temperature $T_{d}$ in the range
 between 26 and 40\,K (\citetalias{Ueta:2014aa}). 
 Moreover, after removing the contribution to the continuum flux in the far-IR 
 by fine-structure lines and molecular emission lines
 (amounts to 8-20\,\% of the total flux), spectral fitting of the
 integrated far-IR fluxes yielded $T_{d}$ = \mbox{$37 \pm 5$\,K} and
 $\beta$ = \mbox{$0.9 \pm 0.3$}.
 Indeed, the \emph{Spitzer}/IRS spectrum (Fig.\,\ref{S:spt}, inset)
 shows PAH bands and featureless dust continuum,
 This is consistent with 
 the dusty nebula of NGC\,6781 containing more amorphous carbon dust and PAHs 
 than amorphous silicate dust.

\subsection{\label{S:CS} The central star}

\subsubsection{\label{S:D}%
Distance, luminosity, and effective temperature}


 \begin{table*}
  \footnotesize
  \renewcommand{\arraystretch}{0.85}
  \centering
  \caption{\label{T:N6720}%
  Similarities between NGC\,6781 and NGC\,6720.}
   \begin{tabularx}{\textwidth}
    {@{}lccccccccccccl@{}}
    \hline\hline
    PNe     & $\epsilon$(He)  &$\epsilon$(C$_{\rm RL}$)
    &$\epsilon$(C$_{\rm CEL}$)&$\epsilon$(N)
   &$\epsilon$(O$_{\rm RL}$)&$\epsilon$(O$_{\rm CEL}$)&$\epsilon$(Ne)
    &$\epsilon$(S) &$\epsilon$(Cl)   &$\epsilon$(Ar)&$T_{\rm eff}$ (kK)
    &$\log\,g$ (cm s$^{-2}$)&References\\
    \hline
    NGC\,6781 &11.06&9.61 &8.56-9.00  &8.15&$\cdots$     &8.76  &8.15&6.91&5.16
   &6.49&80 -- 123&6.0 -- 7.0&(1), (2), (3), (4)\\
    NGC\,6720 &11.05&9.10 &8.59  &8.22&9.18 &8.80  &8.23&6.86&5.19
   &6.54&80 -- 135&6.9 -- 7.0&(5), (6), (7)\\
    \hline
   \end{tabularx}
   \begin{minipage}{\textwidth}
    \tablerefs{\footnotesize
    (1) This work for abundances (see \S\,\ref{S:Abund});
    (2) \citet{Schwarz:2006aa} 
    for $T_{\rm eff}$ and $\log\,g$ 
    via photoionization modeling; 
    (3) \citet{Rauch:2004aa} 
    for $T_{\rm eff}$ and $\log\,g$
    via stellar absorption fitting;
    (4) \citet{Liu:2004ab} for abundances;
    (5) \citet{McCarthy:1997aa} 
    for $T_{\rm eff}$ and $\log\,g$
    via stellar absorption fitting;
    (6) \citet{Napiwotzki:1999aa} 
    for $T_{\rm eff}$ and  $\log\,g$ 
    via stellar absorption fitting;
    (7) \citet{van-Hoof:2010aa} 
    for $T_{\rm eff}$
    via {Cloudy} photoionization modeling.
   }
   \end{minipage}
 \end{table*}


 A vast variety of distance estimates are proposed for NGC\,6781, 
 including
 0.3\,kpc \citep{Tajitsu:1998aa,Phillips:2002ab}, 
 0.7\,kpc \citep{Stanghellini:2008aa,Frew:2016aa},
 0.9\,kpc \citep{Maciel:1984aa}, 
 0.95\,kpc \citep{Schwarz:2006aa}, and
 1.27\,kpc \citep{Ali:2013aa}, to name a few.
 For the present study, rather than adopting any of the previous
 investigations, we elect to determine our own value
 by comparing the observed photometry of the CSPN (Fig.\,\ref{S:spt},
 Table\,\ref{T:phot}) with the post-AGB evolutionary tracks produced
 by \citet{Vassiliadis:1994ab} augmented with a grid of synthesised
 spectra by \citet{Rauch:2003aa}. 
 Although several new evolutionary tracks have been produced 
 since then, there has been no AGB nucleosynthesis models constructed
 based on such new tracks. 
 In comparing observed data with theoretical models, 
 we would regard internal consistencies between models more important.
 Especially when we aim at determining the state of evolution of 
 the CSPN of NGC\,6781, the most critical is adopting AGB nucleosynthesis models 
 that are consistent with evolutionary tracks.
 Therefore, in the following discussion, 
 we adopt the AGB nucleosynthesis models by \citet{Karakas:2010aa} 
 based on \citet{Vassiliadis:1994ab}.

 We start by estimating the CSPN luminosity $L_{\ast}$ using
 a grid of non-LTE line-blanketed plane-parallel hydrostatic atmospheric
 models generated by \citet{Rauch:2003aa} as templates.
 We adopt the solar abundance ($Z = 0.02$) models for the CSPN based
 on the results of our own nebular abundance analysis presented in
 \S\,\ref{abundmass}.
 
 To characterize the stellar atmosphere fully, we also need 
 the effective temperature $T_{\rm eff}$ and surface gravity
 $\log\,g$ of the CSPN. 
 Previously, \citet{Rauch:2004aa} suggested \mbox{$T_{\rm eff} =
 80\,000$\,K} and \mbox{$\log\,g = 6.0$\,cm\,s$^{-2}$} based on the
 stellar absorption line fitting. 
 If this $T_{\rm eff}$ were true,
 the CSPN would have been still burning hydrogen in a thin
 surface layer while increasing $T_{\rm eff}$. 
 However, detection
 of strong He\,{\sc ii}\,4686\,{\AA} and [O\,{\sc iv}]\,25.88\,{\micron}
 lines in the ISIS and \emph{Spitzer}/IRS spectra, respectively
 (Fig.\,\ref{S:spt} and Table\,\ref{T:linelist}) requires
 \mbox{$T_{\rm eff} > 80\,000$\,K}, refuting the previous suggestion.
 The noisy spectrum due to the faintness of the CSPN might have
 compromised the previous absorption line fitting analysis.

 Thus, we decide to look for the appropriate $T_{\rm eff}$
 and $\log\,g$ values in a PN similar to NGC\,6781
 in terms of nebula and CSPN properties. 
 Amongst Galactic PNe,
 NGC\,6720 is very similar to NGC\,6781 in many respects,
 especially in their abundance pattern as shown in Table\,\ref{T:N6720}.
 Spectroscopically, both PNe show PAH features and pure
 rotational H$_{2}$ lines in their \emph{Spitzer}/IRS spectra
 \citep{Phillips:2011aa,Cox:2016aa} as well as
 rotational-vibrational H$_{2}$ emission
 \citep[e.g.,][]{Hiriart:2005aa,van-Hoof:2010aa}.
 Both PNe possess a structure due to a heavy equatorial concentration 
 (i.e., a generic bipolar/barrel shape) viewed nearly pole-on 
 \citep{Schwarz:2006aa,Sahai:2012aa,Ueta:2014aa}.

  The CSPN of NGC\,6720 has a $T_{\rm eff} > 100$\,kK based on
  the absorption line analysis done by 
  \citet{McCarthy:1997aa} and \citet{Napiwotzki:1999aa}.
  Thus, based on the similarities listed above
 we adopt \mbox{$T_{\rm eff} = 110 - 140$\,kK} and 
 \mbox{$\log\,g = 6.9$\,cm\,s$^{-2}$} for the CSPN of
 NGC\,6781 as well. 
 Consistent results were previously obtained from
 detailed SED fitting with {Cloudy} photoionization models of 
 NGC\,6720
 \citep[][also see \S\,\ref{S:CLOUDY}]{van-Hoof:2010aa}.

\begin{figure}
  \centering
  \includegraphics[width=\columnwidth,clip]{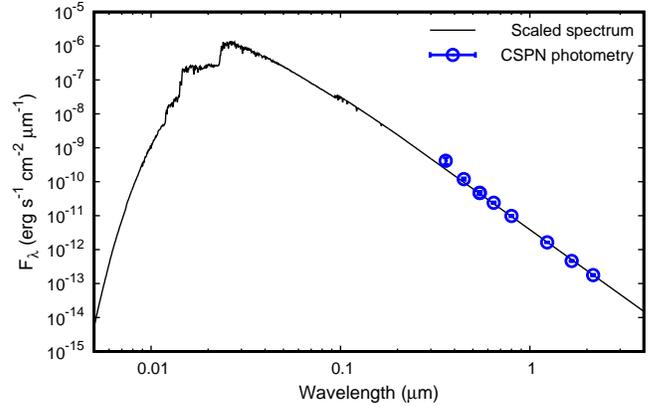}
  \caption{\label{F:cspn}
  The synthesised spectrum of a star with
  $Z$ = 0.02, $T_{\rm eff}$ = 120\,kK, and
  $\log\,g$ = 6.9\,cm s$^{-2}$
  \citep[][the black line]{Rauch:2003aa}
  fit with the observed photometry points
  of the CSPN (the blue circles; Table\,\ref{T:phot}).}
 \end{figure}

 Then, we scale the synthesized Rauch model spectra of the adopted
 CSPN characteristics of $T_{\rm eff} = 110-140$\,kK with a constant
 10\,kK step
 with \mbox{$\log\,g = 6.9$\,cm\,s$^{-2}$} fixed so that
 the observed photometry from the WFC $u$-band to WFCAM $K$-band
 (see Table\,\ref{T:phot}) matches with the model spectra
 (Fig.\,\ref{F:cspn}, showing the $T_{\rm eff} = 120$\,kK case).
 The scaled spectra are integrated to yield $L_{\ast}$, which
 is then parameterized with $T_{\rm eff}$ and the distance $D$ in the form of
 \begin{equation}
  L_{\ast}(D, T_{\rm eff}) = {\bigl[}2.29(-7) \cdot
   T_{\rm eff}^{2} - 4.39(-2)
   \cdot T_{\rm eff} + 2510{\bigr]} \cdot D^{2},
   \label{E:L}
 \end{equation}
 where $D$ is in kpc and $T_{\rm eff}$ is in K.
 Note that $L_{\ast}$ is not very sensitive to $\log\,g$.
 For instance, $L_{\ast}$ increases only by $\sim$0.8\,\%
 when $\log\,g$ is reduced from the adopted 6.9\,cm\,s$^{-2}$
 to 6.6\,cm\,s$^{-2}$. 
 Thus, our choice of single $\log\,g$ value is warranted.

 
 \begin{figure}
  \centering
  \includegraphics[width=\columnwidth,clip]{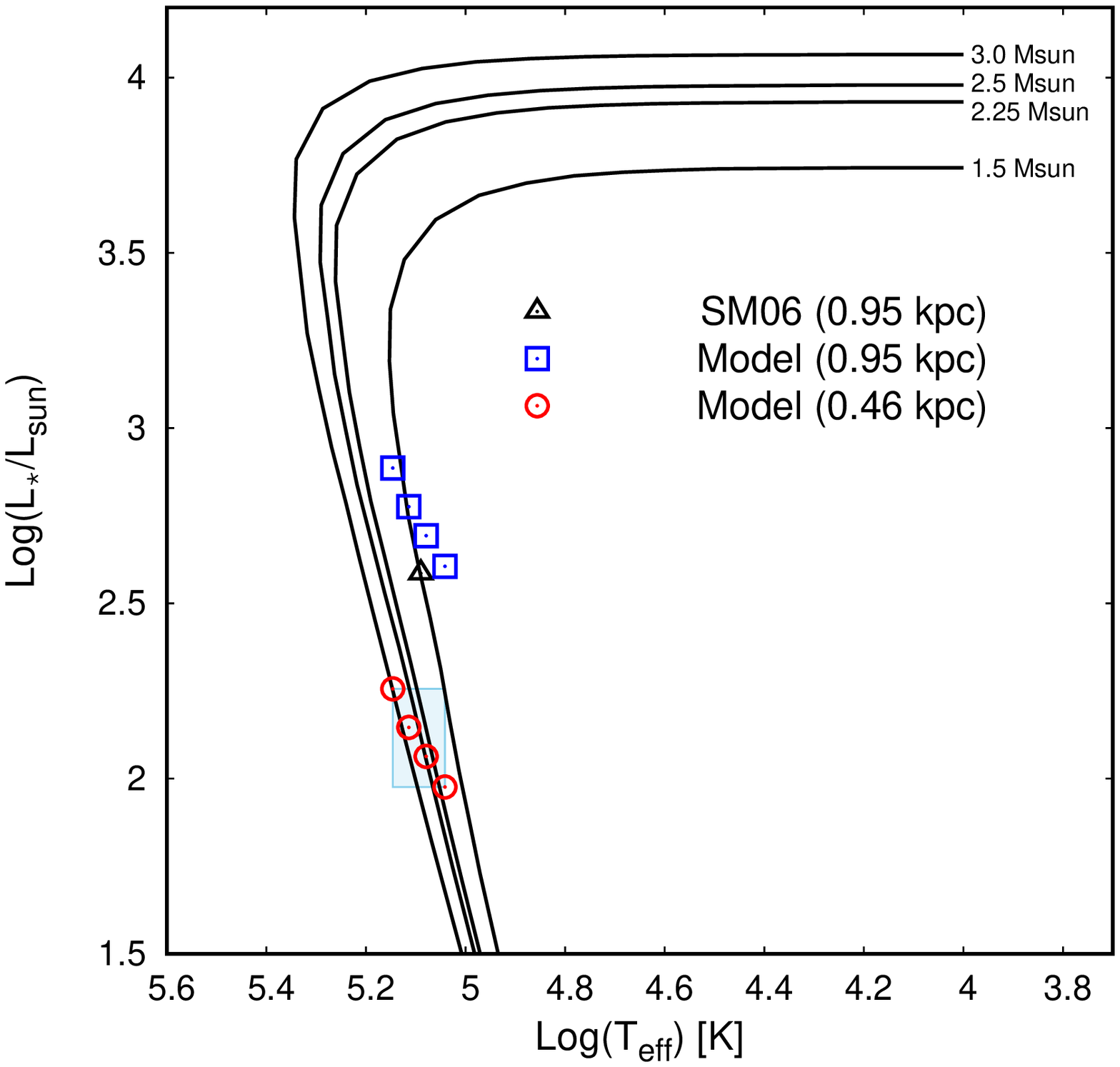}
  \caption{\label{F:HR}%
  The distance-fitting 
  comparison among the post-AGB
  evolutionary model tracks of 1.5, 2.25, 2.5, and 3.0\,$M_{\sun}$
  initial-mass stars \citep{Vassiliadis:1994ab}
  and the CSPN luminosity, $L_{\ast}(D, T_{\rm eff})$, computed for
  $D$ = 0.95 and 0.46\,kpc (blue squares and red circles, respectively)
  and $T_{\rm eff} = 110, 120, 130$, and 140\,kK
  (from right to left, respectively),
  when $\log\,g = 6.9$\,cm\,s$^{-2}$.
  Also shown is the $L_{\ast}-T_{\rm eff}$ pair adopted
  by \citet{Schwarz:2006aa}, with which
  they concluded $D$ =  0.95\,kpc (black triangle).
  The light-blue box indicates the
  $L_{\ast}-T_{\rm eff}$ parameter range
  based on our {Cloudy} model calculations (\S\,\ref{S:CLOUDY})
  at $D = 0.46$\,kpc. See, also, Fig.\,\ref{F:HR-last}.}
 \end{figure}


 Finally, we compute $L_{\ast}(D, T_{\rm eff})$ at $T_{\rm eff}$
 = $110-140$\,kK and for a range of $D$,
 and plot the resulting ($L_{\ast}$, $T_{\rm eff}$)
 pairs over the post-AGB evolutionary tracks of the 1.5,
 2.25, 2.5, and 3.0\,$M_{\sun}$ initial mass stars 
 produced by
 \citet{Vassiliadis:1994ab}, as shown in Fig.\,\ref{F:HR}. 
 Our choice of the initial mass
 of the adopted post-AGB evolutionary tracks is dictated by the
 results of our abundance analysis that indicated the CSPN
 initial mass being between 2.25 and 3.0\,$M_{\sun}$ (\S\,\ref{abundmass}).
 Also, the previous analysis by \citet{Schwarz:2006aa}
 suggested the CSPN initial mass of 1.5\,$M_{\sun}$.

 We find that $D = 0.34-0.52$\,kpc fits
 the initially $2.25-3.0\,M_{\sun}$ post-AGB evolutionary 
 tracks the best for the adopted $T_{\rm eff}$ range
  (light-blue box in Fig.\,\ref{F:HR}).
 Therefore, we adopt $D = 0.46$\,kpc, which is the
 the intermediate value between 0.34 and 0.52\,kpc
 (red circles in Fig.\ref{F:HR}).
 Accordingly, we find $L_{\ast} = 104-196$\,$L_{\sun}$.
 This evolutionary track fitting suggests
 that the CSPN of NGC\,6781 is in the cooling phase.
 The results of the fitting are not significantly altered 
 even when we adopt more recent post-AGB evolution tracks
 such as the ones computed by \citet{Miller-Bertolami:2016aa}
 ($D = 0.46$\,kpc, using the post-AGB
 evolutionary tracks for 2.0\,$M_{\sun}$
 and 3.0\,$M_{\sun}$ stars with $Z = 0.02$; see also Fig.\,\ref{F:HR-last}).

 Previously, \citet{Schwarz:2006aa} concluded that the progenitor 
 of NGC\,6781 was a $1.5 \pm 0.5$\,$M_{\sun}$ initial mass star 
 based on their derived $L_{\ast}$ and $T_{\rm eff}$ values,
 provided $D = 0.95$\,kpc suggested from their photoionization 
 model fitting (black triangle in Fig.\,\ref{F:HR};
 also suggesting that NGC\,6781 was in cooling phase).
 At $D=0.95$\,kpc, our $L_{\ast}$ estimates would be consistent
 with the 1.5\,$M_{\sun}$ evolutionary track (blue squares
 in Fig.\,\ref{F:HR}). 
 However, 
 the progenitor CSPN mass of NGC\,6781 would most likely exceed 
 1.5\,$M_{\sun}$ because of 
 its empirically determined elemental abundances (\S\,\ref{S:Abund})
 and 
 H$_{2}$ detection in this object (\S\,\ref{S:H2}). 
 
 With a survey of \mbox{H$_{2}$ $v= 1-0$ S(1)} emission in Galactic PNe,
 \citet{Kastner:1996aa} suggested that H$_{2}$-rich PNe evolved
 from relatively massive progenitors because H$_{2}$ was
 exclusively
 detected in bipolar PNe \citep[see also e.g.,][]{Guerrero:2000aa}.
 Bipolar PNe are known to be associated with massive 
 ($\ge 1.5$\,$M_{\sun}$) progenitors based on the
 distribution of bipolar PNe
 in the Milky Way 
 with respect to that of elliptical PNe \citep{Corradi:1995aa}.
 Hence, the detection of H$_{2}$ supports 
 our adaptation of the $2.25-3.0$\,$M_{\sun}$ initial mass 
 for the CSPN of NGC\,6781
 and the distance of 0.46\,kpc based on the
 $L_{\ast}(D, T_{\rm eff})$ fitting.
 
 The filamentary appearance of the nebula (Fig.\,\ref{F:H2CFHT})
 and low {\Ne} even in the
 central ionized regions (\S\,\ref{S-plasma}) are also suggestive
 that NGC\,6781 is a highly evolved PN.
 Referring back to the similarity to NGC\,6720, 
 comparisons between $L_{\ast}$ and $T_{\rm eff}$,
 where $L_{\ast}$ is based on {Cloudy} model fitting
 of the SED by \citet{van-Hoof:2010aa} with 
 the evolutionary tracks by \citet{Vassiliadis:1994ab}
 for initially 3.0\,$M_{\sun}$ stars of $Z = 0.02$,
 also suggest that NGC\,6720 is in the cooling phase.

 If the CSPN of NGC\,6781 were still in the final H-burning phase,
 the distance estimate would have to be $\gtrsim 3.6$\,kpc. 
 According to
 \citet{Vassiliadis:1994ab}, $L_{\ast}$ is
 nearly constant at $\sim$6300\,$L_{\sun}$ along the horizontal part 
 of the post-AGB track for a 2.5\,$M_{\sun}$ initial mass star with $Z = 0.02$.
 In this case, the number of the ionizing photons is 4.25(+47)\,s$^{-1}$
 for $T_{\rm eff} = 120\,000$\,K and $\log\,g = 6.9$\,cm\,s$^{-1}$.
 The Str\"{o}mgren radius for this radiation field in a constant 
 hydrogen density of 300\,cm$^{-3}$ 
 (see Table\,\ref{T:diagno_table}, Fig.\ref{F:GEO}) 
 with a filling factor ($f$) of unity 
 would be $\sim$0.41\,pc.
 This corresponds to the apparent radius of 23$\farcs$7
 at $D = 3.6$\,kpc, which disagrees with the
 observed ionization radius of $\sim$55$^{\prime\prime}$.
 Because the Str\"{o}mgren radius is proportional to $f^{-1/3}$, 
 it would be consistent with the observed ionization radius 
 at $D = 3.6$\,kpc if $f$ were 0.12.
 However, according to the empirical method introduced 
 by \citet{Mallik:1988aa},
 the $f$ value of NGC\,6781 is estimated to be $\sim$0.4 at $D =
 3.6$\,kpc and almost unity at 0.46\,kpc.
 Therefore, we conclude that the CSPN of NGC\,6781 already 
 evolved off to the cooling track presently with 
 $L_{\ast} = 104-196\,L_{\sun}$ and $T_{\rm eff} =110-140$\,kK
 at $D = 0.46$\,kpc.

\subsubsection{\label{S:binary}%
Possibility of the presence of a binary companion}
 
 At present, binary evolution would appear to be 
 one of the most viable explanations for the formation of bipolar nebulae 
 via the inevitable equatorial density enhancement
 (e.g., \citealt{Jones2017}).
 Our motivation to collect photometry measurements of the CSPN
 exhaustively in the UV to near-IR is also intended to 
 establish the presence or absence of a near-IR excess, 
 which would suggest the presence of a cooler binary companion.

  From a comparison between the observed colors ($V-I$ and $I-J$) 
  and the grid of theoretical color indices as a function of 
  $T_{\rm eff}$,
  \citet{Douchin:2015aa} argued that CSPN of NGC\,6781 shows near-IR
  excess owing to an M1-M7 type companion star.
  However, we do not observe any IR excess in the SED of the CSPN
  (Figs.\,\ref{S:spt} and \ref{F:cspn}).

  It is true that the IR excess detection can be influenced by the way
  the interstellar extinction is corrected for.
  With our adopted $c(${\hb}$) = 1.007$, the extinction corrected $V-I$
  and $I-J$ colors of the CSPN were $-0.44$ and $-0.90$, respectively. 
  If we used $E(B-V) = 0.56$ (corresponding to $c(${\hb}$) = 0.82$)
  as adopted by \citet{Douchin:2015aa}, the respective $V-I$ and $I-J$
  colors would become redder, $-0.27$ and $-0.61$, which would be in
  perfect agreement with \citet{Douchin:2015aa}. This would negate the
  necessity for a M1-M7 type companion star.

  Thus, whether NGC\,6781 possesses a binary central system is still
  an open question because the evolutionary effects from the secondary,
  even if it existed, would still be negligible at this point, based
  on the observed spectra and photometry. 
  Therefore, we would simply keep the adopted $D = 0.46$\,kpc and other
  quantities for which there is distance dependency, 
  in our analyses as outlined in the previous sections and 
  in the subsequent modeling section.

\section{Cloudy dusty photoionization models}
\label{S:CLOUDY}%

\subsection{Modeling approach}

 In the previous sections, we outlined how we mustered the most
 comprehensive observational data set yet assembled for NGC\,6781 (\S\,\ref{S2})
 and performed various analyses to determine empirically
 the CSPN and nebula characteristics for this object (\S\,\ref{S3}).
 In this section, we outline how we construct a
 \emph{realistic} input numerical model of NGC\,6781
 for {Cloudy}
 \citep[version C13.03,][]{Ferland:2013aa}, 
 comprising the CSPN and the nebula, the latter of which consists 
 of the ionized/neutral/molecular gas and dust components,
 based on the collected data.

 Our aim is to converge on \emph{self-consistent}
 physical conditions of the entire NGC\,6781 system 
 from the highly-ionized region to the PDR
 through iterative model fitting that 
 \emph{comprehensively} reproduces 
 \emph{all} of the observational data that we collected: 
 the spatially-integrated fluxes and flux densities
 from to UV to radio (37 broadband photometry fluxes, 
 19 flux densities, and
 78 emission line fluxes) plus 
 8 elemental abundances.
 The empirically derived quantities of the CSPN and nebula
 provide the input parameters, while the observational data from
 the UV to radio provide the vital constraints in
 iterative fittings of the model parameters.
 For the sake of consistency,
 we substituted the same transition probabilities and effective
 collision strengths of CELs used in our plasma diagnostics
 and nebular abundance analyses in the {Cloudy} code.

\subsection{The input model}

\subsubsection{SED of the CSPN}

 As the incident SED from the CSPN, we adopt the theoretical
 atmospheric model grid by \citet{Rauch:2003aa} for a star
 with $Z = 0.02$ and $\log\,g = 6.9$\,cm\,s$^{-2}$
 (see Fig.\,\ref{F:cspn} for the case of $Z = 0.02$,
 $\log\,g = 6.9$\,cm\,s$^{-2}$, and $T_{\rm eff} = 120$\,kK).
 We keep the distance of 0.46\,kpc to NGC\,6781, and
 vary $T_{\rm eff}$ and $L_{\ast}$ within the possible ranges,
 $L_{\ast} = 104-196\,L_{\sun}$ and $T_{\rm eff} = 110-140$\,kK,
 as determined in \S\,\ref{S:CS}, during the iterative model
 fitting to search for the best-fit model parameters that would
 reproduce the observational data.

\subsubsection{Nebular elemental abundances}

 For the elemental abundances of the nebula, 
 we adopt the empirically-determined abundances
 (Table\,\ref{T:abund2}; \S\,\ref{S-3.1}) as the input values.
 The nebular abundances are then refined via 
 model iterations within $\pm$3-$\sigma$ of the input
 values so that the best-fit abundances would reproduce 
 the observed emission line intensities.

 It should be pointed out here that the metal abundances 
 would affect cooling of the nebula, and hence, would 
 alter the nebula's temperature and ionization structures.
 As we saw in \S\,\ref{S:Abund}, the derivation of the C 
 abundance is definitely a source of uncertainties.
 The only option of the empirical derivation available to us 
 suggests the expected CEL C abundance $\epsilon(\mathrm{C})$ 
 of $8.56 - 9.00$ (Table\,\ref{T:abund2}).
 Hence, for the purpose of the present modeling, 
 we set $\epsilon(\mathrm{C})$ to be at the lower limit of 8.56 
 and keep it fixed during the model iteration.
 This will ensure that the best-fit model always satisfies 
 at least the lower limit of the progenitor mass 
 of 2.25\,$M_{\sun}$ (see \S\,\ref{abundmass}).

 The expected CEL $\epsilon(\mathrm{C})$ of 8.56 is also 
 consistent with the AGB nucleosynthesis model for the
 $2.25\,M_{\sun}$ stars \citep{Karakas:2010aa}.
 As we demonstrated in \S\ref{S:D},
 NGC\,6781 is very similar to NGC\,6720 in terms of 
 the elemental abundance pattern of the nebula and 
 evolutionary state of the CSPN (Table\,\ref{T:N6720}).
 The adopted CEL $\epsilon(\mathrm{C})$ of 8.56 for 
 NGC\,6781 is indeed very much consistent with that
 of 8.59 for NGC\,6720.
 In addition, we adopt a constant $^{12}$C/$^{13}$C ratio of
 20 determined by \citet{Bachiller:1997aa}.

 As for the unobserved elements including heavy metals, 
 we adopt the abundance values predicted with the AGB 
 nucleosynthesis model of the 2.5\,$M_{\sun}$ initial 
 mass star with
 $Z = 0.02$ \citep{Karakas:2010aa}.
 However, the Fe abundance is another exception, 
 because we overpredict the Fe lines when setting
 $\epsilon(\mathrm{Fe}) = 7.53$ as determined by 
 \citet{Karakas:2010aa}.
 The model $I$([Fe\,{\sc ii}]\,17.9\,{\micron}) and
 $I$([Fe\,{\sc iii}]\,4880\,{\AA}) line fluxes turn out to be
 31.2 and 2.6 (with respect to $I(${\hb}$) = 100$), respectively.
 
 Nevertheless, such strong Fe lines are seen neither 
 in the WHT/ISIS spectrum nor in the \emph{Spitzer}/IRS spectrum. 
 Therefore, we must adopt a lower Fe abundance. 
 Previously, \citet{Liu:2004ab} measured
 $\epsilon({\rm Fe}) = 6.20$ in NGC\,6720. 
 Thus, we adopt
 $\epsilon(\mathrm{Fe}) = 6.20$, following the same similarity 
 argument between NGC\,6781 and NGC\,6720 as in \S\,\ref{S:D}.
 For other Fe-peak elements such as Cr, Mn, Co, and Ni, 
 we adopt their solar values simply because their
 elemental abundances are unknown in NGC\,6781.

\subsubsection{Geometry of the nebula}

  
  \begin{figure}
   \centering
   \includegraphics[width=0.8\columnwidth,angle=-90,clip]{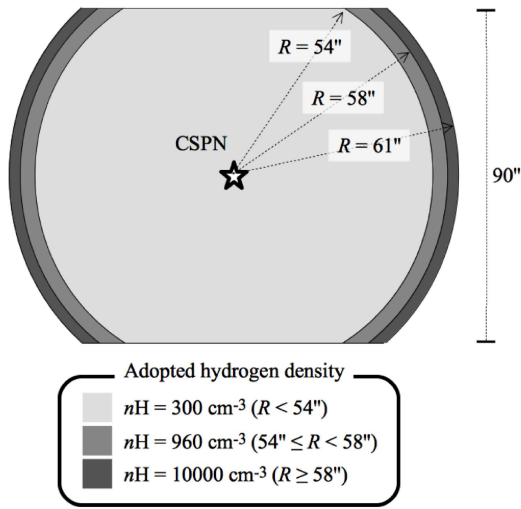}
   \caption{\label{F:GEO}%
   The adopted geometry and hydrogen density ($n$(H)) profile of
   NGC\,6781 in Cloudy model.
   }
  \end{figure}


 Many authors suggested that NGC\,6781 possessed a nearly pole-on
 cylindrical barrel structure,
 which surrounds the central cavity filled with tenuous highly ionized gas
 (e.g., \citealt{Bachiller:1993aa,Hiriart:2005aa,Schwarz:2006aa,Bergstedt:2015aa},
 as well as \citetalias{Ueta:2014aa}). 
 Hence, with the 1-D code {Cloudy}, 
 we represent the barrel wall structure by 
 thin, concentric layers of ionized gas and dusty PDR.
 Such an ``onion skin'' configuration naturally explains
 the observed co-spatial distributions of 
 various components at different temperature
 by the projection effect (Fig.\,\ref{F:H2CFHT}a).
 While clumps/filaments of H$_{2}$ surviving 
 in the ionized region would be plausible (Fig.\,\ref{F:H2CFHT}b), 
 we simply adopt this ``onion skin'' configuration 
 for the sake of 1-D model calculations,
 assuming that such molecular clumps/filaments would 
 not significantly alter the nebular energetics.

 However, we do take into account the barrel geometry of NGC\,6781 
 by invoking the ``cylinder'' option of {Cloudy}, 
 which approximates the cylindrical structure by removing polar caps 
 from a hollow sphere (which is the default 1-D spherically symmetric
 configuration). 
 We set the polar height of the cylinder to 90{\arcsec}, 
 which is the average value between 
 72{\arcsec}  
 \citep[suggested from the velocity channel maps in H$_{2}$;][]{Hiriart:2005aa}
 and 
 117{\arcsec} 
 \citep[suggested from the velocity channel maps taken in CO
 $J=3-2$ at 345.796\,GHz (866.96\,{\micron});][]{Bergstedt:2015aa},
 assuming that the H$_{2}$ and CO emission 
 arose from the same regions because of the similarities between H$_{2}$ and
 CO maps \citep{Bachiller:1993aa,Bergstedt:2015aa}.
  Fig.\,\ref{F:GEO} shows a schematic of the adopted geometry.

\subsubsection{Hydrogen density radial profile of the nebula}

 The input radial hydrogen density profile, $n_{\rm H}(R)$
 (where $R$ is the distance from the CSPN), 
 is adopted from our previous analysis \citepalias[][]{Ueta:2014aa}.
 In the central cavity surrounded by the barrel wall structure
 \mbox{($R < 54${\arcsec})}
 $n_{\rm H}(R)$ = 300\,cm$^{-3}$,
 whereas in the barrel wall \mbox{($54{\arcsec} \leq R < 58{\arcsec}$)}
 $n_{\rm H}(R)$ = 960\,cm$^{-3}$
 (Fig.\,\ref{F:GEO}).

 Unfortunately, $n_{\rm H}(R)$ beyond $58{\arcsec}$ cannot be 
 determined directly from the observed data, 
 because this radial region is where the surface brightness of the nebula
 decreases sharply to the detection limit in the narrow- and broad-band
 images of the object
 (and hence, the observational constraints are scarce). 
 Hence, as discussed in \S\,\ref{exdiag}, 
 we simply adopt a constant density of $n_{\rm H}(R) = 10^{4}$\,cm$^{-3}$
 beyond $58{\arcsec}$.
 The outer radius is then determined iteratively
 by increasing the thickness of this dusty PDR layer 
 until the model flux at 170\,{\micron} would reproduce the observed value, 
 which is one of our model calculation termination criteria.
 In the end, the outer radius is set to $61{\arcsec}$.
 The radial density profile of the nebula is also provided
 in Fig.\,\ref{F:GEO}.

 \subsubsection{Constant pressure model}

 One might surmise that the adopted $n_{\rm H}(R)$ radial profile
 would allow for a constant gas pressure model. 
 Therefore, we test a constant gas pressure model, 
 for which we adopt the average $\log_{10}$({\te}{\Ne})
 = 6.81\,K\,cm$^{-3}$ based on the radial {\te} and {\Ne} 
 profiles measured previously (\citetalias{Ueta:2014aa}). 
 The result is similar to the non-constant gas pressure model, 
 except for He\,{\sc ii} and [O\,{\sc iv}] lines.
 In order to avoid a collapse of the nebula, 
 the inner radius of the nebula has to be set larger.
 This correspondingly results in underestimates of 
 the line fluxes of these high I.P.\ lines.
 Also, NGC\,6781 does not seem to be embedded in a
 dense ISM region.
 Because of these reasons, we conclude that 
 the non-constant gas pressure model that we adopt 
 in the present investigation is a better approximation 
 to NGC\,6781 than a constant gas pressure model.

\subsubsection{Dust grains and PAH molecules}

 As we summarized in \S\,\ref{dust}, NGC\,6781 is 
 determined to be a PN rich in amorphous carbon. 
 Thus, the nebula's dusty PDR is expected to consist largely 
 of amorphous carbon (AC) plus neutral (and possibly ionized) PAHs,
 even though the C-richness of the nebula remains
 uncertain (see \S\,\ref{S-compabun}).
 \citet{Rouleau:1991aa} provided two types of optical constants 
 measured from samples 
 ``BE'' (soot produced from benzene burned in air) and 
 ``AC'' (soot produced by striking an arc between two
 amorphous carbon electrodes in a controlled Ar atmosphere).
 We test both of these BE and AC amorphous carbon grain models, 
 and we find that the AC type grain models yield generally better 
 overall fit to the observed mid-IR to far-IR dust continuum. 
 Thus, we adopt the AC type grain optical constants by \citet{Rouleau:1991aa}.
  We assume spherical grains and adopt the modified interstellar
 size distribution \citep[i.e., $n(a) \propto a^{-3.5}$,][]{Mathis:1977aa}
 with $a = 0.005-0.50$\,{\micron}, which are divided into 20 bins in model
 calculations.

 For PAHs, we adopt the radius
 $a$ in the range of 0.0004\,{\micron} (30 C-atoms)
 to 0.0081\,{\micron} (250 C-atoms) with the same size
 distribution as dust \citep[$a^{-3.5}$,][]{Mathis:1977aa},
 approximating the overall shape by a sphere
 (separated into the same 20 size bins). 
 We include both the neutral and charged PAH grains. 
 The optical constants for PAH-Carbonaceous grains are
 adopted from the theoretical work by \citet{2007ApJ...657..810D}.
 We permit the stochastic heating mechanism of PAH molecules
 in model calculations.

\subsubsection{Density-bounded vs.\ ionization-bounded}

\begin{figure}
  \centering
  \includegraphics[width=\columnwidth,clip]{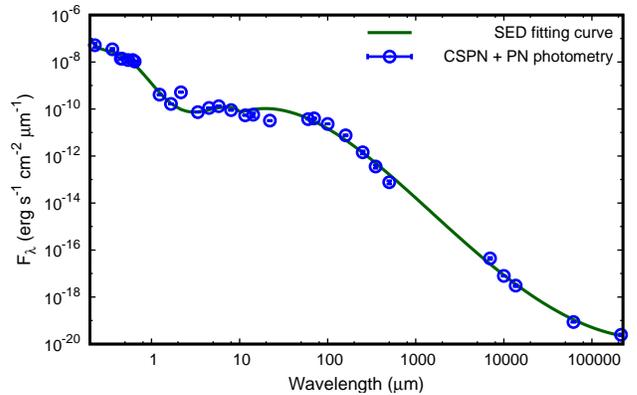}
  \caption{\label{F:pnplots}%
  The empirical SED of the CSPN plus PN
  (blue circles; Table\,\ref{T:obslog}) with the polynomial fitting
  (green curve).
  See Fig.\,\ref{S:spt} and Table\,\ref{T:obslog} for the origins of
  the empirical data.}
 \end{figure}

 Fig.\,\ref{F:pnplots} shows the SED of the CSPN plus PN based
 on the observed photometry from \emph{GALEX} 0.22\,{\micron}
 to radio 1.4\,GHz (Table\,\ref{T:obslog}; Fig.\,\ref{S:spt}).
 Using this empirical SED, we measure the integrated luminosity
 of 114\,$L_{\sun}$ at $D = 0.46$\,kpc for the CSPN plus PN. 
 The contribution to this SED only from the CSPN for the wavelength
 range of $\gtrsim 0.2$\,{\micron} is estimated to be
 4.6\,$L_{\sun}$. 
 Hence, the remainder has to come from the nebula, i.e., 
 $L_{\rm Neb} \approx 110\,L_{\sun}$.

 As for the luminosity of the CSPN, we already determined
 the empirical value of $L_{\ast} = 104-196\,L_{\sun}$ 
 based on Equation\,\ref{E:L} (\S\,\ref{S:D}). 
 Thus, NGC\,6781 could be a density-bounded PN
 (i.e., $L_{\rm Neb} < L_{\ast}$) as previously claimed
 by \citet{Schwarz:2006aa}. 
 However, the fact that NGC\,6781 possesses massive 
 molecular gas and dust components indicates that 
 it is more likely an ionization-bounded PN
 (i.e., $L_{\rm Neb} \approx L_{\ast}$).
 Realistically speaking, 
 whether a PN is density- or ionization-bounded is
 not necessarily straightforward, because both situations 
 could be present in one PN.
 In bipolar PNe such as NGC\,6781, both 
 ionization- and density-bounded conditions 
 are expected to be present in the nebula along the equatorial and polar directions, respectively.
 
 Based on the resemblance between the observed spatial distribution
 of the ionized gas and of the other (molecules and dust) components
 (Fig.\,\ref{F:H2CFHT}; \citealt{Zuckerman:1990aa,Hiriart:2005aa};
 \citetalias{Ueta:2014aa}),
 the transition from the ionized region to the PDR must be happening
 quite rapidly over a small radial range.
 Hence, we start model calculations with
 a nebula that is ionization-bounded at around $R = 55{\arcsec}$, which
 correspond to the outer radius of the central ring structure of the
 nebula and also the intensity peak of H$_{2}$ and CO emission 
 (see \S\,\ref{S-molmass}).
 The use of the cylinder option is also corroborated by the 
 density-bounded nature of the nebula expected in the polar 
 directions of the nebula.

 
\subsubsection{Additional heating source of H$_{2}$\label{tempfloor}}

 We introduce a high-density PDR wall beyond the ionization front
 in the model geometry (Fig.\,\ref{F:GEO})
 to explain the observed molecular emission.
 However, this causes significant underestimates of
 the observed H$_{2}$ and high $J$ CO line fluxes,
 as well as their column densities.
 This failure suggests 
 the presence of an additional heating source in the PDR.

 An obvious extra PDR heating source is the interstellar 
 radiation field (ISRF).
 However, no meaningful heating of the PDR can be achieved 
 by the ISRF in the present model for NGC\,6781:
 only $\lesssim 1$\,\% of the observed H$_{2}$ flux is reproduced
 by the nominal Galactic ISRF.
 Hence, it is unrealistic to expect to generate enough heating 
 to reproduce {\sl all} of the observed H$_{2}$ flux by the ISRF alone
 unless it is unrealistically enhanced.
 Thus, it is reasonable to expect something other than the ISRF 
 for a PDR heating source to explain the observed H$_{2}$ fluxes. 
 By the same token, the Galactic background cosmic-ray cannot 
 possibly work as a PDR heating source unless it is unrealistically 
 enhanced.

 \paragraph{Soft X-rays}

 Another extra heating source is soft X-ray emission 
 from a high-temperature CSPN
 as suggested by the presence of PNe in which 
 X-ray was detected \citep[e.g.,][]{Chu:2001aa,Kastner:2012aa,Montez:2015aa}.
 Soft X-rays (50\,ev -- 10\,keV) from a CSPN of $T_{\rm eff} > 100$\,kK
 can strengthen H$_{2}$ line emission, 
 because such high-energy photons would penetrate into the PDR
 beyond the ionization front \citep{Natta:1998aa}. 
 Using data from the \emph{Chandra} X-ray observatory,
 \citet{Montez:2015aa} examined the X-ray luminosities for a
 group of Galactic PNe including NGC\,6781.
 They found that no X-rays was
 detected from NGC\,6781 in the $0.3-8.0$\,keV energy band,
 while a simple blackbody of $T_{\rm eff} \sim 120-130$\,kK at 0.46\,kpc
 is sufficient for detectable X-ray fluxes in the $0.3-8.0$\,keV
 energy band (their Fig.\,14).
 Hence, the non-detection of X-ray emission in NGC\,6781
 is indicative of strong interstellar extinction or metal
 line-blanketing, either of which can suppress the X-ray
 emission to below the detection limit.

 We examine if X-ray emission possible from the CSPN of
 NGC\,6781 can result in a better fit to the observed H$_{2}$ 
 line fluxes under the following two scenarios:
 (1) the X-ray luminosity ($L_{\rm X}$) of the CSPN were to 
 power the entire observed mid-IR H$_{2}$ luminosity  
 ($\sim 5.59 \times 10^{33}$\,erg\,s$^{-1}$ at
 $D = 0.46$\,kpc; Table\,\ref{T:H2}), but were to be suppressed completely 
 by the extinction, and 
 (2) the CSPN possessed an atmosphere of sub-solar metallicity 
 to circumvent metal line-blanketing.
 The predicted H$_{2}$ line fluxes 
 with these X-ray emission enhancements
 would not reproduce the observed line fluxes 
 even if we adopted
 (1) an extra blackbody emitting
 in the range of $0.27-10.4$\,keV with the luminosity
 of $\sim 10^{33}$\,erg\,s$^{-1}$ at $10^{3}$\,kK or 
 (2) an atmosphere of Galactic halo metallicity for the CSPN.
 Therefore, we conclude that extra soft X-ray would not 
 possibly produce the observed H$_{2}$ line fluxes in NGC\,6781.

 \paragraph{Shock heating in the PDR}

 Yet another extra heating source is a mechanical heat input by shocks
 as suggested from the H$_{2}$ excitation diagram
 analysis (\S\,\ref{S:H2} and \ref{exdiag}).
 This idea, previously used in a study of the C-rich 
 PN NGC\,7027 by \citet{Hasegawa:2000aa},
 can work to excite H$_{2}$ lines in regions far
 enough away from the CSPN. 
 As {Cloudy} does not handle shocks,
 the desired extra heating by shocks is achieved 
 by invoking the ``temperature floor'' option, 
 which forces the predetermined value of 
 the electron temperature $T_{\rm e}$ 
 over a specific region (see \S\,\ref{bf-model}).
 We iteratively search for the optimum floor temperature
 in the PDR ($R \geq 58{\arcsec}$) between 800 and 1600\,K.
 This temperature range is suggested by  
 the H$_{2}$ excitation temperatures derived 
 from the excitation diagram analysis (\S\,\ref{exdiag}).

 While the use of a ``temperature floor'' helps to  
 reproduce the observed warm H$_{2}$ lines (except for 17.04\,{\micron}), 
 as well as high $J$ CO, and OH lines,\footnote{Because OH$^{+}$ is
 not available in {Cloudy}, we are unable to use the observed OH$^{+}$
 line fluxes.},
 the adaptation of the ``temperature floor'' also introduces
 negative side effects such as
 (1) suppression of molecular lines with 
 lower excitation temperatures, and
 (2) overestimation of atomic gas line fluxes
 such as far-IR {\oi} and [{\cii}] lines 
 which have low excitation energy at the upper levels. 
 These side effects would make the mass fraction of the atomic 
 and molecular gas with respect to the neutral (atomic + molecular) 
 gas highly uncertain, primarily because the model would fail 
 to account for the cold molecular component while introducing 
 the corresponding amount of extra atomic gas component
 (as the total amount of neutral gas was practically set by 
 the input hydrogen density profile; Fig.\,\ref{F:GEO}).
 However, the proper amount of the warm and cold molecular 
 components, as well as the atomic gas component can be
 recovered (\S\,\ref{S-mol}).

\subsection{The best-fit model\label{bf-model}}

\subsubsection{Model iteration}

 To find the best-fit model, we vary the following 13 parameters --
 $T_{\rm eff}$, 
 $L_{\ast}$, 
 the inner radius of the shell ($R_{\rm in}$),
 elemental abundances ($\epsilon$(He/N/O/Ne/Si/Cl/Ar), except
 for $\epsilon(\mathrm{C})$, which was fixed), 
 dust and PAH mass fraction, and 
 the floor temperature of the PDR --
 within a given range by using the {\it optimize} command
 available in {Cloudy}.
 We terminate iterative calculations when any one of the predicted flux
 densities, $F_{\nu}$(170\,{\micron}), $F_{\nu}$(250\,{\micron}) or
 $F_{\nu}$(350\,{\micron}), reaches the corresponding observed
 value. 
 Practically, the terminating conditions would determine
 the maximum $R_{\rm out}$, i.e., the thickness of the dense PDR 
 beyond the inner ionized region, by setting the amount of far-IR 
 continuum emission.
 The flux densities at 170, 250, 350\,{\micron} are selected 
 as constraints because there are no strong emission lines in these 
 bands and they can be compared with measurements made in the 
 PACS 160 and SPIRE 250, and 350\,{\micron} bands. 
 In this sense, $R_{\rm out}$ is not a free parameter.

 The best-fit model is determined by the minimum $\chi^{2}$ (16
 for the best-fit model) calculated from the following 136
 observational constraints: 37 broadband fluxes, 78 gas emission
 line fluxes relative to {\hb} as well as $I$({\hb}),
 19 flux densities in mid-IR, far-IR, and radio wavelengths, and
 the ionization boundary radius ($R_{\rm IB}$).
 We define $R_{\rm IB}$ as the radial distance from the CSPN 
 at which $T_{\rm e}$ drops below 4000\,K: 
 below such a temperature, no ionized gas emission
 lines except for [{\cii}] and {\sii} would be measurable.

\begin{figure*}
   \centering
   \includegraphics[width=\textwidth,clip]{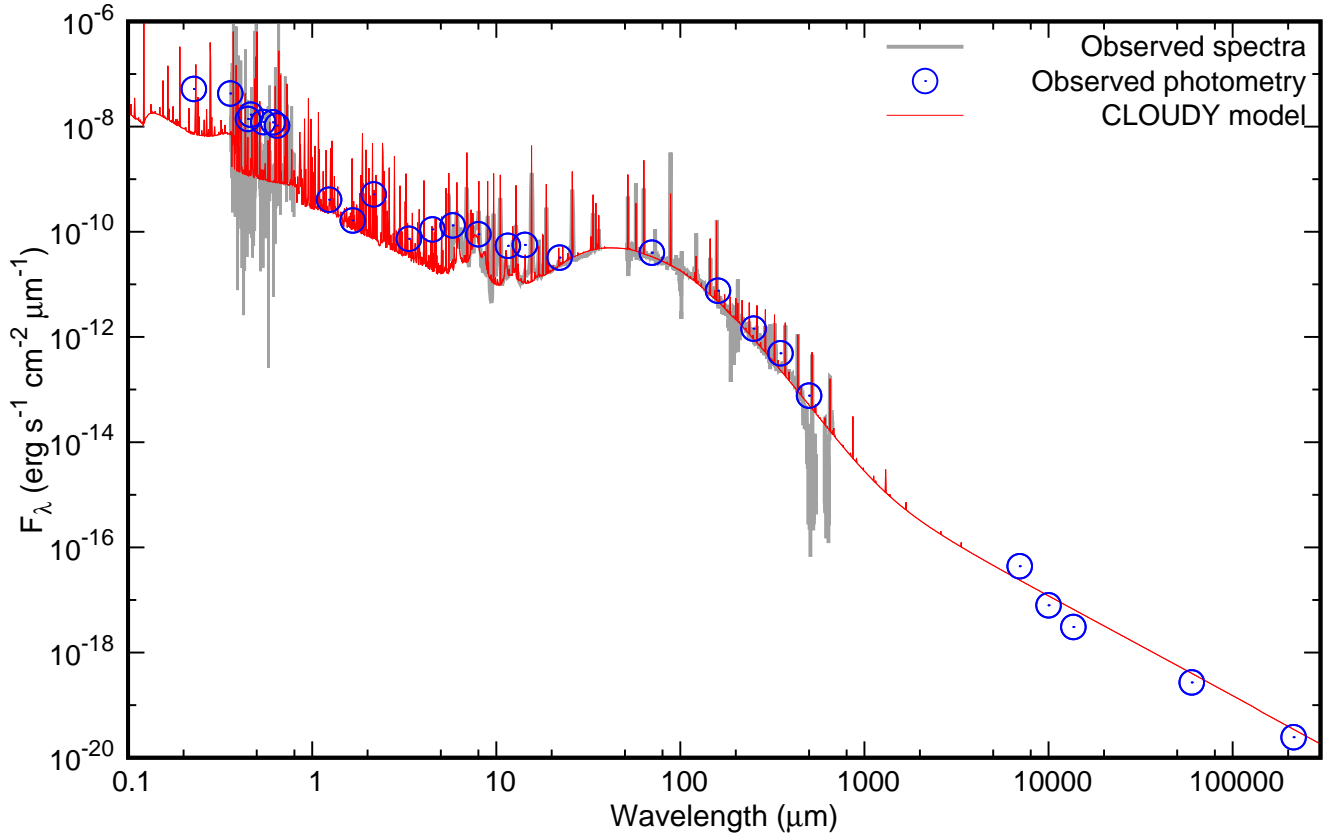}
   \caption{\label{F:SED}%
   The full SED of the best-fit {Cloudy} model of NGC\,6781
   (red line; $R=300$), compared with the observational constraints
   (Table\,\ref{AT:model}):
   photometry data (blue circles) and spectroscopy data (grey line). 
   }
  \end{figure*}

\begin{figure}
   \centering
   \includegraphics[width=\columnwidth,clip]{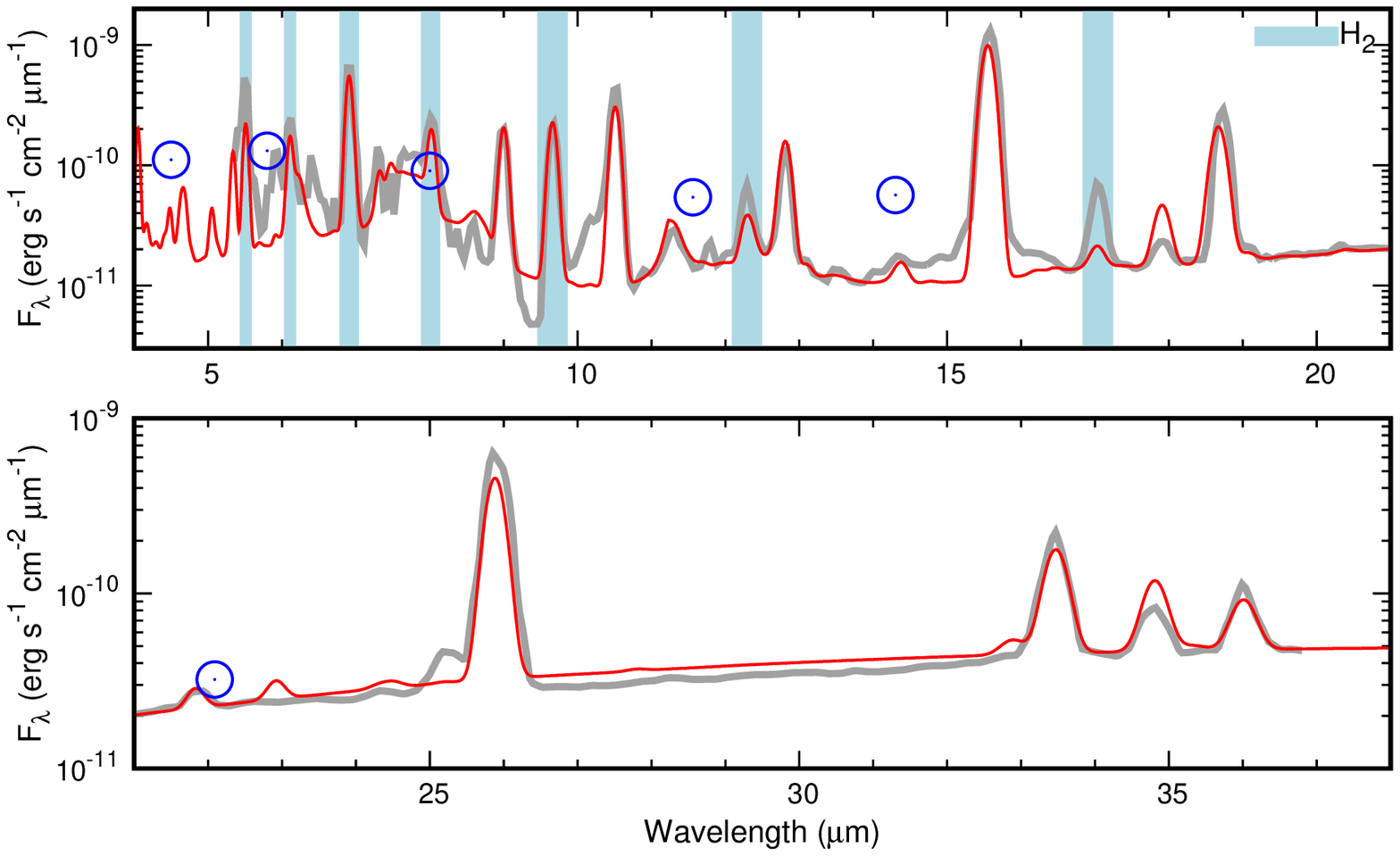}\\
   \includegraphics[width=\columnwidth,clip]{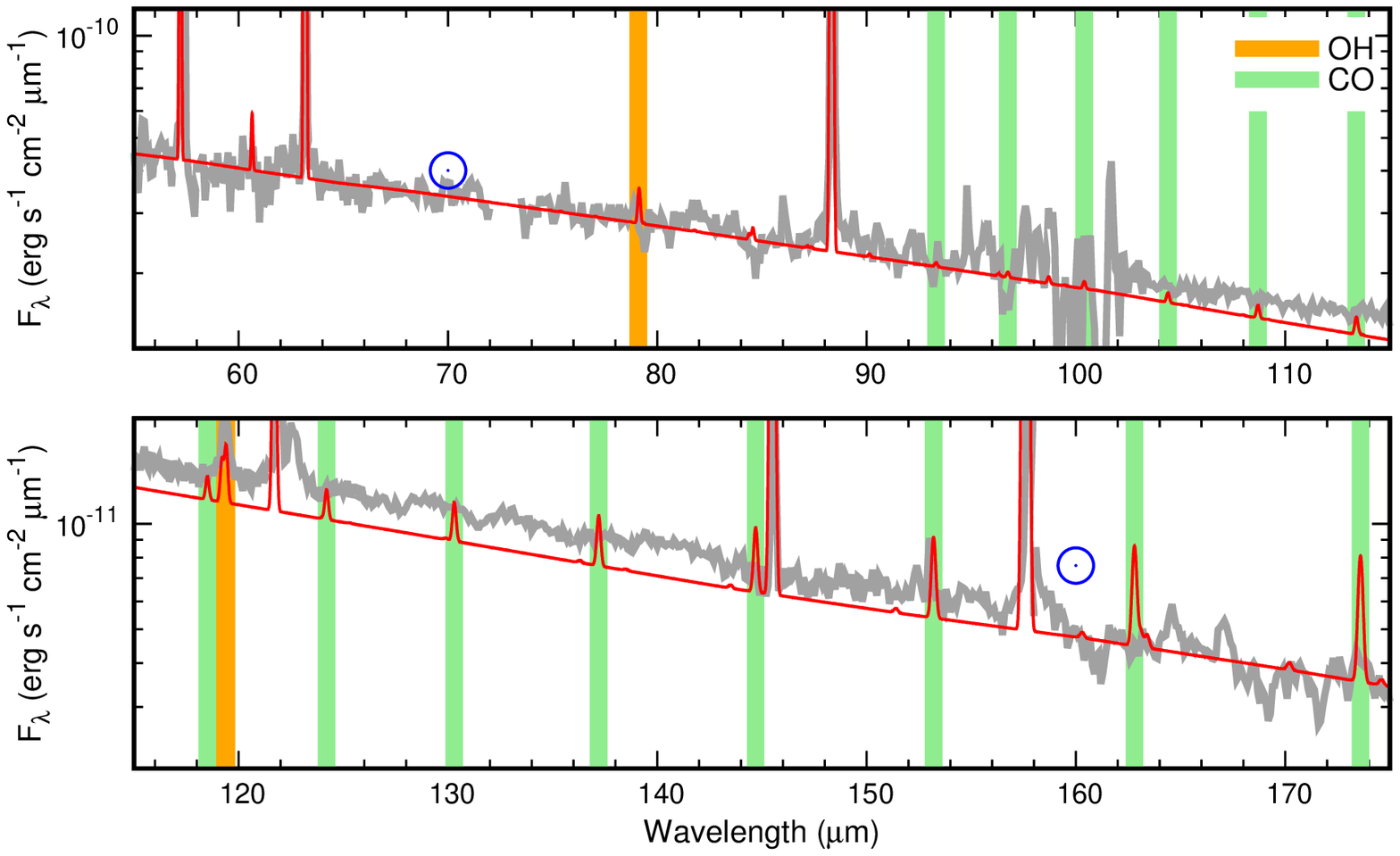}\\
   \includegraphics[width=\columnwidth,clip]{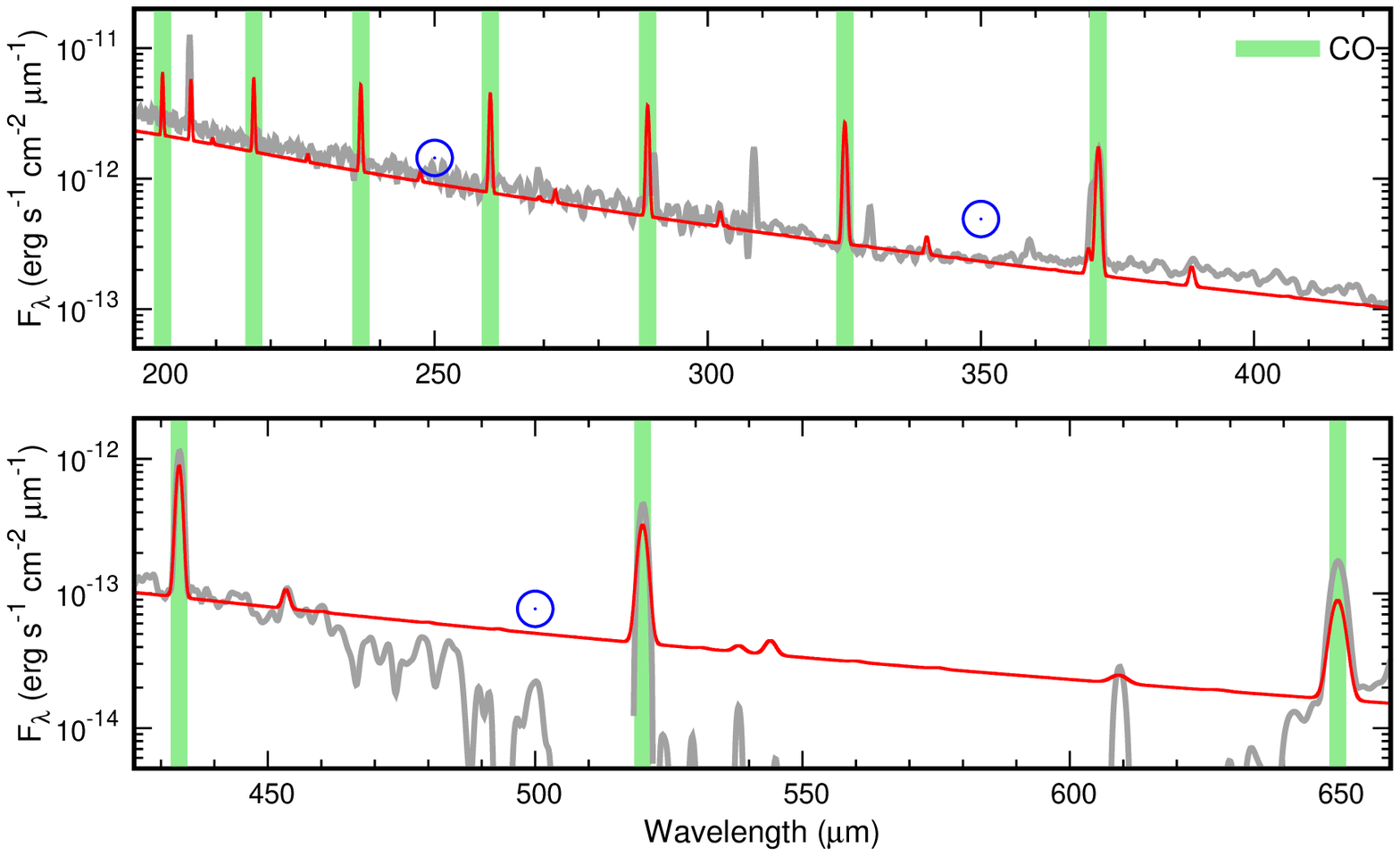}
   \caption{\label{F:SED2}%
   Comparison between the SED of the best-fit {Cloudy} model (red line,
   with undetectable atomic and molecular lines ($<10^{-2}$\,\% of the
   {\hb} flux) removed; $R=100$ in the top 2 frames and $R=480$
   in the other frames, corresponding to the resolution of the
   instrument in the respective bands) and the observational data
   (spectra in grey line and photometry in blue circles) in IR regions 
   (top: \emph{Spitzer}/IRS; middle: \emph{Herschel}/PACS;
   bottom: \emph{Herschel}/SPIRE).
   The positions of molecular line emission are highlighted: 
   rotational H$_{2}$ lines (light blue),
   OH (yellow), and $^{12}$CO (light green).
   See Table\,\ref{AT:model}.}
  \end{figure}

 In Table\,\ref{T:model}, we summarize the best-fit parameters.
 The SED of the best-fit model, in comparison with the observational
 data, is presented in Fig.\,\ref{F:SED}. 
 Fig.\,\ref{F:SED2} is also provided to show the quality of
 the best-fit model with blow-ups of various wavelength ranges
 with major emission lines. 
 In Table\,\ref{AT:model}, we list the best-fit model vs.\ observed quantities of 
 gas emission line fluxes relative to {\hb},
 broadband fluxes relative to {\hb} and flux densities.

 Here, we can retroactively check if the empirical 
 estimates and adaptation of certain quantities 
 in determining the input model parameters
 are actually corroborated by the best-fit model.
 In \S\,\ref{S-plasma}, we used the empirical formulae 
 to estimate the amount of RL contributions to 
 {\oiii}\,4363\,{\AA}, {\oii}\,7320/30\,{\AA}, and {\nii}\,5755\,{\AA}
 lines in deriving {\te}.
 The best-fit model yields 
 $I_{\rm R}$/$I$({\oiii}\,4363\,{\AA}) = $0.67\,\%$, 
 $I_{\rm R}$/$I$({\oii}\,7320/30\,{\AA}) = $1.13\,\%$, and  
 $I_{\rm R}$/$I$({\nii}\,5755\,{\AA}) = $0.31\,\%$,
 which are consistent with the empirical determinations
 adopted
 ($0.73\,\%$, $2.19\,\%$, and $0.54\,\%$, respectively).

 As for the ICFs used in determining the elemental abundances, 
 we can compare the adopted ICFs based on I.P.\ and the ICFs
 calculated by the best-fit Cloudy model based on the ionization 
 fraction of each element in the volume average in
 Table\,\ref{T:ICF}.
 While the values turn out to be consistent in general,
 discrepancies are found in Cl from the uncertain Cl$^{+}$ fraction 
 and in Si from the largely uncertain $\epsilon$(Si) and ICF(I.P.). 
 According to the best-fit model, 
 the fraction of Cl$^{+}$ to Cl is 0.38 and
 of Si$^{+}$ to Si is 0.668.


 \begin{table}
  \centering
  \footnotesize
  \renewcommand{\arraystretch}{0.85}
  \tablewidth{\columnwidth}
  \caption{\label{T:model}%
  Characteristics of the best-fit {Cloudy} model of NGC\,6781}
  \begin{tabularx}{\columnwidth}{@{}ll@{}}
   \hline\hline
   {Parameters of the CSPN}      &{Values}\\
   \hline
   $L_{\ast}$ / $T_{\rm eff}$ / $\log\,g$  &121\,$L_{\sun}$ / 120\,870\,K /
       6.9\,cm s$^{-2}$ \\
   $D$ &0.46\,kpc\\
   \hline
   {Parameters of the Nebula}      &{Values}\\
   \hline
   $\epsilon$(X) &He:11.02, C:8.56, N:8.10, O:8.64, \\
                 &Ne:8.00, Si: 6.25, S:6.82, Cl:5.01, \\
                 &Ar:6.22, Fe:6.20\\
                 &Others: \citet{Karakas:2010aa}\\
   Geometry (Fig.\,\ref{F:GEO})     &``Cylinder'' with height =
       90{\arcsec}
       (0.201\,pc)\\
                 &Inner radius ($R_{\rm in}$)   = 0.52{\arcsec}  (0.001\,pc)\\
   &Ionization boundary ($R_{\rm IB}$) = 55{\arcsec}
       (0.123\,pc)\\
                   &Outer radius ($R_{\rm out}$)  = 61{\arcsec}   (0.135\,pc)\\
   Adopted $n_{\rm H}$ (Fig.\,\ref{F:GEO})
   & Inner Cavity ($R < 54{\arcsec}$): 300\,cm$^{-3}$\\
   & Barrel Wall ($54{\arcsec} \leq R < 58{\arcsec}$): 960\,cm$^{-3}$\\
   & PDR ($58{\arcsec} \leq R < 61{\arcsec}$): 10$^{4}$\,cm$^{-3}$\\
   Temperature {\te} 
   & Inner Cavity ($R < 54{\arcsec}$): $23\,820-10\,260$\,K\\
   & Barrel Wall ($54{\arcsec} \leq R < 58{\arcsec}$): $10\,260-2\,750$\,K\\
   & PDR ($58{\arcsec} \leq R < 61{\arcsec}$): $2\,750-1\,420$\,K\\
   Filling factor ($f$) &1.0\\
   $\log_{10} I$({\hb}) &--9.890\,erg s$^{-1}$ cm$^{-2}$ (de-reddened)\\
   temperature floor &1420\,K\\
   Mass              &ionized gas: 0.094\,$M_{\sun}$\\
   &neutral (atomic $+$ molecular) gas:
       0.31\,$M_{\sun}$$^{\dagger}$\\
   \hline
   Parameters of the Dust      &{Values}\\
   \& PAHs\\
   \hline
   Particle size &PAH (neutral \& ionized): 0.0004-0.011\,{\micron},\\
                 &AC: 0.005-0.50\,{\micron}\\
   Temperature   &PAH (neutral): 71-515\,K, \\
                 &PAH (ionized): 72-367\,K, \\
                 &AC:  22-299\,K\\  
   Mass          &PAH (neutral): 3.30(--7)\,$M_{\sun}$\\
                 &PAH (ionized): 2.46(--6)\,$M_{\sun}$\\
                 &AC: 1.53(--3)\,$M_{\sun}$\\
   GDR           &268\\
   \hline
  \end{tabularx}
    \begin{minipage}{\columnwidth}
     \tablewidth{\columnwidth}
  \tablenotetext{\dagger}{\footnotesize We corrected the molecular gas mass of
     0.11\,$M_{\sun}$ and the atomic gas mass of 0.20\,$M_{\sun}$.
     See \S\,\ref{S-mol} and Table\,\ref{T-comp}.}
   \end{minipage}
  
 \end{table}


  \begin{table}
   \centering
   \footnotesize
  \renewcommand{\arraystretch}{0.85}
  \tablewidth{\columnwidth}
     \caption{The comparison between the ICFs estimated based on
     I.P. (adopted for elemental abundance derivations in \S\,\ref{S:Abund})
   and predicted by Cloudy model.}
    \begin{tabularx}{\columnwidth}{@{}cccccc@{}}
 \hline\hline   
 X  &ICF(I.P.)&ICF(Model) &X  &ICF(I.P.)&ICF(Model)\\
\hline
 He &1.00     &1.00      &Si &6.80 $\pm$ 1.75&1.50\\
 C  &2.03 $\pm$ 0.32&1.89&S  &1.00           &1.01\\
 N  &1.05 $\pm$ 0.06&1.08&Cl &1.17 $\pm$ 0.09&1.66\\
 O  &1.00           &1.00&Ar &1.17 $\pm$ 0.09&1.15\\
 Ne &1.00           &1.03\\
    \hline
    \end{tabularx}
   \label{T:ICF}
  \end{table}

  
 As mentioned in the previous section (\S\,\ref{tempfloor}),
 the best-fit model is achieved by forcing
 the region of constant temperature at 1\,420\,K
 in the PDR. 
 This constant temperature region is established from
 58.06{\arcsec} to 61{\arcsec}, that is, the radial temperature 
 drops precipitously from $2\,750$\,K at 58{\arcsec} to 
 $1\,420$\,K at 58.06{\arcsec}, but is maintained at $1\,420$\,K
 from 58.06{\arcsec} to 61{arcsec}
 to reproduce the observed molecular (H$_{2}$, CO, and OH)  
 line fluxes.
 In this region, the relative proportion of molecular gas
 is maintained.
 So is the relative proportion of atomic gas.
 
 In reality (of the presumed shocked H$_2$ scenario), 
 however, 
 shocked molecular regions are highly localized, and hence, 
 the relative proportion of molecular gas would keep
 increasing radially while that of atomic gas would keep decreasing.
 Therefore, with the presence of this constant temperature PDR, 
 the amount of the atomic gas component is bound to be overestimated
 in the PDR, i.e., the [C\,{\sc ii}] and 
 {\oi} line fluxes are overpredicted 
 (by a factor of 3 to 9; 
 Fig.\,\ref{F:SED2}, Table\,\ref{AT:model}). 

 While our {Cloudy} model extends as far out as 
 $R_{\rm out} = 61\arcsec$, the optical ISIS and far-IR 
 \emph{Herschel}/PACS observations do not detect 
 these [C\,{\sc ii}] and {\oi} lines with a sufficient 
 signal level this far out in the PDR (i.e., the detection 
 limit is reached at $R \approx 55{\arcsec}$).
 If we stopped model calculation at $R_{\rm IB}$ of 55{\arcsec},
 we would obtain reasonable predictions of atomic line fluxes:
 for instance, 
 $I$([O\,{\sc i}]\,63\,${\micron}) = 25.07$ (33.18, observed),
 $I$([O\,{\sc i}]\,145\,${\micron}) = 2.21$ (2.90, observed), and
 $I$([C\,{\sc ii}]\,157\,${\micron}) = 8.25$ (15.9, observed). 
 However, of course, we would not be able to fit molecular lines at all
 (e.g., $I$(H$_{2}$\,9.67\,${\micron}) = 8(-5)$ for the model
 vs.\ 25.79 observed).

 In the present work, we adopt the average [C\,{\sc ii}]
 and {\oi} line fluxes measured in the entire 
 PACS IFU field of view
 (over both of the ``center'' and ``rim'' positions; Fig.\,\ref{S:slit})
 and the model-predicted [C\,{\sc ii}] and {\oi} line fluxes are 
 deemed overestimated as a result. 
 However, we actually measure fluxes as high as 
 $I$({\oi}\,63\,{\micron}) = 103,
 $I$({\oi}\,145\,{\micron}) = 8.69, and 
 $I$([{\cii}]\,157\,{\micron}) = 27.24
 in individual PACS spaxels over the barrel wall.
 Because there is no more data available to fit the model,
 especially the atomic component of the PDR, 
 we have to leave these remaining discrepancies 
 as issues to be resolved in future
 when we obtain more sensitive data of the PDR and beyond.
 We will discuss the molecular component in detail 
 later in \S\,\ref{S-mol}.

\subsubsection{Amorphous silicate grain model} 

 To explore the possible O-rich nature of NGC\,6781 (\S\,\ref{S-compabun}), 
 we also construct the other ``best-fit''
 model with amorphous silicate grains,
 adopting spherical grains of $0.05-0.50$\,{\micron}
 radius (Appendix Fig.\,\ref{F-sil}).
 Overall, the best-fit model with amorphous carbon grains 
 fit the observed continuum much better than the best-fit model 
 with amorphous silicates.
 To fit the observed dust continuum with amorphous silicate grains, 
 we have to reduce the amount of small grains in order not to 
 produce any recognizable 10\,{\micron} silicate feature 
 while achieving reasonable continuum fluxes in the far-IR.
 It is almost impossible to fit the dust continuum 
 both in the mid-IR ($10-40\,\micron$)
 and in the far-IR ($> 70\,\micron$) simultaneously 
 with amorphous silicate grains 
 because amorphous silicates emits continuum only weakly 
 beyond 70\,{\micron}. 
 Therefore, 
 we conclude that NGC\,6781 was more likely C-rich
 in terms of the circumstellar dust composition.

\subsubsection{Evolutionary status and age of the object}


\begin{figure}
\centering
\includegraphics[width=\columnwidth,clip]{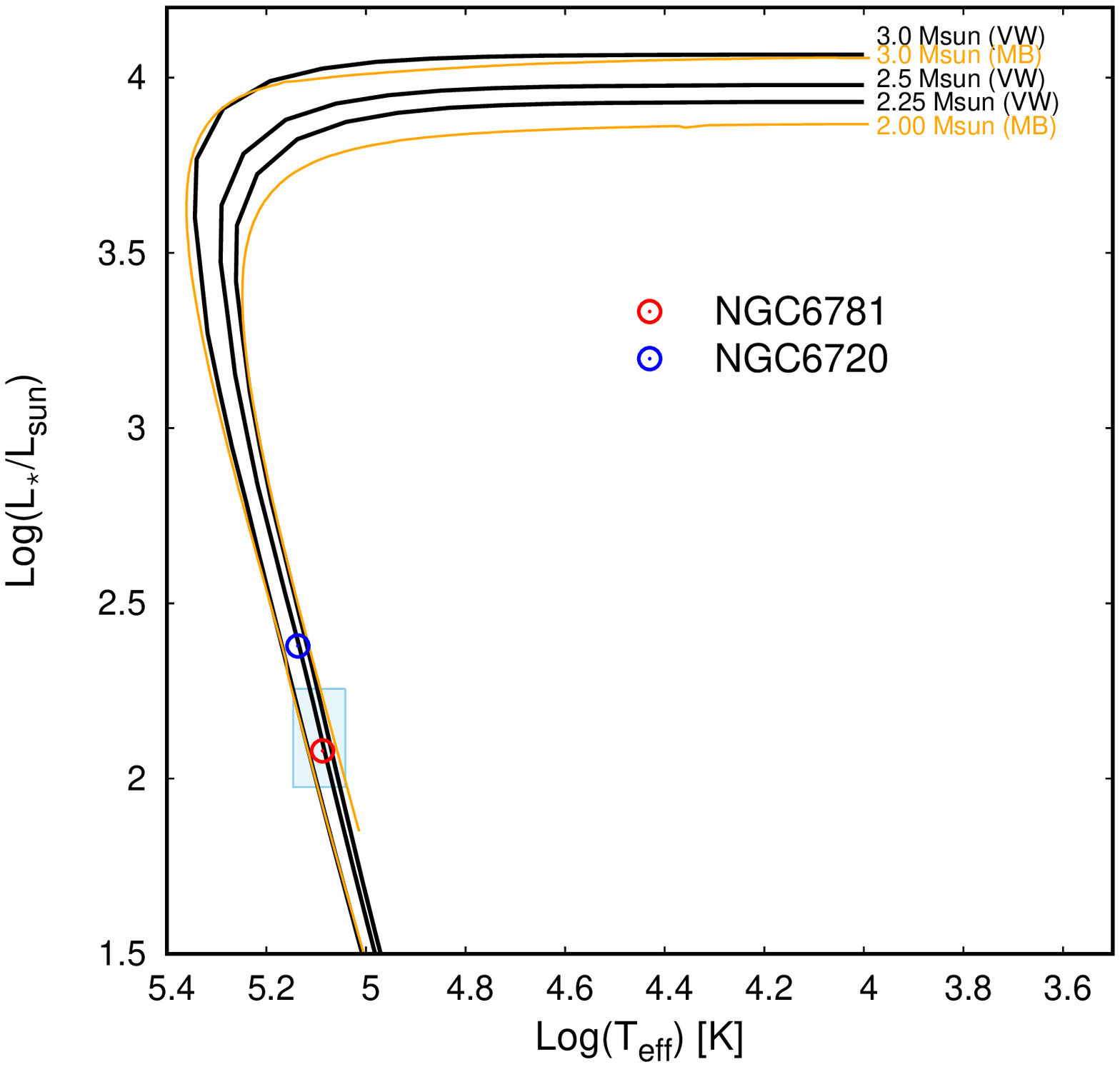}
 \caption{\label{F:HR-last}%
 Comparison between the best-fit {Cloudy} model of NGC\,6781 
 (red circle; $L_{\ast}$ and $T_{\rm eff}$ of the CSPN) and the 
 the post-AGB evolutionary tracks (black lines) of 
 2.25, 2.5, and 3.0\,$M_{\sun}$ 
 initial-mass stars \citep[][also shown in
 Fig.\,\ref{F:HR}]{Vassiliadis:1994ab}.
 We also plot the post-AGB evolutionary tracks (orange lines) of
 2.0 and 3.0\,$M_{\sun}$ stars with $Z = 0.02$ by
 \citet{Miller-Bertolami:2016aa}.
 The light-blue box indicates the empirical
 $L_{\ast}-T_{\rm eff}$ parameter range as discussed in \S\,\ref{S:D}. 
 The best-fit {Cloudy} model of NGC\,6720 (blue circle; 
 $L_{\ast}$ and $T_{\rm eff}$ of the CSPN;
 \citealt{van-Hoof:2010aa}) is also plotted for comparison.}
\end{figure}


 Fig.\,\ref{F:HR-last} shows how the best-fit model compares with 
 the adopted post-AGB evolutionary tracks of \citet{Vassiliadis:1994ab}.
 In the same plot, the best-fit model of NGC\,6720 
 by \citet{van-Hoof:2010aa} is also displayed to confirm the similarity 
 between the two in terms of the evolutionary status.
 A comparison between the evolutionary tracks implies that
 the progenitor of both NGC\,6781 and NGC\,6720 is a $\sim2.5\,M_{\sun}$ 
 star of $Z = 0.02$ and that the post-AGB age (i.e., the
 time since the cessation of AGB mass loss) is $\sim$9400\,yrs for NGC\,6781.

 In addition, we plot in Fig.\,\ref{F:HR-last} 
 the evolutionary tracks of
 \citet[][orange tracks of 2.0 and 3.0\,$M_{\sun}$ stars]{Miller-Bertolami:2016aa}. 
 These newer tracks are computed to address the shorter-than-expected
 timescales for Galactic bulge PNe.
 Their models with $Z = 0.01$ would take
 $\sim$3000, $\sim$2700, and $\sim$8000\,yrs 
 to reach  $T_{\rm eff} = 120\,870$\,K 
 for 2.0, 2.5, and 3.0\,$M_{\sun}$ stars, respectively,
 while models with $Z=0.02$ would take 
 $\sim$2600 to $\sim$12\,000\,yrs
 to reach the same temperature for 
 the 2.0\,$M_{\sun}$ and 3.0\,$M_{\sun}$ models (with), respectively
 (no model track is given for 2.5\,$M_{\sun}$).
 Thus, the post-AGB age of a $2.5\,M_{\sun}$ progenitor 
 with $Z = 0.02$ would be $\sim$3000\,yrs.

 Following the method suggested by \citet{ODell:2007aa}, 
 the empirical dynamical age of a PN can be approximated 
 simply by
 \begin{equation}
  t_{\rm dyn} \simeq 
  \frac{R}{\frac{V_{\rm exp}({\rm today}) + V_{\rm exp}({\rm AGB})}{2}},
 \end{equation}
 where $V_{\rm exp}({\rm today})$ is the present-day shell
 expansion velocity and $V_{\rm exp}({\rm AGB})$ is the shell expansion
 velocity at the beginning the AGB phase. 
 In this formulation, the shell
 expansion velocity is taken to be the rough ``average'' between 
 the AGB wind velocity and the fast wind velocity. 
 Assuming \mbox{$V_{\rm exp}({\rm AGB}) = 16-22$\,{\kms}} 
 (corresponding to the observed expansion velocity of the cold 
 CO gas, \citealt{Bachiller:1993aa,Bergstedt:2015aa}), 
 \mbox{$V_{\rm exp}({\rm today}) = 12$\,{\kms}} 
 (from the {\nii} line; \citealt{Arias:2002aa}),
 and $R_{\rm IB} = 55{\arcsec}$ (the ionization front radius), 
 the dynamical age would be roughly $7100-8600$\,yrs.

 \citet{Gesicki:2016aa} suggested 
 $t_{\rm dyn} \simeq (5/7) \times (R_{\rm IB}/V_{\rm exp})$ 
 based on hydrodynamical model calculations.
 Adopting $V_{\rm exp} = 12$\,{\kms} as above,
 the hydrodynamical age would be 7140\,yrs. 
 Thus, the theoretical post-AGB age inferred from 
 the {Cloudy} best-fit model and the evolutionary tracks
 by \citet{Vassiliadis:1994ab}
 is comparable to these (hydro)dynamical age estimates.
 Meanwhile, 
 the much-shorter post-AGB evolutionary time suggested by 
 the evolutionary tracks by \citet{Miller-Bertolami:2016aa} 
 is more problematic to reconcile
 because the observed PN size 
 would not be consistent with the observed expansion velocity,
 provided that the best-fit distance is 0.46\,kpc 
 (\S\,\ref{S:D})

\subsubsection{Molecular gas components \label{S-mol}}

 Here, we look into the molecular component 
 of the best-fit model, especially into the PDR. 
 We begin by comparing the model-predicted and empirically-derived 
 molecular column densities of H$_{2}$, CO, and OH$^{+}$
 (Table\,\ref{T-molec}).
 The model-predicted results are derived by taking into account
 all of the gas components (i.e., molecular, atomic, and ionized)
 self-consistently allowing molecular formation processes (e.g,
 formation on dust grain surfaces and in the gas phase, and so on).

 As discussed above (\S\,\ref{tempfloor}), 
 we introduced the warm temperature component in the PDR 
 as a necessary extra heating source
 to reproduce the observed H$_{2}$, CO, and OH lines.
 However, the achieved general agreement between the model and empirical 
 column densities (Table\,\ref{T-molec})
 and line intensities (Fig.\,\ref{F:SED2}; Table\,\ref{AT:model})
 permit qualitative characterization of the PDR in NGC\,6781.

 The best-fit floor-temperature of 1420\,K is consistent
 with the empirical estimates of $T(\mathrm{H}_2) = 1279 \pm 109$\,K
 and $1161 \pm 72$\,K by the single- and two-temperature
 excitation diagram fitting, respectively (\S\,\ref{exdiag}).
 This suggests that H$_{2}$ is most likely in LTE
 and its kinetic temperature is about 1420\,K. 
 With this kinetic temperature, CO and OH lines are fit 
 reasonably well. 
 If we are to fit just the high $J$ CO lines, 
 the best-fit floor-temperature for CO would be 680\,K.
 Either way (fitting with or without H$_{2}$), 
 the (kinetic) temperature of CO gas  
 would still be very much higher than excitation temperature of
 $\sim 60$\,K
 (\citetalias{Ueta:2014aa}).
 This discrepancy can be mitigated if CO is assumed to be in non-LTE.
 Given the difference in the number density between H$_2$ and CO, 
 CO could yet be being thermalized while H$_2$ already is.

 Thus, 
 we examine the excitation temperature of each CO line 
 using the 1-D non-LTE radiative transfer code RADEX 
 \citep{van-der-Tak:2007aa}.
 In RADEX calculations, we adopt 
 the kinetic temperature of 1420\,K,
 a constant $n$(H) = 10$^{4}$\,cm$^{-3}$,
 and $\log_{10}$\,$N$(CO) = 15.13\,cm$^{-2}$ 
 as in the {Cloudy} model.
 The RADEX results (Table\,\ref{T-RADEX}) suggest
 that the excitation temperature of high $J$ CO lines
 is $70-80$\,K on average, supporting the non-LTE
 condition for CO.
 We, therefore, conclude that the best-fit {Cloudy} model 
 properly account for the presence of the warm component.

 The best-fit model predicts the amount of molecular 
 gas in the PDR to be $4.15(-3)$\,$M_{\sun}$, 
 which accounts only for the warm H$_{2}$ component 
 (i.e., there is no other ``cold'' molecular components 
 in the best-fit model).
 Meanwhile, this model prediction is actually consistent with 
 the empirical estimate of $2.5(-3)$\,$M_{\sun}$ for the warm
 component (\S\,\ref{S-molmass}).
 However, the presence of the cold molecular component is
 very much expected based on the excitation diagram analysis 
 (\S\,\ref{exdiag}) as well as the non-LTE analysis 
 we just saw above.
 In reality, there is probably a temperature gradient
 in the PDR along the polar direction, which empirically 
 manifests itself as the multi-temperature fit of 
 the excitation diagram analysis and the non-LTE nature 
 of the CO distribution.

 Now, given that the best-fit model already properly accounts for 
 the amount of ionized and neutral (atomic $+$ molecular) gas,
 the cold molecular component that should exist 
 must have been treated as part of the atomic gas component,
 as mentioned earlier (\S\,\ref{tempfloor}).
 Here, by adopting the ratio of the empirically-determined
 cold H$_{2}$ mass to warm H$_{2}$ mass
 ($24.8 = 6.2(-2)\,M_{\sun}/2.5(-3)\,M_{\sun}$; \S\,\ref{S-molmass}),
 we can infer the amount of the cold molecular component 
 to be expected in the best-fit model,
 $1.12(-1)\,M_{\sun}$ $(= 4.15(-3)\,M_{\sun} \times 24.8)$.
 From this, we conclude that 
 the modified best-fit model estimates of the mass of 
 the cold molecular, warm molecular, and atomic gas components
 are 
 $1.12(-1)\,M_{\sun}$, 
 $4.15(-3)\,M_{\sun}$, and
 $1.99(-1)\,M_{\sun}$ $(= 3.11(-1)\,M_{\sun} - 1.12(-1)\,M_{\sun}$),
 respectively (see also Table\,\ref{T-comp}).

 We end the discussion on the molecular component in NGC\,6781
 by pointing out two lesser issues to be resolved that are
 beyond the scope of the present work.
 One is obviously the presence of the extra heating source.
 We incorporated the warm-temperature component in the model PDR
 assuming that shock interactions between the slower AGB wind and
 faster PN wind
 would provide sufficient extra heating to the PDR at the required level.
 Nonetheless, this extra heating source should be identified and 
 self-consistently incorporated in the future.
 The other issue is the discrepancy in the OH$^{+}$ column densities.
 This may well be due to a relatively more uncertain chemical network
 around OH$^{+}$ and/or outdated reaction parameters in the 
 astrochemistry network installed in {Cloudy}.
 However, the cause of the OH$^{+}$ column density discrepancy is also 
 unclear at this moment.

 
 \begin{table}
  \centering
  \footnotesize
  \renewcommand{\arraystretch}{0.85}
  \caption{Comparison  
  between the best-fit model-predicted and
  empirically-derived molecular column densities. 
  \label{T-molec}}
  \begin{tabularx}{\columnwidth}{@{}lD{.}{.}{-1}D{.}{.}{-1}l@{}}
   \hline\hline
   Molecule &
   \multicolumn{1}{c}{$\log_{10}$$N$(Model)} & 
   \multicolumn{1}{c}{$\log_{10}$$N$(Obs)} & 
   Obs.\ References \\ 
    &
    \multicolumn{1}{c}{(cm$^{-2}$)}&
    \multicolumn{1}{c}{(cm$^{-2}$)}\\
   \hline
   H$_{2}$ &
   18.18 & 
   \multicolumn{1}{c}{$18.36 \pm 0.09$}  & 
   This work \\ 
   CO   &
   15.13& 
   \multicolumn{1}{c}{$14.70 - 15.08$} &
   \citetalias{Ueta:2014aa} \\
   OH$^+$  & 
   13.00 & 
   10.54   & 
   \citet{Aleman:2014aa} \\ 
  \hline
  \end{tabularx}
 \end{table}



   \begin{table}
   \centering
    \renewcommand{\arraystretch}{0.85}
  \tablewidth{\columnwidth}
   \footnotesize
   \caption{The RADEX non-LTE CO model results, 
    and comparison with the observed line intensities.
\label{T-RADEX}
  }
 \begin{tabularx}{\columnwidth}{@{}c@{\hspace{8pt}}c@{\hspace{8pt}}
  r@{\hspace{8pt}}c@{\hspace{8pt}}c@{}}
\hline\hline
$J$    &({\micron}) & $T_{\rm ex}$ (RADEX) &
Intensity (RADEX) &Intensity (Obs)\\
 &            &(K) & (erg s$^{-1}$ cm$^{-2}$ sr$^{-1}$)
	 &(erg s$^{-1}$ cm$^{-2}$ sr$^{-1}$)\\
\hline      
4 -- 3 &650.3& 209 & 3.51(--7) & 3.73(--7) $\pm$ 6.35(--8)\\
5 -- 4 &520.2&  85 & 7.71(--7) & 7.67(--7) $\pm$ 2.03(--8)\\
6 -- 5 &433.6&  70 & 1.19(--6) & 1.17(--6) $\pm$ 1.51(--7)\\
7 -- 6 &371.7&  70 & 1.47(--6) & 1.99(--6) $\pm$ 2.49(--7)\\
8 -- 7 &325.2&  74 & 1.58(--6) & 9.71(--7) $\pm$ 1.38(--8)\\
9 -- 8 &289.1&  82 & 1.55(--6) & 1.08(--6) $\pm$ 2.91(--7)\\
 \hline
\end{tabularx}
  \end{table}


\subsubsection{Comparison between theoretical and observed gas masses}

 It is of interest to compare the amount of mass ejected 
 during the AGB phase that is empirically accounted for 
 with the adopted panchromatic data set
 (observational detection $+$ model fitting via the present 
 analyses) to 
 our previous estimates based on an incomplete data set
 and to a theoretical prediction.
 As summarized in Table\,\ref{T-comp},
 the total gas mass empirically accounts for in NGC\,6781 
 was $0.41\,M_{\sun}$, comprising of 
 $0.09\,M_{\sun}$ of ionized gas,
 $0.20\,M_{\sun}$ of atomic gas, and
 $0.11\,M_{\sun}$ of molecular gas.
 These values are based on the adopted volume filling 
 factor $f$ of unity (\S\,\ref{S:D}).

 Previously, using almost exclusively far-IR line data 
 and under the assumption of $D = 0.95$\,kpc, 
 the total gas mass was estimated to be $0.86$\,$M_{\sun}$, 
 which consisted of
 0.54\,$M_{\sun}$ of ionized gas (only H$^{+}$, He$^{+}$, and He$^{2+}$), 
 0.12\,$M_{\sun}$ of atomic gas, and
 0.20\,$M_{\sun}$ of molecular gas (only H$_{2}$ based on $N$(H$_{2}$)
 calculated from the excitation diagram), while adopting $f = 0.5$
 \citepalias{Ueta:2014aa}.
 With the updated distance of $D = 0.46$\,kpc and $f = 1$, 
 these previous estimates correspond to the total gas mass 
 of 0.40\,$M_{\sun}$.
 While the total mass turns out to be consistent with the 
 present result,
 the relative proportion of the individual gas components
 in the previous result is very different.
 This is of course because of the fact that we 
 have to scale the relative proportion to fill 
 gaps of the absence of sufficiently constraining observational data.

 According to \citet{Karakas:2007aa} and 
 \citet{Karakas:2010aa}, a 2.5\,$M_{\sun}$ initial-mass star
 with $Z = 0.02$ would experience 25 AGB thermal pulse (TP) episodes
 while ejecting the total mass of $\sim1.25$\,$M_{\sun}$.
 However, the predicted amount of the mass-loss ejecta
 would remain small ($<0.01$\,$M_{\sun}$) until the 22nd TP episode.
 Over the last three TP episodes, the amount of the ejecta
 would increase precipitously reaching $\simeq 0.70$\,$M_{\sun}$ 
 during the last TP episode. 
 Hence, our best-fit model accounts for roughly 60\,\% of the 
 amount of mass theoretically predicted to have been ejected 
 during the last TP episode.

 Meanwhile, the total gas mass within the ionization bound,
 $R_{\rm IB} = 55{\arcsec}$, is $0.12\,M_{\sun}$ (consisting
 of $0.09\,M_{\sun}$ and $0.03\,M_{\sun}$ ionized and atomic
 gas, respectively), accounting for about 23\,\% of the total 
 gas mass.
 This proportion is consistent with a previous theoretical 
 prediction made by \citet{Villaver:2002aa}, in which the 
 evolution of the ejecta was modeled based on the stellar
 evolution tracks by \citet{Vassiliadis:1993aa}.
 They concluded that the bright ionized shell would contain 
 about 0.5\,$M_{\sun}$ of gas for a 2.5\,$M_{\sun}$ initial 
 mass (their Fig.\,25), which roughly translates to 25\,\%
 of the total ejecta mass.

 Comparisons among these quantities indicate that the bulk 
 of the nebular mass is found to be in the PDR of the nebula 
 beyond $R_{\rm IB}$ in the form of neutral (atomic/molecular) 
 gas.
 This finding is quite intriguing given the fact that PNe are 
 generally known as the hallmark of the presence of ionized gas 
 as H$^{+}$ regions.
 The present work also demonstrates that PNe would provide a unique
 window of opportunities to investigate the mass loss history of
 the progenitor star, 
 because PNe should allow 
 (1) access to a significant fraction of
 the AGB mass loss ejecta when observed with sufficiently 
 sensitive instruments (as opposed to AGB stars themselves) and
 (2) spatially-resolved investigations more into the past 
 (i.e., regions of larger radii) due to much larger
 energy input by the central star to illuminate the PDR of 
 the nebula (as opposed to proto-PNe).


 \begin{table}
  \centering
  \footnotesize
   \renewcommand{\arraystretch}{0.85}
  \caption{Comparison of each of the mass components
  between this work and the \citetalias{Ueta:2014aa} results \label{T-comp}}
  \begin{tabularx}{\columnwidth}{@{}lccc@{}}
   \hline\hline
   Parameters      &This work &\citetalias{Ueta:2014aa} & \citetalias{Ueta:2014aa}\\
   &&&(scaled)\\  
   \hline
   $D$ (kpc)                                   &$0.46$     &$0.95$    &$0.46$    \\
   filling factor                              &$1.0$      &$0.5$     &$1.0$ \\
   total gas ($M_{\sun}$)                      &$0.41$     &$0.86$    & $0.40$      \\ 
   ~ionized gas ($M_{\sun}$)                    &$0.09$     &$0.54$    & $0.25$      \\ 
   ~atomic gas ($M_{\sun}$)                     &$0.20$     &$0.12$    & $0.05$    \\
   ~total molecular gas ($M_{\sun}$)            &$0.11$     &$0.20$    & $0.09$   \\
   ~~warm molecular gas ($M_{\sun}$)             &$4.15(-3)$ &$\cdots$  & $\cdots$   \\
   ~~cold molecular gas ($M_{\sun}$)$^{\dagger}$ &$1.12(-1)$ &$\cdots$  & $\cdots$     \\
   ~total dust mass ($M_{\sun}$)                &$1.53(-3)$ &$\cdots$  & $\cdots$\\    
   ~~dust mass beyond $R_{\rm IB}$ ($M_{\sun}$)  &$1.06(-3)$ &$4.0(-3)$$^{\ddagger}$ & $9.4(-4)$$^{\ddagger}$\\
   GDR                                         &268        &335 (median) & 335 (median)\\
   \hline
  \end{tabularx}
    \begin{minipage}{\columnwidth}
     \tablewidth{\columnwidth}
     \tablenotetext{\dagger}{\footnotesize The model-predicted
     cold molecular mass was scaled from the model-predicted
     warm molecular mass in this work. 
     In \citetalias{Ueta:2014aa}, the molecular component was not
     sub-divided by temperature.}
     \tablenotetext{\ddagger}{\footnotesize The empirical dust
     mass estimate was for the cold dust of $20-40$\,K.}
   \end{minipage}
 \end{table}


\subsubsection{The far-IR/cold dust component of the nebula}

 The best-fit model yields the dust mass ($m_{\rm dust}$) of
 $1.53(-3)$\,$M_{\sun}$, while the empirically-determined value
 obtained by fitting far-IR broadband images \citepalias{Ueta:2014aa},
 scaled to the present distance estimate of $D = 0.46$\,kpc is
 $9.4(-4)$\,$M_{\sun}$. 
 In both estimates, dust grain properties
 are the same (i.e., AC grains). 
 This discrepancy is expected
 because the previous empirical estimate considered only the cold
 dust component detected in the far-IR ($\sim 20-40$\,K;
 \citetalias{Ueta:2014aa}), missing the higher-temperature
 component emitting mainly in the shorter wavelength (e.g.,
 mid-IR). 
 The present best-fit model includes the entire (warm $+$ cold) dust
 component ($\sim 22-299$\,K).

 To assess the consistency between
 the best-fit model and the empirical measurements, 
 we estimate the mass of the cold/far-IR dust component in the best-fit model.
 Similar to the discussion in the previous section, 
 we consider the cold dust component existing in 
 the PDR beyond the IB, over which the model-predicted dust temperature
 would be $23-38$\,K. 
 In the best-fit model, 
 the dust mass beyond $R_{\rm IB}$ is $1.06(-3)$\,$M_{\sun}$,
 which is consistent with 
 the empirical cold dust mass of 9.4(--4)\,$M_{\sun}$.

 The circumstellar dust mass is typically estimated via SED fitting
 of the thermal dust excess in the near- and mid-IR wavelengths. 
 However, the present study reveals that 
 there is a larger amount of cold dust (of $1.06(-3)$\,$M_{\sun}$)
 than warm dust (of $4.61(-4)$\,$M_{\sun}$) around NGC\,6781.
 This finding suggests that  
 the far-IR/cold dust component could take
 up a significant portion of the circumstellar dust in PNe 
 ($\sim 69$\,\% for the case of NGC\,6781),
 and hence,
 far-IR fluxes must always be incorporated in studying PNe
 especially when considering the energetics in
 the whole volume of the nebula (especially the PDR and beyond).

\subsubsection{Gas-to-Dust mass ratio}

 In {Cloudy} model calculations, the presence of dust is scaled with
 the hydrogen density profile by the gas-to-dust mass ratio (GDR).
 The dust radiative transfer is done at each radial bin taking into
 account all the radiation available locally for dust heating (i.e.,
 radiation from the ambient gas as well as from the CSPN).
 However, there is no mechanism to produce/destroy dust grains in the
 code.
 The best-fit model yields the ``mean'' GDR of 268 over the entire volume. 
 The derived GDR is comparable with the average GDR of $386 \pm 90$
 among 18 C-rich evolved stars \citep{Knapp:1985ab} based on the
 direct comparison between the gas component (via CO $J=1-0$
 observations in the radio, i.e., the cold gas component) and the dust
 component (via SED fitting of IR excess in the $N$-band, i.e., the warm
 dust component). From our discussion in the previous section, it is
 likely that the Knapp work may have missed the cold dust component and
 hence their GDR may have been overestimated.

 In our previous empirical estimate \citepalias{Ueta:2014aa},
 the GDR distribution in NGC\,6781 shows a tenfold decrease of the
 GDR from around 500 near the inner radius of the barrel wall
 to around 50 beyond the IB into the PDR with the median of 335.
 Caution needs to be exercised to compare these numbers because
 the empirical GDR distribution is susceptible to the projection
 effect (i.e., the gas and dust components being ratio-ed may not
 be present at the same location along the line of sight).
 Nevertheless, the median value is certainly consistent with the
 modeling results.

\subsubsection{3-D effects on the dusty photoionization models}

 \citet{Gesicki:2016aa} reported that 3-D photoionization 
 models could reproduce the observed emission line fluxes 
 with ionized gas mass that is several times less than 1-D 
 models may suggest.
 This is because in 3-D models there is usually a greater 
 amount of ``surfaces'' at which ionization can happen. 
 In 1-D models, radiation would always have to be attenuated 
 before penetrating into the next/outer radial layer of the 
 nebula.
 However, in 3-D models, attenuation may not even occur along 
 some lines of sight (e.g., along the polar direction vs.\ 
 equatorial directions in the case of a bipolar nebula), 
 providing means to ionize the outer parts of the nebula
 to a greater extent.
 Indeed, we already saw some indication of the 3-D effects
 especially in the PDR based on the multi-temperature fit of 
 the excitation diagram analysis and the non-LTE nature 
 of the CO distribution, suggesting a temperature gradient
 along the polar direction of the nebula.

 While 3-D photoionization codes are available, 
 we adopt the 1-D {Cloudy} code because at this point 
 no 3-D photoionization codes would satisfactorily incorporate 
 lower temperature components (i.e., the dusty PDR)
 to be fit with the broad array of the adopted constraining observational data.
 For the case of NGC\,6781 in particular, 
 this 1-D/3-D issue implies that there could be a distribution of 
 ionized gas extending along the polar directions 
 (i.e., the regions of the polar caps and beyond),
 which would alter the overall proportion of the ionized gas 
 in terms of the total mass of the nebula.
 However, this 3-D effects on the ionized gas mass are considered
 to be minor in the present work.
 This is because 
  model parameters that are critical in determining
  line fluxes, and hence, masses,
  such as the
  hydrogen density profile $n_{\rm H}$($R$), 
  $D$, 
  $L_{\ast}$, 
  $T_{\rm eff}$, 
  nebular elemental abundances, and 
  spatial distributions of various gas/dust components,
  were fixed to empirically-derived values based on the spatially-resolved data 
  and not treated as free parameters,
  which is often the case in typical 1-D models
  based on spatially-unresolved data.

\section{Conclusions}
\label{summary}%

 We have investigated the physical conditions and evolution of a bipolar
 PN NGC\,6781 by 
 (1) collecting the most comprehensive panchromatic data set for 
 the object ever assembled sourced from 
 various data archives covering from UV to radio 
 including our own \emph{Herschel} data (Fig.\,\ref{S:spt}, 
 Tables\,\ref{T:phot} and \ref{T:linelist}), and 
 (2) performing dusty photoionization pseudo-2-D model SED/line 
 fitting with the {Cloudy} code using this panchromatic data set 
 which yielded 136 constraints.
 The primary aim of the investigation was therefore to generate 
 the best-fit model that satisfies {\sl all} of the 
 adopted panchromatic data self-consistently.

 Using nebular lines detected in the optical, mid-IR, and far-IR, 
 we have performed detailed plasma diagnostics and derived 
 {\Ne} and {\te} for 
 9 diagnostic lines based on 15 different line ratios computed 
 from 28 individual line fluxes (Fig.\,\ref{F:diagno}, 
 Tables\,\ref{T:diagno_table} and \ref{T:teane}),
 ionic abundances for 19 species (Table\,\ref{T:ionic}) and
 elemental abundances for 9 species (Table\,\ref{T:abund2}).

 By comparing the empirically-derived elemental abundances
 (Table\,\ref{T:abund2}) with the theoretically-predicted abundances
 of the AGB nucleosynthesis models \citep{Karakas:2010aa},
 the progenitor of NGC\,6781 has been determined as a
 $2.25-3.0$\,$M_{\sun}$ initial-mass star of $Z \simeq 0.02$. 
 By fitting the CSPN luminosity (Fig.\,\ref{F:cspn}) 
 as a function of the distance ($D$) and
 effective temperature ($T_{\rm eff}$) with the post-AGB evolutionary tracks
 of $2.25-3.0$\,$M_{\sun}$ initial-mass stars \citep{Vassiliadis:1994ab}, 
 we have derived the best-fit $D$ of 0.46\,kpc and 
 $L_{\ast}$ of $104-196\,L_{\sun}$ (Fig.\,\ref{F:HR}).

 We have also performed the excitation diagram analysis to probe the
 physical conditions of the H$_{2}$ emitting PDR of the nebula. 
 The excitation diagram for the observed H$_{2}$ lines can be fit
 reasonably with a single- and double-temperature model 
 at around $1300$\,K and $1200\,$K$/240$\,K, respectively 
 (Table\,\ref{T:H2}, Fig.\,\ref{F:h2}). 
 Comparisons with theoretical shock models by \citet{Flower:2010aa}
 indicated that H$_{2}$ could be excited by shocks caused by
 interactions between the remnant AGB circumstellar envelope and the
 fast wind emanating from the CSPN.

 The results of our analyses of the observational data suggest that
 the apparent ring shape of NGC\,6781 was best represented by 
 a pole-on spherical cylinder structure (of $54\arcsec$ inner radius and 
 the ``barrel'' height of $90\arcsec$,)
 with a physically-thin (of $4\arcsec$ thickness)
 but dense ($n_{\rm H} = 960$\,cm$^{-3}$) wall
 surrounding a tenuous ionized gas ($n_{\rm H} =
 300$\,cm$^{-3}$), all of which is surrounded by an
 even denser PDR ($n_{\rm H} = 10^4$\,cm$^{-3}$; Fig.\,\ref{F:GEO}).

 Armed with 
 the empirically-established CSPN characteristics
 and input model of the nebula, 
 plus the most comprehensive panchromatic observational
 constraints ever compiled
 (37 broadband fluxes from UV to mid-IR,
 19 flux densities from mid-IR to radio, 
 78 emission lines in 4 spectra ranges and 
 8 elemental abundances,
 totaling 136 constraints; Tables\,\ref{AT:model} and \ref{T:abund2}),
 we have arrived at the best-fit photoionization model of NGC\,6781 
 using the {Cloudy} code
 \citep{Ferland:2013aa} through iterative model fitting
  (Table\,\ref{T:model}, Figs.\,\ref{F:SED}, \ref{F:SED2},
  and \ref{F:HR-last}).

 The best-fit model indicates that the circumstellar nebula of NGC\,6781 
 is illuminated by the CSPN of 
 $L_{\ast} = 121$\,$L_{\sun}$ and $T_{\rm eff} = 121$\,kK 
 so that the ionization front is settled at $R_{\rm IB} = 55\arcsec$
 (i.e., the nebula is ionization-bounded along the equatorial direction, but
 density-bounded along the polar directions)
 with the outer radius of the PDR at $61\arcsec$. 
 To explain the observed H$_{2}$ and CO line fluxes,
 the PDR would have to possess an extra heating source 
 to keep the PDR temperature at about $1400$\,K.
 However, there must also be a component of cold molecules in the PDR,
 suggested by the excitation diagram analysis of H$_2$ and CO
 and by non-LTE radiative transfer calculations of CO,
 which could not be simultaneously modeled in the present study
 because of lack of observational data that probe/constrain the even
 colder part of the PDR.
 It is likely that a temperature gradient in the PDR along the polar
 direction contributes to the multi-temperature characteristic of the PDR
 that was not fully constrained by the present pseudo 2-D model.

 This best-fit model can account for about 60\,\% of the
 theoretically predicted gas mass of $\sim 0.70$\,$M_{\sun}$ 
 (Table\,\ref{T-comp}) ejected
 during the last AGB thermal pulse episode of
 a 2.5\,$M_{\sun}$ initial-mass star
 of $Z = 0.02$ \citep{Karakas:2007aa,Karakas:2010aa}. 
 Of which, only 20\,\% of the total mass appears to be contained 
 within the ionized region of the nebula.
 This finding emphasizes that, 
 while PNe are known as the hallmark of ionized gas in
 H$^{+}$ regions, 
 the colder dusty PDR that surrounds the ionized gas carries
 greater significance 
 in terms of the progenitor's mass loss history and cannot be neglected
 to account for the full energetics of the nebula. 
 Nonetheless,
 the present work has demonstrated that PNe can indeed serve as
 (1) empirical constraints for stellar evolutionary models
 because empirically-derived CSPN and nebula parameters can now
 comprehensively confront theoretical predictions (and the present 
 AGB models are shown to be correct in general),
 and
 (2) important probes of mass recycling and chemical evolution in galaxies
 because PNe would permit one of the most thorough mass accounting
 of the mass loss ejecta in the circumstellar environments.
 
 Our present investigation has also demonstrated 
 that detailed dusty photoionization PN models can explain a wide variety
 of observational data self-consistently and 
 that the PDR is critically important to characterize PNe comprehensively.
 However, our work has also revealed that there is still a considerable 
 lack of observational data to constrain the input parameters, especially 
 those that probe the PDR (i.e., the coldest realm of PNe) and 
 the X-ray emission properties of the CSPN and 
 highly ionized gas in its vicinity (i.e., the hottest realm of PNe).
 Moreover, ideally 3-D models would have to be used.
 In future, critical issues to be investigated in PNe will be
 (1) far-IR and sub-millimeter spatially-resolved spectroscopy 
 of the cold molecular component
 with ALMA, EVLA, and SKA, as well as \emph{SPICA}, 
 (2) mid-IR spatially-resolved spectroscopy 
 of the warm molecular component with \emph{JWST},
 (3) optical spatially-resolved spectroscopy of the atomic gas 
 component
 and
 (4) X-ray/far-UV observations to better characterize the CSPN and
 possible accompanying extra high-energy sources.

 \section*{Acknowledgments}

 This work is partly based on observations made with 
 the \emph{Herschel} Space Observatory, 
 a European Space Agency (ESA) Cornerstone Mission
 with significant participation by National Aeronautics
 and Space Administration (NASA),
 the \emph{Spitzer} Space Telescope, obtained from the
 NASA/IPAC Infrared Science Archive, both of which are operated
 by the Jet Propulsion Laboratory (JPL), California Institute
 of Technology (Caltech) under a contract with NASA, and
 the Infrared Space Observatory (\emph{ISO}), an ESA project
 with instruments funded by ESA Member States and with the
 participation of the Institute of Space and Aeronautical
 Science/Japan Aerospace exploration Agency (ISAS/JAXA) and NASA.

 Some of the data presented in this paper were 
 obtained from the Wide-field Infrared Survey Explorer,
 which is a joint project of the University of California, Los Angels,
 and JPL/Caltech, funded by NASA,
 the European Southern Observatory Science Archive Facility,
 the Isaac Newton Group Archive,
 which is maintained as part of the CASU Astronomical Data Centre at
 the Institute of Astronomy, Cambridge, U.K., and
 the Mikulski Archive for Space Telescopes (MAST) at the Space
 Telescope Science Institute (STScI), which is operated by the
 Association of Universities for Research in Astronomy, Inc., under
 NASA contract NAS5-26555. Support for MAST for GALEX data is provided
 by the NASA Office of Space Science via grant NNX09AF08G and by other
 grants and contracts. When some of the data reported here were
 acquired, UKIRT was operated by the Joint Astronomy Centre on behalf
 of the Science and Technology Facilities Council of the U.K.
 A portion of this work was based on 
 the use of the ASIAA clustering computing system. 

 We are grateful to the anonymous referee for a careful reading and
 valuable suggestions.
 MO was supported by the research fund 104-2811-M-001-138 and
 104-2112-M-001-041-MY3 from the Ministry of Science and Technology
 (MOST), R.O.C. 
 MO sincerely expresses his thanks to Drs.\ Naomi Hirano
 and Tatsuhiko Hasegawa for
 fruitful discussion on molecular gas excitation. 
 TU was partially supported 
 by an award to the original \emph{Herschel} observing program
 (OT1\_tueta\_2) under Research Support Agreement (RSA) 1428128
 issued through JPL/Caltech, and by the NASA under Grant NNX15AF24G
 issued through the Science Mission Directorate. 
 PvH was funded by the Belgian Science Policy
 Office under contract no.\ BR/154/PI/MOLPLAN.
 I.A. acknowledges the support of CNPq, Conselho Nacional de
 Desenvolvimento Cient\'ifico e Tecnol\'ogico - Brazil, process
 number 157806/2015-4.
 AAZ was supported
 by the UK Science and Technology Facility Council, through grant
 ST/L000768/1. Y.-H.C. was supported by the research fund
 104-2112-M-001-044-MY3 from the MOST.
 E.V. acknowledges support from Spanish
 Ministerio de Econom\'{i}a y Competitividad under grant
 AYA2014-55840-P. MLL-F was supported by CNPq, Conselho
 Nacional de Desenvolvimento Cient\'ifico e Tecnol\'ogico - Brazil,
 process number 248503/2013-8.

  \facilities{
  \emph{Herschel, Spitzer, WISE, ISO, HST, GALEX},
  ING/INT 2.5-m,
  ING/WHT 4.2-m,
  ESO/NTT 3.6-m,
  UKIRT 3.8-m,
  CFHT 3.6-m}

  \software{Cloudy \citep[v C13.03;][]{Ferland:2013aa},
  IRAF (v.2.16),
  SMART \citep[v.8.2.9;][]{Higdon:2004aa},
  IRSCLEAN \citep[v.2.1.1;][]{Ingalls:2011},
  Multidrizzle \citep{Koekemoer:2003aa}
  }

\appendix

\section{Photometry data and measurements \label{A:PHOT}}

\subsection{\label{S:INT1}INT 2.5-m/WFC photometry}

 We downloaded raw broadband imaging data at RGO $U$, Sloan $g$
 and Sloan $r$ and narrowband imaging data at IPHAS {\ha}
 ($\lambda_{\rm c} =6568.2$\,{\AA} with the $93.97$\,{\AA} equivalent
 width), taken with the Wide Field Camera (WFC) mounted on the 2.5-m
 Isaac Newton Telescope (INT) at the Roque de Los Muchachos Observatory,
 La Palma, Spain, from the Cambridge Astronomical Survey Unit (CASU)
 Astronomical Data Centre.

 We reduced the downloaded raw data using {\sc IRAF} following the standard
 procedure (i.e., bias subtraction, flat-fielding, bad-pixel masking,
 cosmic-ray removal, detector distortion correction, and sky
 subtraction), and performed PSF fitting and aperture photometry
 using the {\sc IRAF} {\tt noao.digiphot} package.
 The gain and readout noise of the detector, determined from the
 {\sc IRAF} task {\tt findgain}, were 0.65\,e$^{-}$\,ADU$^{-1}$
 and 1.48\,e$^{-}$, respectively.

 Photometry was performed for the CSPN and two standard stars
 SA110$-$246 and BD$+$28\,4211 ($m_{u} = 14.521$,
 $m_{g} = 10.277$, $m_{r} = 13.103$ and $m_{u} = 9.977$,
 $m_{g} = 10.277$, $m_{r} = 14.440$, respectively, in the SDSS
 system; \citealt{Ahn:2012aa}), of which the standard stars
 were used to do flux calibration as well as PSF fitting.
 Then, we removed field stars in the vicinity of NGC\,6781
 and carried out photometry of the entire nebula (CSPN plus PN)
 using the residual images. In the end, the respective instrumental
 magnitudes of $m_{U}$, $m_{g}$, and $m_{r}$ were converted into
 the SDSS magnitudes of $u$, $g$, and $r$ with the following formulae;
 \begin{eqnarray}
  m_{u} &=& -26.878  + r + 2.844 \sec(z_{U})  + 10.924 (u - g),\\
  m_{g} &=& -27.724  + r + 2.573 \sec(z_{g})  + 0.816 (g - r),\\
  m_{r} &=& -26.794  + r + 1.534 \sec(z_{r})  + 0.653 (g - r), 
 \end{eqnarray}

 \noindent where $z_{band}$ stands for the airmass at the time of observations.

 To obtain the flux density in the IPHAS {\ha} band, we made
 measurements in the count rates (i.e., ${\rm e}^{-}$ per second),
 while the measurement procedure itself was the same as the other broadbands.
 The count rate to flux density conversion factor was calculated by
 (1) measuring the count rate of the standard star BD+17$^{\circ}$\,4708
 in the IPHAS {\ha} image, and 
 (2) computing the flux density per count rate in this band using the
 spectrum of BD+17$^{\circ}$\,4708 from the \emph{HST} {\sc CALSPEC}
 Calibration Database
 \footnote{\url{http://www.stsci.edu/hst/observatory/crds/calspec.html}},
 taking into account the filter transmission curve of the {\ha} band.
 Then, we converted the {\ha} photometry of NGC\,6781 in count
 rates into the flux density using this conversion factor.

\subsection{ESO NTT 3.6-m/EFOSC2}

 We downloaded raw broadband imaging data at Bessel $B$, $V$, and $R$,
 taken with the ESO Faint Object Spectrograph and Camera 2 (EFOSC2)
 mounted on the 3.58-m New Technology Telescope (NTT) at the La Silla
 Observatory, Chile, from the ESO Science Archive Facility. 

 We reduced the data and performed photometry of the CSPN and CSPN
 plus PN with the standard star PG1657+078 and nearby four field
 stars PG1657+078A, B, C, and D \citep{Landolt:2009aa} as calibration
 standards using {\sc IRAF} packages in the same procedure used for
 the INT/WFC data. 
 The gain and readout noise were measured to be
 1.26\,${\rm e}^{-}$\,ADU$^{-1}$ and 8.27\,${\rm e}^{-}$ in the
 NGC\,6781 images and 1.22\,${\rm e}^{-}$\,ADU$^{-1}$ and
 11.55\,${\rm e}^{-}$ in the standard star images, respectively. 

 We converted the respective instrumental magnitudes of
 $m_{B}$, $m_{V}$, and $m_{R}$ into the Landolt system $B$, $V$,
 and $R$ band magnitudes with the following formulae;
 \begin{eqnarray}
 m_{B} &=& -26.659 + V -0.242\,sec(z_{B})  +0.967\,(B - V),\\
 m_{V} &=& -25.746 + V -0.425\,sec(z_{V})  -0.016\,(B - V),\\
 m_{R} &=& -25.780 + V -0.483\,sec(z_{R})  -0.968\,(V - R).
 \end{eqnarray}

\subsection{UKIRT 3.8-m/WFCAM}

 We downloaded raw broadband imaging data products at $J$, $H$, and $Ks$,
 taken with the Wide Field Camera (WFCAM) mounted on the 3.8-m United
 Kingdom Infra-Red Telescope (UKIRT) at the Mauna Kea Observatory,
 Hawai'i, from the UKIRT WFCAM Science Archive (WSA).

 We measured $J$, $H$, and $Ks$ band magnitudes of the CSPN and CSPN
 plus PN based on our own photometry of 96 nearby field stars, and  
 converted the respective instrumental magnitudes of $m_{J}$, $m_{H}$,
 and $m_{Ks}$ into the 2MASS system $J$, $H$, and $Ks$ band magnitudes
 with the following formulae;
 \begin{eqnarray}
 m_{J} &=& -26.091 + J -0.047\,(J - H),\\
 m_{H} &=& -26.444 + J -0.859\,(J - H),\\
 m_{K} &=& -25.477 + J -0.886\,(J - Ks).
 \end{eqnarray}

\subsection{HST/WFPC2 photometry}

 We downloaded raw broadband imaging data at F555W and F814W
 (roughly corresponding to Johnson-Cousins $V$ and $I_{\rm c}$,
 respectively), taken with the Wide-Field Planetary Camera 2 (WFPC2)
 on-board the 2.4-m Hubble Space Telescope (HST), from the Mikulski
 Archive for Space Telescopes (MAST).
 The raw data sets were processed with the {\tt stsdas.multidrizzle}
 package \citep{Koekemoer:2003aa} included in {\sc PyRAF}. 
 We performed aperture photometry for the CSPN after we subtracted
 the nearby stars by the PSF fitting using the {\sc IRAF} packages
 {\tt noao.digiphot}.

 \setcounter{table}{0}

 \begin{table*}
  \renewcommand{\thetable}{A\arabic{table}} 
  \footnotesize
  \centering
   \renewcommand{\arraystretch}{0.85}
  \caption{\label{T:phot}
  The broadband flux densities of NGC\,6781 adopted in
  the present study. The flux densities at $K$
  or shorter wavelengths are corrected for the interstellar reddening.}
 \begin{tabular}{@{}cccD{p}{\pm}{-1}D{p}{\pm}{-1}D{p}{\pm}{-1}@{}}
  \hline\hline
  \multicolumn{6}{c}{CSPN}\\
  Instruments & $\lambda$ & Band & m
  & \multicolumn{1}{c}{$F_{\lambda}$}
  & \multicolumn{1}{c}{$I_{\lambda}^{\ast}$}   \\
  & ({\micron}) &  &
	      & \multicolumn{1}{c}{(erg s$^{-1}$ cm$^{-2}$ {\micron}$^{-1}$)}
  & \multicolumn{1}{c}{(erg s$^{-1}$ cm$^{-2}$ {\micron}$^{-1}$)}\\  
\hline
WFC & 0.3595 & $u$ & 16.67~p~0.21 & 1.82(-11)~p~3.52(-12) & 4.14(-10)~p~8.94(-11) \\
EFOSC2 & 0.4481 & $B$ & 17.15~p~0.02 & 9.07(-12)~p~1.51(-13) & 1.20(-10)~p~9.69(-12) \\
WFC & 0.4640 & $g$ & 16.82~p~0.21 & 9.47(-12)~p~1.83(-12) & 1.11(-10)~p~2.31(-11) \\
EFOSC2 & 0.5423 & $V$ & 16.96~p~0.01 & 6.21(-12)~p~5.72(-14) & 4.70(-11)~p~2.95(-12) \\
WFPC2 & 0.5443 & F555W & 16.90~p~0.11 & 6.21(-12)~p~6.47(-13) & 4.66(-11)~p~5.64(-12) \\
EFOSC2 & 0.6441 & $R$ & 16.75~p~0.02 & 4.54(-12)~p~7.54(-14) & 2.40(-11)~p~1.29(-12) \\
WFPC2 & 0.7996 & F814W & 16.52~p~0.04 & 2.97(-12)~p~1.05(-13) & 9.79(-12)~p~4.99(-13) \\
WFCAM & 1.235 & $J$ & 16.32~p~0.02 & 9.24(-13)~p~1.70(-14) & 1.64(-12)~p~4.18(-14) \\
WFCAM & 1.662 & $H$ & 16.34~p~0.05 & 3.25(-13)~p~1.45(-14) & 4.64(-13)~p~2.13(-14) \\
WFCAM & 2.159 & $K$ & 16.21~p~0.05 & 1.41(-13)~p~7.09(-15) & 1.77(-13)~p~9.04(-15) \\
\hline
\multicolumn{6}{c}{CSPN+PN}\\
Instruments & $\lambda$ & Band & m   & \multicolumn{1}{c}{$F_{\lambda}$}   &\multicolumn{1}{c}{$I_{\lambda}^{\ast}$}   \\
 & ({\micron}) &  &     & \multicolumn{1}{c}{(erg s$^{-1}$ cm$^{-2}$ {\micron}$^{-1}$)}   &\multicolumn{1}{c}{(erg
		      s$^{-1}$ cm$^{-2}$ {\micron}$^{-1}$)}\\  
\hline
GALEX & 0.2274 & NUV &    & 1.46(-10)~p~1.74(-11) & 5.19(-8)~p~1.12(-8) \\
WFC & 0.3595 & $u$ & 11.63~p~0.01 & 1.87(-9)~p~1.88(-11) & 4.27(-8)~p~4.13(-9) \\
EFOSC2 & 0.4481 & $B$ & 11.97~p~0.04 & 1.07(-9)~p~3.87(-11) & 1.40(-8)~p~1.22(-9) \\
WFC & 0.4640 & $g$ & 11.37~p~0.01 & 1.44(-9)~p~1.43(-11) & 1.69(-8)~p~1.29(-9) \\
EFOSC2 & 0.5423 & $V$ & 10.93~p~0.02 & 1.61(-9)~p~2.96(-11) & 1.22(-8)~p~7.88(-10) \\
WFC & 0.6122 & $r$ & 10.37~p~0.01 & 2.07(-9)~p~1.15(-11) & 1.21(-8)~p~6.61(-10) \\
EFOSC2 & 0.6441 & $R$ & 10.15~p~0.03 & 1.98(-9)~p~5.29(-11) & 1.05(-8)~p~6.06(-10) \\
WFCAM & 1.235 & $J$ & 10.33~p~0.01 & 2.30(-10)~p~1.06(-12) & 4.08(-10)~p~7.42(-12) \\
WFCAM & 1.662 & $H$ & 9.96~p~0.01 & 1.15(-10)~p~1.38(-12) & 1.64(-10)~p~2.66(-12) \\
WFCAM & 2.159 & $K$ & 7.55~p~0.01 & 4.09(-10)~p~3.55(-12) & 5.16(-10)~p~5.81(-12) \\
WISE & 3.353 & W1 &    &    & 7.38(-11)~p~1.16(-12) \\
IRAC & 4.500 & Band2 &    &    & 1.11(-10)~p~3.33(-12) \\
IRAC & 5.800 & Band3 &    &    & 1.32(-10)~p~3.97(-12) \\
IRAC & 8.000 & Band4 &    &    & 8.99(-11)~p~2.70(-12) \\
WISE & 11.56 & W3 &    &    & 5.41(-11)~p~7.71(-13) \\
ISOCAM & 14.30 & LW3 &    &    & 5.65(-11)~p~1.13(-11) \\
WISE & 22.09 & W4 &    &    & 3.23(-11)~p~5.78(-13) \\
PACS & 70.00 & BLUE &    &    & 4.01(-11)~p~2.01(-12) \\
PACS & 160.00 & RED &    &    & 7.60(-12)~p~3.84(-13) \\
SPIRE & 250.00 & PSW &    &    & 1.44(-12)~p~2.21(-13) \\
SPIRE & 350.00 & PMW &    &    & 4.90(-13)~p~5.51(-14) \\
SPIRE & 500.00 & PLW &    &    & 7.69(-14)~p~1.22(-14) \\
Radio& 6972  &43GHz & & &4.38(-17)~ \\
Radio& 9993  &30GHz & & &7.93(-18)~p~2.13(-19) \\
Radio& 13627 &22GHz & & &3.07(-18)~ \\
Radio& 59959 &5GHz & & &2.70(-19)~ \\
Radio& 214138&1.4GHz & & &2.46(-20)~p~7.85(-22) \\
\hline
 \end{tabular}
  \begin{minipage}{\textwidth}
   \tablenotetext{\ast}{\footnotesize We corrected the observed flux 
   densities $F_{\lambda}$ in the fifth column by the method explained 
   in \S\,\ref{S-spec} to obtain the de-reddened flux densities
   $I_{\lambda}$ in the sixth column. 
   $A$($B$) means $A\times10^{-B}$.}
  \end{minipage}
 \end{table*}

\section{Spectroscopy data and measurements \label{A:SPEC}}

\subsection{WHT 4.2-m/ISIS optical spectrum}

 We downloaded raw long-slit spectroscopic data in the optical taken with
 the Intermediate-dispersion Spectrograph and Imaging System (ISIS)
 mounted on the 4.2-m William Herschel Telescope (WHT) at the
 Observatorio del Roque de los Muchachos, La Palma, Spain, from the
 CASU Astronomical Data Centre.

 The observations covered spatially the bulk of the nebula by scanning
 the central part of the nebula with the $79.6\arcsec \times 1.0\arcsec$
 slit during integration (Fig.\,\ref{S:slit}).
 The spectral coverage was 
 $4170-4970/5190-6670$\,{\AA} and $3600-4400/6520-8010$\,{\AA}
 with the R600B (blue) and R316R (red) gratings, respectively,
 at the airmass of $\sim$1.1 with the seeing of $0.7-0.8\arcsec$,
 according to the observation log.
 Before and after observing NGC\,6781, the CuAr+CuNe lamp frames
 were taken for the wavelength calibration. 
 The standard star BD+28$^{\circ}$\,4211 was observed with
 the 6.1$\arcsec$-wide slit at the airmass of $\sim$1.0.

 Plasma-diagnostics and chemical abundance analyses based on this data
 in conjunction with data taken with \emph{ISO} were presented by
 \citet{Liu:2004aa} and \citet{Liu:2004ab}. 
 We re-reduced the data by ourselves so that we could perform our
 own calculations of ionic and elemental abundances with measurements
 made with the \emph{Spitzer}/IRS and \emph{Herschel}/PACS spectra in
 terms of the line fluxes per arcsec$^{2}$.
 Data reduction was done with the two-dimensional spectra reduction
 package {\sc noao.twodspec} in {\sc IRAF} following the standard
 procedure, i.e., bias subtraction, flat-fielding, spectra aperture
 alignment, distortion correction along the spatial direction,
 wavelength calibration, and cosmic-ray subtraction.

 We corrected the count rates reduced by airmass extinction using the
 atmospheric extinction table provided by the La Palma Observatory,
 and performed the flux calibrations. We extracted 199 and 181 spatial
 pixels in the blue and red arm, respectively, and summed up all the
 spatial pixels. In the end, we obtained a single 3600-8010\,{\AA}
 spectrum of a $79.6\arcsec \times 1.0\arcsec$ region of the nebula.

\subsection{\label{S:INT2}The {\ha} and {\hb} line fluxes of the entire nebula}

 Because the filter transmission of the IPHAS H$\alpha$ band includes
 contributions from the H$\alpha$ and neighboring
 {\nii}\,6527/6548/6583\,{\AA} lines as well as the nebular and stellar
 continuum, we have to subtract the contributions other than the
 H$\alpha$ line itself as much as possible in order to obtain the
 clean H$\alpha$ line flux. We used the ISIS spectrum to estimate
 contributions to the {\ha} band line flux by the neighboring lines. 
 Taking into account the IPHAS {\ha} filter transmission, we compared
 the {\ha} line flux of NGC\,6781 measured from the IPHAS image of
 the entire nebula, $F_{\lambda}$(IPHAS,{\ha}), with that measured
 from the ISIS spectrum covering a $79.6\arcsec \times 1.0\arcsec$
 region, $F_{\lambda}$(ISIS,{\ha}). The resulting scaling factor
 $F_{\lambda}$(IPHAS,{\ha})/$F_{\lambda}$(ISIS,{\ha}) turned out to be 133.33.
 Using this factor, the ISIS spectrum over $3600-8010$\,{\AA}
 was scaled to represent the spectrum of the entire nebula, and
 the clean {\ha} and {\hb} line fluxes of the entire nebula,
 $F$({\ha}) of 6.95(--11)\,$\pm$\,8.61(--13)\,erg\,s$^{-1}$\,cm$^{-2}$
 and $F$({\hb}) of
 1.22(--11)\,$\pm$\,1.59(--12)\,erg\,s$^{-1}$\,cm$^{-2}$
 were determined. We used these {\ha} and {\hb} line fluxes of
 the entire NGC\,6781 nebula to normalise the line fluxes detected
 in the \emph{Spitzer}/IRS and \emph{Herschel}/PACS and SPIRE spectra.

\subsection{Spitzer/IRS mid-IR spectrum \label{S:Spitzer}}

 We downloaded long-slit spectroscopic data in the mid-IR taken
 with the Infra-Red Spectrograph (IRS) on-board the 0.85-m Spitzer
 Space Telescope (Spitzer) as part of the IRS Calibration Program
 (AORKEY:16099072), from the Spitzer Heritage Archive
 \footnote{\url{http://sha.ipac.caltech.edu/applications/Spitzer/SHA/}} (SHA).

 As indicated in Fig.\,\ref{S:slit}, we only used the spectra taken
 from the central parts of the nebula, covering the $57\arcsec
 \times 3.7\arcsec \times 2$ regions along the N-S direction and
 $168\arcsec \times 10.7\arcsec$ region along the E-W direction
 in the Short-Low ($5.1 - 14.3\,\micron$) and Long-Low
 ($13.9-39.9\,\micron$) bands, respectively.
 We reduced the adopted raw data using the data reduction
 packages {\sc SMART} v.8.2.9 \citep{Higdon:2004aa} and IRSCLEAN
 v.2.1.1 \citep{Ingalls:2011}, provided by the \emph{Spitzer} Science Centre.

 Then, we scaled the measured flux densities of the single
 $5.2-39.9\,{\micron}$ spectrum by a constant factor of 14.40, which
 was determined to match the flux densities of the entire PN
 (cf.\ Fig.\,\ref{S:slit}) at the \emph{Spitzer}/IRAC Band-4
 ($\lambda_{\rm c} = 8.0$\,{\micron}, 1.92 $\pm$ 0.058\,Jy),
 \emph{WISE} W3 ($\lambda_{\rm c} = 11.56$\,{\micron}, 2.41
 $\pm$ 0.034\,Jy), the \emph{ISO}/ISOCAM 14.3\,{\micron}
 (3.85 $\pm$ 0.77\,Jy), and \emph{WISE} W4 ($\lambda_{\rm c}
 = 22.1$\,{\micron}, 5.25 $\pm$ 0.094\,Jy).

\subsection{Herschel far-IR spectrum \label{S:Herschel}}

 We adopted \emph{Herschel} far-IR spectra presented by
 \citet{Ueta:2014aa}, especially those that covered the
 central part of the nebula (Fig.\,\ref{S:slit}).
 To scale the line fluxes detected by PACS and SPIRE for
 the entire nebula, we synthesised 
 the {\hb} image based on the the {\ha} image taken with
 the Andalucia Faint Object Spectrograph and Camera (ALFOSC)
 mounted on the 2.5-m Nordic Optical Telescope (NOT) at the
 Observatorio del Roque de los Muchachos, La Palma,
 Spain, presented by \citet{Phillips:2011aa}. 
 Because the ALFOSC {\ha} filter (IAC$\#$40\footnote{%
 \url{http://www.iac.es/telescopes/pages/en/home/filters.php?lang=ES}})
 whose central wavelength is 6567\,{\AA} with the bandwidth of 8\,{\AA},
 the contributions from the {\nii}\,6548/6583\,{\AA} lines and the
 underlying continuum are considered to be negligible. 
 After field stars overlapped with the nebula were removed by PSF
 fitting, we scaled the {\ha} map so that photometry of the entire
 nebula would yield $I$({\hb}).
 This scaled {\ha} map would represent the {\hb} map under the
 assumption that the emitting
 regions of {\ha} and {\hb} are the same. 
 Using this synthesised {\hb} image, we measured the counts in
 the regions covered by the PACS and SPIRE observations and scaled
 the measured line fluxes according to the {\hb} fluxes.

  \begin{table*}
   \setcounter{table}{0}
  \renewcommand{\thetable}{B\arabic{table}} 
   \renewcommand{\arraystretch}{0.85}
   \footnotesize
   \centering
   \caption{\label{T:linelist}%
   The relative emission line fluxes of NGC\,6781 adopted in the present
   study.
   }
\begin{tabular}{@{}ccD{p}{\pm}{-1}D{.}{.}{-1}cD{p}{\pm}{-1}@{}}
\hline\hline
$\lambda$&Line&\multicolumn{1}{c}{$I$($\lambda$)}  & \multicolumn{1}{c}{$\lambda$} & Line & \multicolumn{1}{c}{$I$($\lambda$)} \\ 
({\AA})  &    &\multicolumn{1}{c}{($I$({\hb})=100)}& \multicolumn{1}{c}{({\micron})} & &\multicolumn{1}{c}{ ($I$({\hb})=100)}\\ 
\hline
\multicolumn{3}{c}{ING/WHT ISIS} & \multicolumn{3}{c}{\emph{Spitzer} IRS} \\
\hline
3726.03 & $[$O\,{\sc ii}$]$ & 268.177~p~8.139 & 5.51 & H$_{2}$ 0-0\,S(7) & 30.757~p~7.120 \\ 
3728.82 & $[$O\,{\sc ii}$]$ & 316.109~p~8.809 & 6.11 & H$_{2}$ 0-0\,S(6) & 17.269~p~2.594 \\ 
3750.15 & H12 & 5.048~p~0.678 & 6.91 & H$_{2}$ 0-0\,S(5) & 52.628~p~12.762 \\ 
3770.63 & H11 & 3.885~p~0.677 & 8.02 & H$_{2}$ 0-0\,S(4) & 19.413~p~2.378 \\ 
3797.90& H10 & 5.055~p~0.880 & 8.99 & $[$Ar\,{\sc iii}$]$ & 22.867~p~1.731 \\ 
3835.38 & H9 & 9.414~p~1.037 & 9.67 & H$_{2}$ 0-0\,S(3) & 25.792~p~4.127 \\ 
3869.07 & $[$Ne\,{\sc iii}$]$ & 125.764~p~3.150 & 10.51 & $[$S\,{\sc iv}$]$ & 49.677~p~3.516 \\ 
3888.86 & H8+He\,{\sc i} & 27.314~p~1.517 & 11.30 & PAH+H\,{\sc i} & 3.119~p~0.282 \\ 
3967.79 & $[$Ne\,{\sc iii}$]$ & 37.177~p~1.290 & 12.29 & H$_{2}$ 0-0\,S(2) & 6.314~p~0.619 \\ 
3970.07 & H7 & 20.340~p~0.882 & 12.81 & $[$Ne\,{\sc ii}$]$ & 14.802~p~1.017 \\ 
4026.32 & He\,{\sc i} & 2.932~p~1.099 & 15.55 & $[$Ne\,{\sc iii}$]$ & 234.571~p~16.147 \\ 
4068.60 & $[$S\,{\sc ii}$]$ & 3.230~p~0.706 & 17.04 & H$_{2}$ 0-0\,S(1) & 9.241~p~0.657 \\ 
4101.74 & H6(H$\delta$) & 31.011~p~0.846 & 17.88 & $[$P\,{\sc iii}$]$+[Fe\,{\sc ii}]? & 1.736~p~0.131 \\ 
4267.26 & C\,{\sc ii} & 2.070~p~0.495 & 18.71 & $[$S\,{\sc iii}$]$ & 46.998~p~3.277 \\ 
4340.46 & H5(H$\gamma$) & 47.863~p~1.481 & 20.30 & $[$Cl\,{\sc iv}$]$ & 0.333~p~0.061 \\ 
4363.21 & $[$O\,{\sc iii}$]$ & 5.225~p~0.343 & 21.82 & $[$Ar\,{\sc iii}$]$ & 1.622~p~0.131 \\ 
4471.46 & He\,{\sc i} & 5.099~p~0.407 & 25.88 & $[$O\,{\sc iv}$]$ & 174.473~p~12.154 \\ 
4641.10 & N\,{\sc iii} & 0.943~p~0.634 & 33.47 & $[$S\,{\sc iii}$]$ & 50.073~p~3.480 \\ 
4685.76 & He\,{\sc ii} & 8.201~p~0.284 & 34.81 & $[$Si\,{\sc ii}$]$ & 12.287~p~1.143 \\ 
4712.62 & He\,{\sc i} & 1.341~p~0.198 & 36.00 & $[$Ne\,{\sc iii}$]$ & 17.011~p~1.387 \\ 
\cline{4-6}
4740.17 & $[$Ar\,{\sc iv}$]$ & 0.671~p~0.262 & \multicolumn{3}{c}{\emph{Herschel} PACS} \\
\cline{4-6}
4861.33 & H4({\hb}) & 100.000~p~1.562 & 57.32 & $[$N\,{\sc iii}$]$ & 78.829~p~9.712 \\ 
4958.91 & $[$O\,{\sc iii}$]$ & 274.612~p~4.087 & 63.17 & $[$O\,{\sc i}$]$ & 33.175~p~4.186 \\ 
5198.84 & $[$N\,{\sc i}$]$ & 6.341~p~0.782 & 88.33 & $[$O\,{\sc iii}$]$ & 190.944~p~23.417 \\ 
5517.72 & $[$Cl\,{\sc iii}$]$ & 0.838~p~0.381 & 119.20 & OH & 0.590~p~0.105 \\ 
5537.89 & $[$Cl\,{\sc iii}$]$ & 0.577~p~0.467 & 119.40 & OH & 0.655~p~0.116 \\ 
5577.95 & $[$O\,{\sc i}$]$ & 3.510~p~0.599 & 121.73 & $[$N\,{\sc ii}$]$ & 7.880~p~0.973 \\ 
5754.64 & $[$N\,{\sc ii}$]$ & 6.674~p~0.194 & 145.50 & $[$O\,{\sc i}$]$ & 2.904~p~0.367 \\  
5875.58 & He\,{\sc i} & 16.406~p~0.552 & 153.00 & OH$^{+}$ & 0.296~p~0.070 \\ 
5888.49 & $[$Mn\,{\sc v}$]$? & 0.576~p~0.203 & 157.64 & $[$C\,{\sc ii}$]$ & 15.915~p~1.955 \\  
\cline{4-6}
6300.28 & $[$O\,{\sc i}$]$ & 32.959~p~0.765 & \multicolumn{3}{c}{\emph{Herschel} SPIRE} \\
\cline{4-6}
6312.10 & $[$S\,{\sc iii}$]$ & 1.698~p~0.450 & 205.40 & $[$N\,{\sc ii}$]$ & 1.607~p~0.257 \\ 
6363.79 & $[$O\,{\sc i}$]$ & 10.804~p~0.301 & 289.10 & CO $J$=9-8 & 0.528~p~0.142 \\ 
6548.04 & $[$N\,{\sc ii}$]$ & 132.950~p~4.709 & 290.20 & OH$^{+}$ & 0.539~p~0.143 \\  
6562.80 & H3({\ha}) & 286.124~p~7.646 & 308.40 & OH$^{+}$ & 0.495~p~0.067 \\ 
6583.46 & $[$N\,{\sc ii}$]$ & 410.904~p~10.523 & 325.30 & CO $J$=8-7 & 0.473~p~0.067 \\  
6678.14 & He\,{\sc i} & 4.405~p~0.322 & 329.70 & OH$^{+}$ & 0.083~p~0.051 \\ 
6716.44 & $[$S\,{\sc ii}$]$ & 26.242~p~0.738 & 370.30 & $[$C\,{\sc i}$]$ & 0.354~p~0.045 \\ 
6730.82 & $[$S\,{\sc ii}$]$ & 21.854~p~0.620 & 371.60 & CO $J$=7-6 & 0.969~p~0.121 \\ 
7065.33 & He\,{\sc i} & 3.844~p~0.321 & 433.50 & CO $J$=6-5 & 0.572~p~0.074 \\ 
7135.80 & $[$Ar\,{\sc iii}$]$ & 22.349~p~0.707 & 520.30 & CO $J$=5-4 & 0.374~p~0.099 \\  
7281.72 & He\,{\sc i} & 0.688~p~0.078 & 650.30 & CO $J$=4-3 & 0.182~p~0.031 \\ 
7320.03 & $[$O\,{\sc ii}$]$ & 6.490~p~0.279\\ 
7330.27 & $[$O\,{\sc ii}$]$ & 5.396~p~0.261 \\ 
7751.10 & $[$Ar\,{\sc iii}$]$ & 5.351~p~0.277\\ 
\hline
   \end{tabular}
\end{table*}

\newpage

\section{Ionic abundance derivations}

 \setcounter{table}{0}
  \begin{table}
   \renewcommand{\thetable}{C\arabic{table}}
    \footnotesize
   \renewcommand{\arraystretch}{0.85}
   \tablewidth{\columnwidth}
   \centering
  \caption{The adopted {\te} and {\Ne} pairs for
  ionic abundance calculations.
  \label{T:teane}
   }
   \begin{tabular}{@{}cD{p}{\pm}{-1}D{p}{\pm}{-1}l@{}}
    \hline\hline
    Type of &\multicolumn{1}{c}{{\te}}&\multicolumn{1}{c}{{\Ne}}&Ions\\
    line    &\multicolumn{1}{c}{(K)} &\multicolumn{1}{c}{(cm$^{-3}$)}\\
    \hline
    RL &7070~p~1880     &\multicolumn{1}{c}{100}&He$^{+}$, He$^{2+}$\\
    RL &9350~p~400      &\multicolumn{1}{c}{10\,000}&C$^{2+}$\\ 
    CEL& 9350~p~400     & 220~p~50  & Ne$^{+}$, S$^{2+}$, Cl$^{2+}$, Ar$^{2+}$\\
    CEL& 9650~p~200     & 260~p~80  & O$^{+}$\\
    CEL& 10\,050~p~210  & 220~p~50  & O$^{2+}$\\
    CEL& 10\,050~p~210  & 1020~p~300& S$^{3+}$\\
    CEL& 10\,340~p~250  & 220~p~50  & O$^{3+}$, Ne$^{2+}$, Cl$^{3+}$\\
    CEL& 10\,520~p~1820 & 260~p~80  & C$^{+}$, N$^{0}$, O$^{0}$, Si$^{+}$, S$^{+}$\\
    CEL &10\,800~p~170  & 260~p~80  & N$^{+}$\\
    \hline
   \end{tabular}
  \end{table}

  \begin{table*}
   \renewcommand{\thetable}{C\arabic{table}}
   \footnotesize
   \renewcommand{\arraystretch}{0.85}
   \caption{\label{T:ionic}%
   Ionic abundances and elemental abundance derivations using
   the ionization correction factors (ICFs).}
   \centering
    \begin{tabular}{@{}cccD{p}{\pm}{-1}D{p}{\pm}{-1}ccc
     D{p}{\pm}{-1}D{p}{\pm}{-1}@{}}
     \hline\hline
X & X$^{\rm +m}$ & \multicolumn{1}{c}{$\lambda$} & \multicolumn{1}{c}{$I$($\lambda$)} & 
\multicolumn{1}{c}{ X$^{\rm m+}$/H$^{+}$}&
 X & X$^{\rm +m}$ & \multicolumn{1}{c}{$\lambda$} &
 \multicolumn{1}{c}{$I$($\lambda$)} &
\multicolumn{1}{c}{X$^{\rm m+}$/H$^{+}$} \\ 
\hline
He  &  He$^{+}$  &  4026.32\,{\AA}  & 2.932~p~1.099 &  1.26(-1)~p~6.89(-2)  &  Ne  &  Ne$^{+}$  &  12.81\,{\micron}  & 14.802~p~1.017 &  2.06(-5)~p~1.50(-6)   \\ 
  &    &  4471.46\,{\AA}   & 5.099~p~0.407 &  1.01(-1)~p~3.95(-2)  &    &  Ne$^{2+}$  &  3869.07\,{\AA}   & 125.764~p~3.15 &  1.22(-4)~p~1.22(-5)   \\ 
  &    &  4712.62\,{\AA}   & 1.341~p~0.198 &  3.06(-1)~p~9.55(-2)  &    &     &  3967.79\,{\AA}   & 37.177~p~1.29 &  1.19(-4)~p~1.23(-5)   \\ 
  &    &  5875.58\,{\AA}   & 16.406~p~0.552 &  1.14(-1)~p~4.48(-2)  &    &     &  15.55\,{\micron}  & 234.571~p~16.147 &  1.48(-4)~p~1.03(-5)   \\ 
  &    &  6678.14\,{\AA}   & 4.405~p~0.322 &  1.07(-1)~p~4.50(-2)  &    &     &  36.00\,{\micron}  & 17.011~p~1.387 &  1.20(-4)~p~9.87(-6)   \\ 
  &    &  7065.33\,{\AA}   & 3.844~p~0.321 &  1.86(-1)~p~5.45(-2)  &    &    &    &  &        \textit{1.20(--4)}~p~\textit{6.50(--6)}   \\ 
  &    &  7281.72\,{\AA}   & 0.688~p~0.078 &  1.07(-1)~p~3.41(-2)  &    &    &         & \multicolumn{1}{c}{ICF(Ne)}   & \multicolumn{1}{c}{1.00}      \\ 
  &    &    &    &      \textit{1.08(--1)}~p~\textit{1.92(--2)}  &    &    &     &    &      {\bf 1.41(-4)} ~p~{\bf 6.67(-6)}   \\ 
  &  He$^{2+}$  &  4685.76\,{\AA}   & 8.201~p~0.284 &  6.48(-3)~p~2.58(-3)  &  Si  &  Si$^{+}$  &  34.81\,{\micron}  & 12.287~p~1.143 &  1.59(-6)~p~1.49(-7)   \\ 
  &    &     &\multicolumn{1}{c}{ICF(He)}        &\multicolumn{1}{c}{1.00} &     &    &    & \multicolumn{1}{c}{ICF(Si)}    &         6.80~p~1.75 \\ 
  &     &     &    &      {\bf 1.15(-1)}~p~{\bf 1.94(-2)}  &    &    &     &    &            {\bf 1.08(-5)}~p~{\bf 2.96(-6)}   \\ 
C  &  C$^{+}$  &  157.64\,{\micron}  & 15.915~p~1.955 &  2.70(-4)~p~5.13(-5)  &  S  &  S$^{+}$  &  4068.60\,{\AA}   & 3.23~p~0.706 &  1.31(-6)~p~8.23(-7)   \\ 
  &  C$^{2+}$  &  4267.26\,{\AA}   & 2.070~p~0.495 &  2.00(-3)~p~4.95(-4)  &    &     &  6716.44\,{\AA}   & 26.242~p~0.738 &  1.17(-6)~p~4.31(-7)   \\ 
  &   &     &\multicolumn{1}{c}{ICF(C)}    &     2.03~p~ 3.19(-1)  &    &     &  6730.82\,{\AA}   & 21.854~p~0.62 &  1.17(-6)~p~4.75(-7)   \\ 
  &     &     &    &      4.06(-3)~p~1.19(-3)^{\dagger}  &    &    &    &    &      \textit{1.19(--6)}~p~\textit{2.97(--7)}   \\ 
  &     &    &    &      {\bf 9.89(-4)}~p~{\bf 3.14(-4)}^{\dagger}  &    &  S$^{2+}$  &  6312.10\,{\AA}   & 1.698~p~0.45 &  5.78(-6)~p~1.12(-6)   \\ 
N  &  N$^{0}$  &  5198/200\,{\AA}   & 6.341~p~0.782 &  4.90(-5)~p~2.95(-6)  &    &     &  18.71\,{\micron}  & 46.998~p~3.277 &  5.87(-6)~p~4.37(-7)   \\ 
  &  N$^{+}$  &  5754.64\,{\AA}   & 6.638~p~0.194 &  6.57(-5)~p~5.50(-6)  &    &     &  33.47\,{\micron}  & 50.073~p~3.48 &  5.80(-6)~p~6.12(-7)   \\ 
  &     &  6548.04\,{\AA}   & 132.950~p~4.709 &  6.35(-5)~p~3.34(-6)  &    &    &    &    &      \textit{5.84(--6)}~p~\textit{3.50(--7)}   \\ 
  &     &  6583.46\,{\AA}   & 410.904~p~10.523 &  6.63(-5)~p~3.10(-6)  &    &  S$^{3+}$  &  10.51\,{\micron}  & 49.677~p~3.516 &  1.04(-6)~p~7.40(-8)   \\ 
  &     &  121.73\,{\micron}  & 7.880~p~0.973 &  5.36(-5)~p~1.20(-5)  &    &    &     & \multicolumn{1}{c}{ICF(S)}       & \multicolumn{1}{c}{1.00}       \\ 
  &     &  205.40\,{\micron}  & 1.607~p~0.257 &  5.44(-5)~p~2.00(-5)  &    &    &     &    &      {\bf 8.09(-6)}~p~{\bf 4.65(-7)}   \\ 
  &    &    &    &      \textit{6.46(--5)}~p~\textit{2.96(--6)}  &  Cl  &  Cl$^{2+}$  &  5517.72\,{\AA}   & 0.838~p~0.381 &  1.07(-7)~p~5.03(-8)   \\ 
  &  N$^{2+}$  &  57.32\,{\micron}  & 78.829~p~9.712 &  7.01(-5)~p~9.08(-6)  &    &  Cl$^{3+}$  &  20.30\,{\micron}  & 0.333~p~0.061 &  1.57(-8)~p~2.89(-9)   \\ 
  &   &     & \multicolumn{1}{c}{ICF(N)}       & 1.05~p~ 5.76(-2)  &    &   &     &\multicolumn{1}{c}{ICF(Cl)}    &     1.17~p~ 9.07(-2)   \\ 
  &    &     &    &      {\bf 1.42(-4)}~p~{\bf 1.27(-5)}  &    &    &  &  &   {\bf 1.43(-7)}~p~{\bf 6.01(-8)} \\ 
O  &  O$^{0}$  &  6300.28\,{\AA}   & 32.959~p~0.765 &  7.05(-5)~p~4.03(-5)  &  Ar  &  Ar$^{2+}$  &  7135.80\,{\AA}   & 22.349~p~0.707 &  2.45(-6)~p~2.73(-7)   \\ 
  &     &  6363.79\,{\AA}   & 10.804~p~0.301 &  7.23(-5)~p~4.14(-5)  &    &     &  7751.10\,{\AA}   & 5.351~p~0.277 &  2.45(-6)~p~2.90(-7)   \\ 
  &     &  145.50\,{\micron}  & 2.904~p~0.367 &  5.38(-4)~p~1.05(-4)  &    &     &  8.99\,{\micron}  & 22.867~p~1.731 &  2.43(-6)~p~1.94(-7)   \\ 
  &    &    &    &      \textit{1.04(--4)}~p~\textit{2.78(--5)}  &    &     &  21.82\,{\micron}  & 1.622~p~0.131 &  2.56(-6)~p~2.19(-7)   \\ 
  &  O$^{+}$  &  3726.04\,{\AA}   & 268.177~p~8.139 &  2.74(-4)~p~3.09(-5)  &    &    &    &    &      \textit{2.44(--6)}~p~\textit{1.39(--7)}   \\ 
  &     &  3728.82\,{\AA}   & 316.109~p~8.809 &  2.72(-4)~p~1.56(-5)  &    &  Ar$^{3+}$  &  4740.20\,{\AA}   & 0.671~p~0.262 &  2.08(-7)~p~8.32(-8)   \\ 
  &     &  7320/30\,{\AA}   & 11.625~p~0.382 &  2.77(-4)~p~5.01(-5)  &    &    &     &\multicolumn{1}{c}{ICF(Ar)}    &     1.17~p~ 9.07(-2)   \\ 
  &    &    &    &      \textit{2.72(--4)}~p~\textit{1.34(--5)}  &    &    &  &  &   {\bf 3.10(-6)}~p~{\bf 3.06(-7)} \\ 
  &  O$^{2+}$  &  4363.21\,{\AA}   & 5.187~p~0.343 &  2.79(-4)~p~4.35(-5)   \\ 
  &     &  4958.91\,{\AA}   & 274.612~p~4.087 &  2.78(-4) ~p~2.07(-5)      \\ 
  &     &  88.33\,{\micron}  & 190.944~p~23.417 &  2.78(-4) ~p~4.30(-5)      \\ 
  &    &    &    &      \textit{2.78(--4)}~p~\textit{1.71(--5)}     \\ 
  &  O$^{3+}$  &  25.88\,{\micron}  & 174.473~p~12.154 &  3.02(-5) ~p~2.10(-6)   \\ 
  &    &     &\multicolumn{1}{c}{ICF(O)}        & \multicolumn{1}{c}{1.00}  \\ 
  &    &     &    &      {\bf 5.81(-4)}~p~{\bf 2.19(-5)}  \\
\hline
   \end{tabular}
  \begin{minipage}{\textwidth}
\tablecomments{\footnotesize
   {The RL C abundance using the RL C\,{\sc ii}\,4267\,{\AA} line is 4.06(--3), and the {\it expected} CEL C abundance using the average C
 $^{2+}$(RL)/C$^{2+}$(CEL) ratio of $4.10\pm0.49$ among 58 PNe
 \citep{Otsuka:2011aa} is $9.89(-4)$.}
   {The ICF(X) value of the element ``X'' and the
   resulting elemental abundance, \mbox{X/H = ICF(X)\,$\times
   (\Sigma_{\rm m=1}$X$^{\rm m+}$/H$^{+})$} is shown \textbf{in bold}.}
   }
  \end{minipage}
  \end{table*}

\newpage

\section{Comparison of relative line fluxes, band fluxes, flux densities
 between the observation and the {Cloudy} model}

  \setcounter{table}{0}

      \begin{table}
      \renewcommand{\arraystretch}{0.85}
     \centering
\renewcommand{\thetable}{D\arabic{table}}
\footnotesize
 \tablewidth{\textwidth}      
\caption{The comparison between the observed and {Cloudy} model
  predicted line fluxes, band fluxes, and band flux
  densities. $\Delta\lambda$ indicates the bandwidth of each band. 
  \label{AT:model}}
\begin{tabular}{@{}ccD{.}{.}{-1}D{.}{.}{-1}ccD{.}{.}{-1}D{.}{.}{-1}@{}}
  \hline\hline
 $\lambda_{\rm lab}$
 & Line
 &\multicolumn{1}{c}{$I_{\rm model}$($\lambda$)}
 &\multicolumn{1}{c}{$I_{\rm obs}$($\lambda$)}
 & $\lambda_{\rm lab}$
 & Line
 &\multicolumn{1}{c}{$I_{\rm model}$($\lambda$)}
 &\multicolumn{1}{c}{$I_{\rm obs}$($\lambda$)}\\ 
 ({\AA})
 &
 &\multicolumn{1}{c}{($I$({\hb})=100)}
 &\multicolumn{1}{c}{($I$({\hb})=100)}
 & ({\micron})
 &
 &\multicolumn{1}{c}{($I$({\hb})=100)}
 & \multicolumn{1}{c}{($I$({\hb})=100)} \\ 
 \hline
 3726 & $[$O\,{\sc ii}$]$ & 269.526 & 268.117 & 2.12 & H$_{2}$ 1-0S(1) & 39.911 & 57.189 \\ 
3729 & $[$O\,{\sc ii}$]$ & 274.987 & 316.109 & 5.51 & H$_{2}$ 0-0S(7) & 13.856 & 30.064 \\ 
3750 & H12 & 3.048 & 5.048 & 6.11 & H$_{2}$ 0-0S(6) & 9.834 & 19.021 \\ 
3771 & H11 & 3.964 & 3.885 & 6.91 & H$_{2}$ 0-0S(5) & 43.037 & 56.328 \\ 
3798 & H10 & 5.293 & 5.055 & 8.02 & H$_{2}$ 0-0S(4) & 13.637 & 31.052 \\ 
3835 & H9 & 7.299 & 9.414 & 8.99 & $[$Ar\,{\sc iii}$]$ & 19.674 & 22.867 \\ 
3869 & $[$Ne\,{\sc iii}$]$ & 138.608 & 125.764 & 9.67 & H$_{2}$ 0-0S(3) & 24.196 & 25.792 \\ 
3889 & H8+He\,{\sc i} & 23.293 & 27.314 & 10.51 & $[$S\,{\sc iv}$]$ & 35.542 & 49.677 \\ 
3967 & $[$Ne\,{\sc iii}$]$ & 41.775 & 37.177 & 12.29 & H$_{2}$ 0-0S(2) & 2.723 & 5.314 \\ 
3970 & H7 & 15.876 & 20.340 & 12.81 & $[$Ne\,{\sc ii}$]$ & 20.408 & 14.802 \\ 
4026 & He\,{\sc i} & 2.693 & 2.932 & 15.57 & $[$Ne\,{\sc iii}$]$ & 164.303 & 234.571 \\ 
4069 & $[$S\,{\sc ii}$]$ & 10.332 & 3.230 & 17.04 & H$_{2}$ 0-0S(1) & 1.281 & 9.241 \\ 
4102 & H$\delta$ & 25.847 & 31.011 & 18.72 & $[$S\,{\sc iii}$]$ & 37.567 & 46.998 \\ 
4267 & C\,{\sc ii} & 0.261 & 2.070 & 20.33 & $[$C\,{\sc iv}$]$ & 0.232 & 0.333 \\ 
4340 & {\hg} & 46.714 & 47.863 & 21.86 & $[$Ar\,{\sc iii}$]$ & 1.430 & 1.622 \\ 
4363 & $[$O\,{\sc iii}$]$ & 6.757 & 5.225 & 25.90 & $[$O\,{\sc iv}$]$ & 102.632 & 174.473 \\ 
4471 & He\,{\sc i} & 5.725 & 5.099 & 33.46 & $[$S\,{\sc iii}$]$ & 39.736 & 50.073 \\ 
4686 & He\,{\sc ii} & 10.954 & 8.201 & 34.79 & $[$Si\,{\sc ii}$]$ & 22.045 & 12.287 \\ 
4713 & He\,{\sc i}+[Ar\,{\sc iv}] & 1.580 & 1.341 & 36.01 & $[$Ne\,{\sc iii}$]$ & 13.943 & 17.011 \\ 
4740 & $[$Ar\,{\sc iv}$]$ & 0.746 & 0.671 & 57.00 & $[$N\,{\sc iii}$]$ & 47.680 & 78.829 \\ 
4861 & {\hb} & 100.000 & 100.000 & 63.00 & $[$O\,{\sc i}$]$ & 282.115 & 33.175 \\ 
4959 & $[$O\,{\sc iii}$]$ & 272.773 & 274.612 & 88.00 & $[$O\,{\sc iii}$]$ & 121.725 & 190.944 \\ 
5199 & $[$N\,{\sc i}$]$ & 32.473 & 6.341 & 119.2 & OH & 0.561 & 0.590 \\ 
5518 & $[$Cl\,{\sc iii}$]$ & 0.783 & 0.838 & 119.4 & OH & 0.784 & 0.650 \\ 
5538 & $[$Cl\,{\sc iii}$]$ & 0.594 & 0.577 & 121.0 & $[$N\,{\sc ii}$]$ & 6.627 & 7.880 \\ 
5578 & $[$O\,{\sc i}$]$ & 1.257 & 3.510 & 146.0 & $[$O\,{\sc i}$]$ & 19.410 & 2.904 \\ 
5755 & $[$N\,{\sc ii}$]$ & 5.897 & 6.674 & 158.0 & $[$C\,{\sc ii}$]$ & 49.076 & 15.915 \\ 
5876 & He\,{\sc i} & 16.425 & 16.406 & 205.0 & $[$N\,{\sc ii}$]$ & 1.161 & 1.607 \\ 
6300 & $[$O\,{\sc i}$]$ & 76.664 & 32.959 & 289.1 & CO $J$=9-8 & 1.611 & 0.528 \\ 
6312 & $[$S\,{\sc iii}$]$ & 2.426 & 1.698 & 325.3 & CO $J$=8-7 & 1.480 & 0.473 \\ 
6364 & $[$O\,{\sc i}$]$ & 24.449 & 10.804 & 370.3 & $[$C\,{\sc i}$]$ & 0.082 & 0.354 \\ 
6548 & $[$N\,{\sc ii}$]$ & 136.031 & 132.950 & 371.6 & CO $J$=7-6 & 1.221 & 0.969 \\ 
6563 & {\ha} & 283.796 & 286.124 & 433.5 & CO $J$=6-5 & 0.861 & 0.241 \\ 
6583 & $[$N\,{\sc ii}$]$ & 401.428 & 410.904 & 520.3 & CO $J$=5-4 & 0.477 & 0.374 \\ 
6678 & He\,{\sc i} & 4.629 & 4.405 & 650.3 & CO $J$=4-3 & 0.185 & 0.182 \\ 
6716 & $[$S\,{\sc ii}$]$ & 72.423 & 26.242 &  &  &  &  \\ 
6731 & $[$S\,{\sc ii}$]$ & 69.778 & 21.854 &  &  &  &  \\ 
7065 & He\,{\sc i} & 3.277 & 3.844 &  &  &  &  \\ 
7136 & $[$Ar\,{\sc iii}$]$ & 21.948 & 22.349 &  &  &  &  \\ 
7282 & He\,{\sc i} & 0.871 & 0.688 &  &  &  &  \\ 
7320 & $[$O\,{\sc ii}$]$ & 8.122 & 6.490 &  &  &  &  \\ 
7330 & $[$O\,{\sc ii}$]$ & 6.481 & 5.396 &  &  &  &  \\ 
7751 & $[$Ar\,{\sc iii}$]$ & 5.296 & 5.351 &  &  &  &  \\ 
\hline
\end{tabular}
      \end{table}
 \begin{table*}
  \centering
\renewcommand{\arraystretch}{0.85}
\renewcommand{\thetable}{D\arabic{table}}
\footnotesize
 \tablewidth{\textwidth}     
 \setcounter{table}{0}
  \caption{Continued.}
  \begin{tabular}{@{}ccD{.}{.}{-1}D{.}{.}{-1}ccD{.}{.}{-1}D{.}{.}{-1}@{}}
\hline
 $\lambda_{c}$($\Delta\lambda$)
 & Band
 & \multicolumn{1}{c}{$I_{\rm model}$($\lambda$)}
 & \multicolumn{1}{c}{$I_{\rm obs}$($\lambda$)}
 & $\lambda_{c}$($\Delta\lambda$)
 & Band
 &\multicolumn{1}{c}{$I_{\rm model}$($\lambda$)}
 &\multicolumn{1}{c}{$I_{\rm obs}$($\lambda$)}\\ 
 ({\micron})
 &
 &\multicolumn{1}{c}{($I$({\hb})=100)}
 &\multicolumn{1}{c}{($I$({\hb})=100)}
 & ({\micron})
 &
 &\multicolumn{1}{c}{($I$({\hb})=100)}
 &\multicolumn{1}{c}{($I$({\hb})=100)} \\ 
\hline
0.2274(0.073) & NUV & 1032.146 & 3045.783 & 13.20(0.30) & IRS-g & 3.243 & 3.244 \\ 
0.3595(0.056) & $u$ & 1285.318 & 1921.023 & 14.00(0.20) & IRS-h & 2.534 & 2.974 \\ 
0.464(0.116) & $g$ & 1793.895 & 1577.160 & 14.65(0.20) & IRS-i & 1.643 & 2.389 \\ 
0.5423(0.088) & $V$ & 1149.396 & 859.786 & 16.50(0.40) & IRS-j & 4.326 & 4.913 \\ 
0.6122(0.111) & $r$ & 1364.283 & 1084.993 & 17.50(0.30) & IRS-k & 3.419 & 3.495 \\ 
0.6441(0.170) & $R$ & 1709.897 & 1434.254 & 18.30(0.20) & IRS-l & 2.342 & 2.626 \\ 
1.235(0.162) & $J$ & 81.943 & 53.230 & 19.75(0.70) & IRS-m & 9.667 & 10.167 \\ 
1.662(0.251) & $H$ & 41.969 & 23.302 & 20.00(0.30) & IRS-n & 4.221 & 4.467 \\ 
2.159(0.262) & $K$ & 65.999 & 34.696 & 21.00(0.30) & IRS-o & 4.773 & 4.956 \\ 
3.353(0.663) & W1 & 36.711 & 39.400 & 22.50(0.40) & IRS-p & 7.625 & 7.643 \\ 
4.50(0.86) & IRAC-2 & 22.251 & 76.892 & 23.50(0.40) & IRS-q & 8.268 & 8.028 \\ 
5.80(1.26) & IRAC-3 & 71.323 & 134.290 & 27.00(0.40) & IRS-r & 10.790 & 9.471 \\ 
8.00(2.53) & IRAC-4 & 194.977 & 183.207 & 28.00(0.50) & IRS-s & 14.412 & 12.840 \\ 
7.70(0.30) & IRS-a & 20.084 & 28.010 & 29.00(0.50) & IRS-t & 15.044 & 13.318 \\ 
8.60(0.20) & IRS-b & 6.274 & 4.248 & 30.00(0.50) & IRS-u & 15.787 & 14.200 \\ 
9.35(0.15) & IRS-c & 1.497 & 0.622 & 31.00(0.50) & IRS-v & 16.379 & 14.469 \\ 
10.90(0.20) & IRS-d & 1.921 & 2.010 & 32.00(0.50) & IRS-w & 16.988 & 15.703 \\ 
11.30(0.50) & IRS-e & 10.094 & 8.517 & 35.40(0.20) & IRS-y & 7.830 & 7.571 \\ 
12.00(0.20) & IRS-f & 2.414 & 2.836 &  &  &  &  \\ 
\hline
 $\lambda$
 & Band
 &\multicolumn{1}{c}{$F_{\nu}$(model)}
 &\multicolumn{1}{c}{$F_{\nu}$(obs)}
 & $\lambda$
 & Band
 &\multicolumn{1}{c}{$F_{\nu}$(model)}
 &\multicolumn{1}{c}{$F_{\nu}$(obs)} \\ 
 &
 &\multicolumn{1}{c}{(Jy)}
 &\multicolumn{1}{c}{(Jy)}
 &
 &
 &\multicolumn{1}{c}{(Jy)}
 &\multicolumn{1}{c}{(Jy)} \\ 
\hline
17.0\,{\micron} & IRS-1 & 1.371 & 2.688 & 300.0\,{\micron} & SPIRE-2 & 13.250 & 14.770 \\ 
20.0\,{\micron} & IRS-2 & 2.386 & 5.421 & 350.0\,{\micron} & SPIRE-3 & 9.506 & 14.560 \\ 
30.0\,{\micron} & IRS-3 & 12.170 & 10.510 & 400.0\,{\micron} & SPIRE-4 & 7.022 & 8.539 \\ 
70.0\,{\micron} & PACS-1 & 54.970 & 55.360 & 450.0\,{\micron} & SPIRE-5 & 5.411 & 5.551 \\ 
80.0\,{\micron} & PACS-2 & 58.710 & 60.520 & 43GHz/7mm & Radio-1 & 0.378 & 0.710 \\ 
100.0\,{\micron} & PACS-3 & 60.610 & 65.810 & 30GHz/1cm & Radio-2 & 0.395 & 0.264 \\ 
110.0\,{\micron} & PACS-4 & 57.910 & 66.010 & 22GHz/1.3cm & Radio-3 & 0.409 & 0.190 \\ 
130.0\,{\micron} & PACS-5 & 50.590 & 61.740 & 5GHz/6cm & Radio-4 & 0.481 & 0.323 \\ 
170.0\,{\micron} & PACS-6 & 36.820 & 36.800 & 1.4GHz/21cm & Radio-5 & 0.531 & 0.377 \\ 
250.0\,{\micron} & SPIRE-1 & 18.940 & 22.410 &  &  &\\
 \hline
\end{tabular}
 \end{table*}

\clearpage

    \begin{figure}
     \setcounter{figure}{0}
     \centering
    \renewcommand{\thefigure}{D\arabic{figure}}
   \includegraphics[scale=0.8]{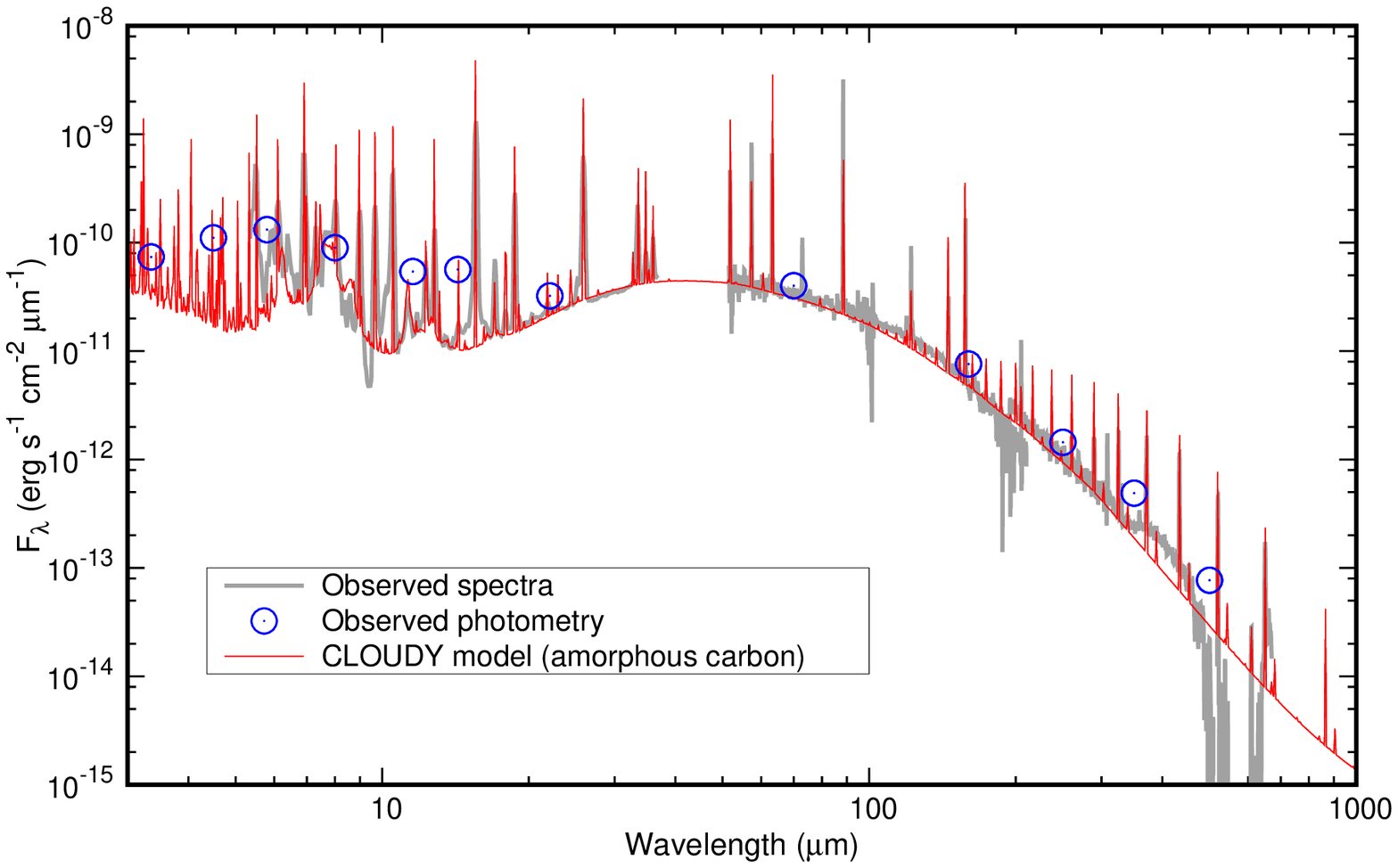}\\
   \includegraphics[scale=0.8]{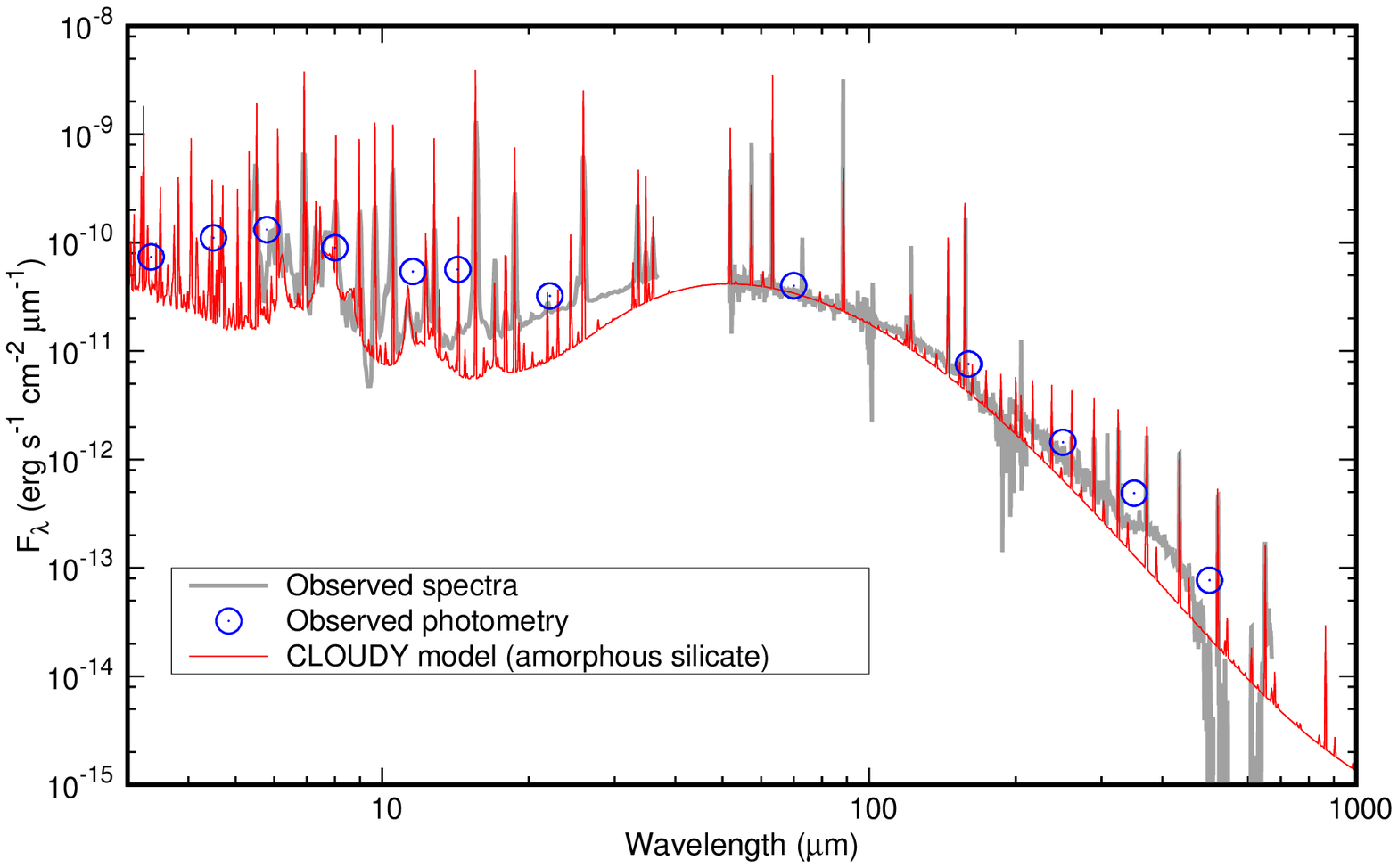}
     \caption{The best-fit model SEDs:
     [Top Panel] with amorphous carbon grains only, and
     [Bottom Panel] with amorphous silicate grains only. 
     The amorphous carbon grain model gives 
     better fitting to the observed continuum fluxes
     than the amorphous silicate grain model, 
     especially in the mid-IR ($10-40\,\micron$)
     and far-IR ($> 70\,\micron$).}
     \label{F-sil}
    \end{figure}

\end{document}